\documentclass[a4paper,12pt,titlepage]{report}
\usepackage[british]{babel}
\usepackage{listings}
\usepackage{xcolor}
\usepackage{pdflscape}

\definecolor{codegreen}{rgb}{0,0.6,0}
\definecolor{codegray}{rgb}{0.5,0.5,0.5}
\definecolor{codepurple}{rgb}{0.58,0,0.82}
\definecolor{backcolour}{rgb}{0.95,0.95,0.92}

\lstdefinestyle{mystyle}{
    backgroundcolor=\color{backcolour},   
    commentstyle=\color{codegreen},
    keywordstyle=\color{magenta},
    numberstyle=\tiny\color{codegray},
    stringstyle=\color{codepurple},
    basicstyle=\footnotesize,
    breakatwhitespace=false,         
    breaklines=true,                 
    captionpos=b,                    
    keepspaces=true,                 
    numbers=left,                    
    numbersep=5pt,                  
    showspaces=false,                
    showstringspaces=false,
    showtabs=false,                  
    tabsize=2
}
 
\usepackage{titlesec}
\usepackage{nameref}
\usepackage{geometry}
\usepackage[pdftex]{graphicx}
\usepackage{comment}
\usepackage{amsmath,amssymb,amsthm}
\usepackage{tabularx}
\usepackage{times}
\usepackage{multirow}
\usepackage{booktabs}
\usepackage{url}
\usepackage[hyphenbreaks]{breakurl}
\usepackage[pdftex]{hyperref}
\fussy

\title{Modelling and Analysing Behaviours and Emotions via Complex User Interactions}
\date{May 2018}
\author{Mohamed Mostafa Mohamed Sayed Ahmed}

\begin{document}
\maketitle

\pagenumbering{roman}

\cleardoublepage
\phantomsection
\addcontentsline{toc}{chapter}{Abstract}
\begin{abstract}
Over the past 15 years, the volume, richness and quality of data collected from the combined social networking platforms has increased beyond all expectation, providing researchers from a variety of disciplines to use it in their research. Perhaps more impactfully, it has provided the foundation for a range of new products and services, transforming industries such as advertising and marketing, as well as bringing the challenges of sharing personal data into the public consciousness. But how to make sense of the ever-increasing volume of big social data so that we can better understand and improve the user experience in increasingly complex, data-driven digital systems. This link with usability and the user experience of data-driven system bridges into the wider field of human-computer interaction (HCI), attracting interdisciplinary researchers as we see the demand for consumer technologies, software and systems, as well as the integration of social networks into our everyday lives. The fact that the data largely posted on social networks tends to be textual, provides a further link to linguistics, psychology and psycholinguistics to better understand the relationship between human behaviours offline and online.\\

\noindent In this thesis, we present a novel conceptual framework based on a complex digital system using collected longitudinal datasets to predict system status based on the personality traits and emotions extracted from text posted by users. The system framework was built using a dataset collected from an online scholarship system in which 2000 students had their digital behaviour and social network behaviour collected for this study. We contextualise this research project with a wider review and critical analysis of the current psycholinguistics, artificial intelligence and human-computer interaction literature, which reveals a gap of mapping and understanding digital profiling against system status.\\

\noindent Through developing and applying a hybrid approach of data science and data analysis techniques to the datasets which ultimately led to the development of the novel conceptual model and {\emph{PMSys}} system. The empirical foundation and validation is underpinned by a chain of experiments exploring the association and interrelations between the key parameters, linking back to the wider literature, which is used to improve the response of the intelligent agents based on the reported errors, as well as predicting the emotions raised by the user and selecting the appropriate answer. By extracting the user's behaviour (\emph{personality traits and emotions}), the proposed conceptual model predicted 68\% of the system statuses (\emph{idle, down, slow and error}). Furthermore, a web-based application was developed to simulate events to users and to verify the framework; this model predicted  61\% of the system statuses.\\

\noindent Alongside the wider academic dissemination of this work, features of this novel model and system are currently being commercialised as part of an intelligent chatbot engine to provides a customer services support to a range of commercial clients across a variety of industrial sectors.
\end{abstract}

\renewcommand{\abstractname}{Acknowledgements}
\begin{abstract}
Firstly, I would like to express my sincere gratitude to my advisor Prof. Tom Crick MBE for his continuous support during my PhD study and related academic work -- for his patience, motivation, and immense knowledge. His guidance helped me throughout the research and writing of this thesis; I could not have imagined having a better advisor and mentor during this journey. Besides my advisor, I would like to thank my director of studies Dr Jason Williams for always jumping in to help anytime and through my academic career and for being a great head of a department. I would like to express my special appreciation to Dr Giles Oatley for his massive support through the early stages of my research and his guidance in my academic life. I would also like to thank Dr Ana Calderon for always being there when needed for her insightful comments and encouragement and continuous support on this journey at all levels.

My sincere thanks also go to Dr Yasser Elshayeb and Dr Ehab Abdelrahmen, for their support during my early career start and for always being there, I am grateful for their advice and support.

I would like to dedicate this thesis to my father, who always believed in me and pushed me forward towards my goals, his guidance and encouragement were always invaluable, I am forever grateful to him (may his soul rest in peace). My sincere thanks to my mother, her prayer to me is what sustained me thus far. My brothers (Hossam and Ahmed) and sisters (Hala and Heba) and all my family members for their continuing support during my PhD journey.

Last but not least,  words cannot express how grateful I am to my beloved wife Marwa and my two darling daughters, I cannot thank her enough for her sacrifices, companionship, love, support and encouragement you have provided in every minute of this journey. I am beyond grateful to you -- without your precious support it would not be possible to finally complete this chapter of my life.
\end{abstract}

\cleardoublepage

\tableofcontents

\listoffigures
\phantomsection
\addcontentsline{toc}{chapter}{List of Figures}

\listoftables
\phantomsection
\addcontentsline{toc}{chapter}{List of Tables}

\renewcommand\lstlistlistingname{List of Code Listings}
\lstlistoflistings
\phantomsection
\addcontentsline{toc}{chapter}{List of Code Listings}

\pagenumbering{arabic}
\cleardoublepage

\chapter{Introduction}

\section{Overview}

Marc Andreessen -- co-author of Mosaic, the first widely used web browser -- boldly stated in 2011 that ``software is eating the world''~\cite{Andreessen2011WhyWorld}, with software and technology playing a huge role in our daily life, from communication, entertainment, education and across the economy. Changes in the software engineering domain have occurred just as rapidly, with the increased requirements on the responsiveness, robustness, and usability of software. Furthermore, we are seeing the emergence of artificial intelligence and machine learning as general-purpose technologies that could transform whole industries and even re-invent the process of invention itself~\cite{brynjolfsson-et-al:2018} -- with potentially serious positive and negative consequences~\cite{brundage-et-al-aireport:2018}; so, does this raise the question: will AI eat software?

In the broader software realm, there is the grand question of ``{\emph{how will users react to this application?}}''. In the user experience and software design research domains, there are some intersecting strands of research and development in an attempt to address this question. Researchers have previously used a range of different approaches to measuring the satisfaction of users and how they behave when using their products. However, this currently does not provide the necessary level of insight into behaviours and emotions of users; more so with the increasingly complex software systems that we encounter today. 

Our world has an ever-increasing economic, educational and socio-cultural dependency on data, technology and computation -- and thus interconnected computer systems -- from smart cities and \{big,open,personal\} data~\cite{cosgrave-et-al:2014,Tryfonas-Crick,tryfonas+crick:petra2018,cooper-et-al:jut2019}; supporting advancements in science and engineering~\cite{crick-et-al:2009a,crick-et-al_wssspe2,crick-et-al_recomp2014,crick-et-al_jors,tennant-et-al:f1000,tennant-et-al:bitss2019}; knowledge representation and reasoning~\cite{devos-et-al:2006,brain-et-al:2006b,crickphd:2009,crick-et-al:2009b}; insight into political systems, processes and policymaking~\cite{tumasjan-et-al:2010,wolfsfeld-et-al:2013,tan-et-al:2013,burnap-et-al-es:2016,delvicario-et-al:2017}; innovation and the economy~\cite{cooper-et-al-gsict2015,nesta:2015,carr+crick-csss2015,whicher+crick:pmm2019}; to developing digitally competent and capable citizens~\cite{wgictreview:2013,brown-et-al-toce2014,murphy-et-al:programming2017,moller+crick:jce2018,davenport-et-al:cep2019}. People use and interact with complex platforms and systems in various ways, providing insight into aspects of their personality, behaviour and emotional responses~\cite{Norman1963,schwartz-et-al:2013}. These multiple interactions can be measured, profiled and modelled to critically analyse the communications and related processes and provide valuable insight into improving system architectures and designs. Furthermore, it provides a predictive capability to better understand digital behaviours -- particularly through big social media datasets and corpora~\cite{lazer-et-al:2009,postsm:2014,mostafa-et-al-ai2016,burnap-et-al-es:2016,albishry-et-al:iccci2017,albishry-et-al:ssei2018,albishry-et-al:iccci2018} -- and thus develop systems that are more resilient and robust against undesirable behaviour and security breaches, such as ``insider threat'' scenarios~\cite{kang-et-al:2013,oatley+crick:2014,tryfonas-etal-has2016}, cyberhate~\cite{burnap+williams:2015,williams+burnap:2016}, as well as more generally for crime informatics~\cite{oatley+crick_asonam2014,burnap-et-al-snam:2014,oatley+crick_fosintsi2014,oatley+crick:2015}.

\begin{figure}[!ht]
\centering
\includegraphics[width=\textwidth,height=400px,keepaspectratio]{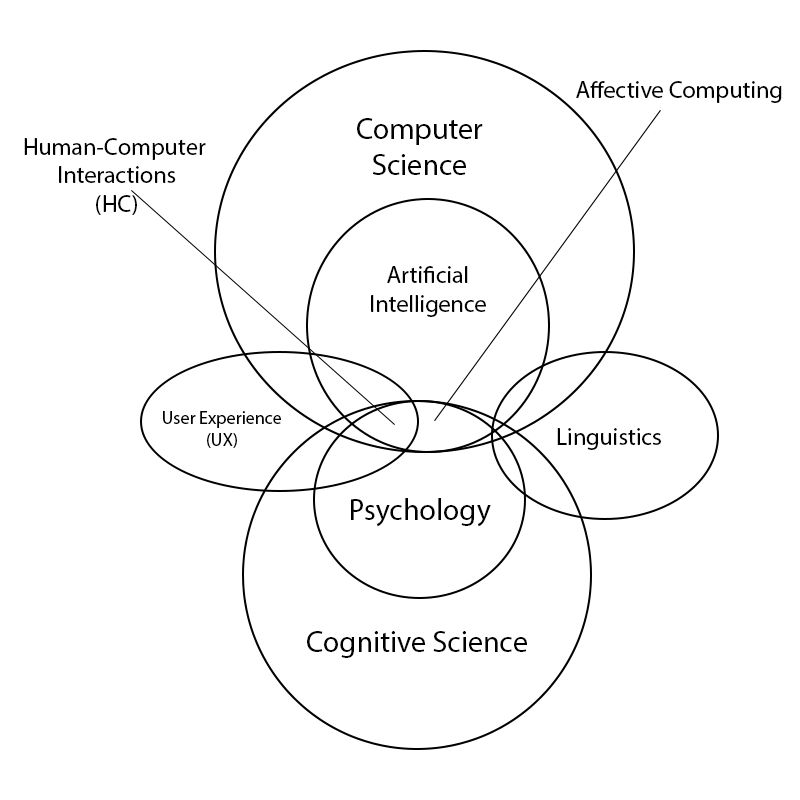}
\caption[Intersection of the key research areas for this thesis]{Intersection of the key research areas for this thesis (adapted from ~\cite{harrison2007three,picard1999affective,fulcher2008computational,isac2013language,card2017psychology})}
\label{fig:vennOverview}
\end{figure}

This research project is grounded in the area of human-computer interaction where it intercepts with the fields of artificial emotional intelligence and behavioural sciences under the umbrella of computer science (see Figure~\ref{fig:vennOverview}). It aims to develop new theoretical modelling of digital behaviour in complex computer systems -- namely, pervasive, data-driven social networks. Data were initially collecting and analysed from Facebook and through complex information systems, we demonstrate the intrinsic relationship between offline behaviour and online behaviour. Moreover, by investigating the relationship between personality and emotions encountered during different events, we can describe the associations between the traits which eventually led to extracting the fine-grained features used in the model. Thus, evidence that user's emotions correlate with different complex computer system status, giving us deeper, predictive insight into behaviours and sentiments. 

The primary objective of the human-computer interaction (HCI) research field is to ultimately make systems more useful and usable. A key part of this research and development improvement process is to understand user's behaviour by developing user-profiling. With the rise of social networks and ``big social data'' providing easily accessible large-scale datasets; furthermore, enhancing the user's interactions of the system, attracted researchers from different fields to develop robust methodologies to achieve this objective. Furthermore, HCI is increasingly an inter-, multi- and trans-disciplinary field, embracing aspects of artificial intelligence, cognitive science, psychology and user experience. This research project is an attempt to provide a solid theoretical foundation -- with practical validation and verification -- of profiling digital behaviour through the intersection of HCI with other key research areas (see Figure~\ref{fig:vennOverview}). This work is based on a unique and rich longitudinal dataset of user's interaction with a complex online computer system, supported by the use of social networking platform (Facebook) data and other modes of interaction (e.g. help desk ticketing system). Thus, with these varied data extracted from the complex system for each user, we are able to apply various hybrid data mining, data science, artificial intelligence and machine learning methodologies.

The large-scale longitudinal dataset consisting of interactions of applicants to a real-world online scholarship mobility platform. These interactions constitute part of the ``digital footprint'' of the users, including the important artefacts and communications, as well as the user interactions and activities during a predetermined system failure event. New conceptual models will be developed using these datasets and associated empirical data, making theoretical contributions to domain knowledge, as well as an adaptive framework for a range of real-world practical applications to some problem domains

\section{Contribution to Knowledge}

The primary contribution to knowledge from this thesis is the development of novel conceptual models and an adaptive framework to improve predictions of user interactions and behaviours. Thus, this can be widely applied to enhance system design and user experience for modern complex information systems across a number of domains. This will provide a deeper critical understanding of the relationship between user emotions, personality and behaviours and computer system interactions.

Specifically, this thesis: 

\begin{itemize}
\item
Critically reviews the state-of-the-art domain research, contextualised by the relevant human-computer interaction, artificial intelligence (including natural language processing, sentiment analysis, and machine learning) and human-computer interaction literature with regards to relationships between user behaviour, emotion and personality and their digital interactions and activities.
\item
Analyses and investigates the key relationships between emotion and personality and online/digital behaviour to develop new conceptual models.
\item
This research study, provides robust empirical validation and evidence for relationships between user's online behaviour and responses and associated personality/emotion, particularly during disruptive system events.
\item
Develops a novel adaptive framework to provide predictive capability for user's behaviour in various computer systems, providing new insight to enhancing the design, usability and user experience of complex computer systems across some domains.
\item
The is study, validates the new theoretical models and adaptive framework against some challenging problem sets in various domains; including user experience, software engineering, and system design.
\end{itemize}

\section{Publications}\label{mypubs}

To further reinforce the novel contributions to knowledge from this research project, both theoretical foundations and applications derived from this work have been published in (or submitted to) the following peer-reviewed journals and international conferences:

\begin{itemize}
\item J. Rafferty, {\textbf{M. Mostafa}}, T. Crick, G. Oatley, C. Ranson, and I. S. Moore. {\emph{Using machine learning to predict concussions in professional rugby union}}. Submitted to Artificial Intelligence in Medicine, 2019~\cite{rafferty-et-al_aiinmed} -- in this paper, various data science/machine learning techniques that have been used in this thesis has been applied to a novel application areas (sports injuries).
\item J. Rafferty, C. Ranson, G. Oatley, {\textbf{M. Mostafa}}, P. Mathema, T. Crick, and I. S. Moore. {\emph{On average, a professional rugby union player is more likely than not to sustain a concussion after 25 matches}}. British Journal of Sports Medicine, 2018\\ {\footnotesize{\url{https://doi.org/10.1136/bjsports-2017-098417}}}~\cite{rafferty-et-al_bjsm} -- in this paper, various data science/machine learning techniques that have been used in this thesis has been applied to a novel application areas (sports injuries).
\item {\textbf{M. Mostafa}}, T. Crick, A. C. Calderon, and G. Oatley. {\emph{Incorporating Emotion and Personality-Based Analysis in User-Centered Modelling}} in Research and Development in Intelligent Systems XXXIII, pp. 383--389, Springer, 2016\\{\footnotesize{\url{https://doi.org/10.1007/978-3-319-47175-4_29}}}~\cite{mostafa-et-al-ai2016} -- this paper directly relates to the studies and outputs from this thesis, in particular as presented in Chapter~\ref{pmsys}.
\item G. Oatley, T. Crick, and {\textbf{M. Mostafa}}. {\emph{Digital Footprints: Envisaging and Analysing Online Behaviour}} in Proceedings of 2015 Symposium on Social Aspects of Cognition and Computing Symposium (SSAISB), 2015\\ {\footnotesize{\url{https://cronfa.swan.ac.uk/Record/cronfa43383}}}~\cite{oatley-et-al-soccogcomp2015} -- this paper directly relates to the building of the conceptual model, as presented in Chapter~\ref{modellingphase}.
\end{itemize}

\section{Ethics Approval}

The main dataset has been approved to be used as part of this research project by the registered data owner: the International Office at Cardiff Metropolitan University. The wider PhD study and in particular, the verification experiment presented in Section~\ref{lbl:verificationData}, was approved by Cardiff Metropolitan University's Ethics Committee (CSM: 2015S0054).

\section{Thesis Outline}

Following on from this introduction in {\textbf{Chapter 1}, the rest of the thesis is structured as follows, contextualised by three literature review chapters:

\begin{description}
\item[Chapter~\ref{BehaviourandEmotions}] provides the theoretical background and critical review of the psycholinguistics and personality theory literature, focusing on the Big Five personality traits as it plays a vital role in the study. Furthermore, it provides an in-depth background to the Linguistic Inquiry and Word Count (LIWC) approach from a computer science perspective, and critically reviews the current literature in the field of LIWC and personality traits. Moreover, we present the key interaction with cognitive science, highlighted the state of art in emotions and temporal behaviour with respect to the wider human-computer interaction field. 

\item[Chapter~\ref{hci}] builds from the previous chapter to provide a deeper discussion of cognitive science's role in human-computer interaction (HCI); this chapter critically reviews the relevant research themes and provides a summary of HCI, user experience and usability literature and background and state of art. Demonstrate and define the complex computer system and how human perceptual to different computer status.

\item[Chapter~\ref{ComputationalIntelligence}] reviews the history of natural language processing (NLP) and provides an in-depth background to data usage and applications. In addition, we provide a summary of relevant regressions, classifiers, and analyses used in this research study.

\end{description}

\noindent Then we present the main methodology and modelling, the underlying system and data architecture, as well as the key experiments:

\begin{description}

\item[Chapter~\ref{methodology}] gives an overview to the complex computer system used as part of this study from a technical and non-technical perspective. In addition, it provides a detailed information about the data sources and how it been extracted and highlights the challenges and the developed application/tools used as part of the extraction process. Furthermore, this chapter explains the algorithm used to identify the status of the complex computer system and matching the users between Facebook and the system.  This chapter also defines the extraction of the personality traits and emotions and providing a rational of usage the Big Five Personality traits as part of the experiments. 

\item[Chapter~\ref{pmsys}] demonstrate list of experiments conducted as part of the study, structured to  provide a rationale and key contributions, as well as presenting summary findings and analysis.

\item[Chapter~\ref{modellingphase}] concludes the features extracted from all experiments presented in Chapter~\ref{pmsys}, focusing on the usage of the random forest tree classifier to build the model, demonstrating the main findings. Furthermore, it provides detailed information about the verification methodology used.

\end{description}

\noindent Finally, we present the key discussions in {\textbf{Chapter~\ref{disc}}} and the concluding remarks in {\textbf{Chapter~\ref{conclusion}}}, providing a summary of the key contributions and findings, as well as an overview of future work.


\newpage
\chapter{Personality, Behaviour and Emotions}\label{BehaviourandEmotions}

\section{Introduction}
This chapter presents a review and analysis of the key literature at the intersection of personality, behaviour and emotions. Personality traits make us who we are and that what attracted researchers from different disciplines to investigate how to describe the personality and human behaviour. For decades psychologists proposed different theories of personality that contributed to towards a better understanding of how to reform a footprint of a person~\cite{mcadams1990person}. In this chapter we present different personality theories~\cite{matthews2003personality}, demonstrating the development of those theories and how computers contributed to this development, drawing on an extraordinarily broad array of research from cognitive science and emotion science. Different research studies suggested that it is possible to infer emotion through the communication between people and computers~\cite{hibbeln2017your}, This chapter presents a broader view of the role of emotions in communication between people and computers in everyday life as~\cite{nass2007emotion}.

\section{Psycholinguistics}

Psycholinguistics is a branch of psychology that deals with and aims at studying the relationship between linguistics and psychology. It primarily studies the cognitive process that forms the foundation of language understanding and the interactions between social cultures, linguistics and psychology in a broader framework~\cite{Harley05howdo}. Researchers in this field try to obtain objective data regarding models of prediction of linguistic behaviours by various users of the language. The study attempt to find out the mental processes that are involved and used during language use~\cite{jodai2011introduction} psycholinguistics is crucial as it creates a platform through which language gets a chance to be processed, developed, used, broken down and interpreted.

\section{Personality Theories}

As with our genetic make-up, our personalities are unique and personal to each of us. Over several decades, the term ``personality" has not only been used to describe features of a person but also denoted the characteristic patterns of thinking coupled with their general feeling and acting towards others. Thus, the consistency of the way people think feels and act derives models. However, the theories of personality view people and try to associate their choices with the hereditary closeness that they share. The theories of personality are thus a tool for not only learning but also understanding and questioning personality as we currently understand it. There are four general personality theories, that is a psychoanalytic theory, trait theory, behavioural theory / social learning theory and lastly the humanistic view~\cite{Kasschau1980}. The discussed personality theories in this section determine the general definition of the personality as we know it; however, from a large population sample, regardless of how one views personality, it has to be denoted by specific elements. These elements are: stability of character, consistency, and uniqueness of personality~\cite{Kasschau1980}.

The psychoanalytic theory of personality was first described by Freud~\cite{blum1953psychoanalytic}, with the foundation of the theory built on human consciousness. According to Freud~\cite{perls1992ego}, human behaviour is a factor of human thoughts, ideas, and wishes to originate in the brain. Freud develops and expands this theory on the importance and position that is held by the unconscious mind, sexual aggression and early life experiences on a person’s personality~\cite{H.1953}. Furthermore, Freud implies that characters are such that some thoughts can be pushed off and not acted upon from the conscious state to the unconscious. He proposes that awareness exists in various forms; unconscious state, conscious state and pre-conscious state. During the conscious level, the experience is limited to present moments while the pre-conscious level reveals information one is currently unaware of but can immediately enter the conscious level.  We are unaware of the ideas and feelings are marked by of the last level. However, has a direct impact on the conscious mind~\cite{fairbairm2001}.

This version of personality was controversial at first, since it implied that the brain ``knew" things that the mind did not. Freud tried to explain the theory further by creating a clear structure to explain it~\cite{fairbairm2001}. This structure bases it is assumptions on the human hormone libido which fuels anger, aggression, and sexual anxiety. He proposes three structures that interact with each other: id, ego and superego. It is the primitive part of people that are susceptible to morality and social expectations; it is a self-centred behaviour and seeks to please itself. Ego, on the other hand, goes hand in hand with the reality principle, implies that it acts as a balance between human emotion and external social demands. The superego is part of the conscious self and is influenced by social upbringing and guidance on morality. Based on these three structures, a person's personality is defined. 

The second personality theory is the trait theory which was developed and analysed by Sheldon (1977)~\cite{Kenneth2013}, Allport (1937)~\cite{Allport1937} and Cattell (1943)~\cite{Cattell1943}. This theory is simple and specific elaborating that human behaviour is a factor of well-known effects from the organisms’ passed on traits and capabilities from the past \cite{Epstein1994}. This theory was labelled as a dependent theory since it relied heavily on past events and characterised personality because humans engage in actions which form patterns and judgements that a form a level of cross-situations uniformity.

The third theory which was the behavioural/social learning theory was from Bandura, Miller and Dollard and is commonly referred to as the social cognitive theory~\cite{bandura1977social}. It is termed and viewed as a factor of cognition, environmental factors, past and present and behaviour. Bandura puts forth the argument that social learning is directly related to experiences that a person has by observing the behaviours of other people. This form of learning is principled by the action-reaction constant. That is, the consequences or rewards derived from behaviour. This means that when confronted by situations, people have to choose what response to offer with regards to either the benefits or the consequence that will come from the after effects. The major limitation of this theory is that it fails to consider the conscious self and biological disposal as factors of personality and eventually behaviour~\cite{bandura1977social}.

The last theory presented here is the humanistic theory. To understand the human behaviour the theory asserts with regards to a person's internal perceptions of his or herself and others leading toward personal fulfilment. It was developed by Rogers and Maslow~\cite{Mahrer1989,Jung1921}. The central features of this theory are the personal conscious choice, freedom and free will which all lead to self-actualisation. This theory took over a new leaf as it addressed human behaviour in a manner suggestive of human growth and meaning. Maslow argues that self-actualisation directly influences human motives which in the end are all with physiological and transcended needs~\cite{Mahrer1989}, eventually presenting us with the now-famous Maslow's hierarchy of needs~\cite{mcleod2007maslow}. Moreover, further leads to the conclusion that self-actualised individuals make decisions with more spontaneity, ease, creativity and enjoy the positive aspects of life more which is in line with Carl’s perspective on the actualisation tendencies. Although, there are other personality theories, the discussion above sets forth the principal theories regarding personality as a front. 

\subsection{Cattell's 16 Personality Factors}

Characters or behaviours associated with specific people define the Human traits. Traits are unique with regards to the degree in which we exhibit them at a personal level. Over 4000 different traits represent the Human behaviour and personality in its nature. In a bid to rationalise all the different traits into a simplified and meaningful manner, Cattell made a list of 16 traits that he referred to as dimensions of the human personality \cite{Cattell2008}. Cattell’s view of personality was that it reflected human being behaviour in certain situations. Leads to the face that, inferred the person's behaviour through a set of behaviours obtained from information about their personality traits. 

Cattell initially carried out statistical analysis to come up with data. He considered the experimental data, life data, and questionnaire data as a basis for his argument on the dimensions of human personality and eventually the 16 Personality Factors (16PF)~\cite{rosenthal1973cattell}. The arbitrary scale whose scores relate to specific personality type is the core of the 16PF is a personality assessment. The first scale traits are; vigilance, sensitivity, perfectionism, tension, reliance, open-mindedness, social assertiveness, warmth, rational, emotional firmness, pensiveness, tension, dominance, rule awareness, privateness, alertness and  perfectionism~\cite{Cattell2008}. Its use is primarily in determining personality types and holds the position of the most used model of personality test with each trait relating to testing subjects in different ways. 

The 16PF is a factor of primary and secondary level traits. The primary traits are the most important and most potent and form the basis of human personality and behaviour. Upon analysis of the first traits, the ``Big Five" or the global order of human traits emerged. The Big Five has been re-defined in recent times to include; control or lack of resistance, tough-mindedness or open-mindedness to ideas, extraversion or introversion to social life, high or low levels of anxiety and independence of self or accommodation towards others~\cite{Cattell2008}. Research suggests that a variety of other personality traits are all housed in the Big Five. For example, high anxiety/low anxiety is an umbrella to bold-shy, self-reliant –group-oriented, private-forthright, lively-serious and warm-reserved. The Big Five personality traits demonstrate into a more straightforward yet broader perspective~\cite{Primi2014}.

\subsection{The Enneagram of Personality}

The Enneagram of personality is a branch of human psychology based upon the nine personality types of humans. The original enneagram bases it is originality on some ancient traditions and can be traced as far as back to the mystical Judaism, Christianity, Islam, Buddhism and ancient Greek philosophy~\cite{riso1996personality}. Despite its broad rooted history, the enneagram’s representation of human nature still stands as all the traits it represents are still relevant. 

As a concept, enneagram was taught and introduced by Ichazo~\cite{ichazo1982interviews} and based on the three groups or three triads. The organisation of these triads is such that nine personality types are birthed from three personality types in each of the three groups. These three namely; 1-the instinctive triad, 2-feeling triad, 3-the thinking triad. As their names suggest, each one of these triads is representative of the associated traits such that a person who is oriented towards feelings and self-image belongs to the feeling triad~\cite{riso2000understanding}.  The feeling triad sets forth the helper; who is encouraging, demonstrative, possessive, the motivator; who is marked by ambition, pragmatism and image consciousness and the individualist; who is sensitive, self-absorbed, and depressive. The thinking triad, on the other hand, is composed of the investigator; perceptive, cerebral, provocative, the loyalist; the active, dutiful and suspicious type and finally the enthusiast who are spontaneous, fun-loving and excessive. The last triad, instinctive, we find the leader who shows assertiveness, self-confidence and very confrontational. This triad also represents the peacemaker who is pleasant, easy-going and very complacent. The last trait representation of this triad is the reformer who shows excellent heights of rationality, idealistic nature and orderly~\cite{riso1996personality}. 

\subsection{Analytical Psychology (Jungian)}

Analytical psychology or analytic psychology is a school of thoughts and teaching that trace its roots from psychiatrist Carl Jung. Carl’s teachings and concepts idealise and emphasise on psychology on a personal front at achieving wholeness~\cite{stevens2001jung}. The most important concepts according to his teachings were archetypes, complexes, the persona, the shadow, the anima and animus, individuation, symbols, collective unconsciousness, and finally the unconscious.  Jung developed the concept of pattern archetypes which elaborated human complexes from his research on the human-response relationship. The relationship bases its assumptions on previous experiments that showed if a person takes a long time to respond to specific words generally read to them in the form of a list, the person was most likely experiencing a complex in the form of experience~\cite{stevens2001jung}). For instance, someone with a mother complex might have faced a lot of early experiences with their mother.  

\subsection{Myers-Briggs Personality Types}

Perception as a concept includes the process of becoming aware of things. The Meyer Briggs Personality Type Indicator is a questionnaire that structured as a way of demonstrating various psychological alignments concerning their environmental perceptions~\cite{Huber2017}. It was developed and constructed by Katherine Cook Briggs and her daughter and considered as an extension of the Jungian theory of human experience is a retrospect of sensation, intuition, feeling and thinking~\cite{Huber2017}.

\section{The ``Big Five'' Personality Traits}\label{bigfive}

In the late 19th century, Galton~\cite{Galton1907} solved one of the fundamental problems at the time in the research of psychology, which is how to represent classification of human character traits based on the ``lexical hypothesis". In 1884, Galton~\cite{Galton1907}, estimated the personality characters traits in respect of the English Dictionary and in 1936, Allport and Odbert, were the first psychologists to put Galton's hypothesis into practice, by extracting adjectives that they believed would describe the personality traits from the language style~\cite{Allport1936}. Cattell continued the work conducted by Allport and Odbert, and eliminated synonyms to reduce the total to 171 instead of 4,504~\cite{Bagby2005}. In 1940, Raymond~\cite{Cattell1943} constructed a self-report methodology for the classifying the personality traits found from the adjectives, which he later called  the \emph{Sixteen Personality Factor Questionnaire}. Cattell limited the personality dimensions into 20 out of 36. Later in 1961, Christal and Tupes suggested that all personality traits can be shorted to only five broad factors which \emph{surgency}, \emph{agreeableness}, \emph{dependability}, \emph{emotional stability} and \emph{culture}~\cite{Tupes1992}. The trait \emph{dependability} has been relabelled to \emph{conscientiousness} according to previous study by Norman in 1936~\cite{Norman1963}.

\begin{figure}[!ht]
\centering
\includegraphics[width=\textwidth,height=400px,keepaspectratio]{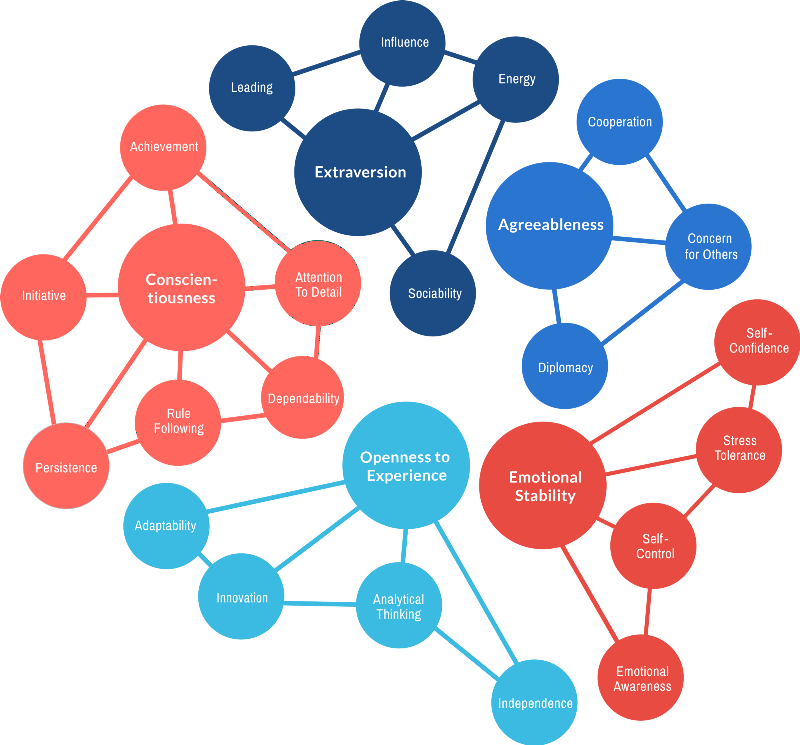}
\caption[The ``Big Five" personality traits, describing individual personality differences]{The ``Big Five" personality traits, describing individual personality differences (adapted from ~\cite{John1999,McCrae1992,Costa1992,costa+mccrae:1992})}
\label{fig:bigfivetraits}
\end{figure}

The ``Big Five" personality traits have been the centre of attention of the psychology used to describe the constituent features of personality in the field of contemporary psychology~\cite{Goldberg1981,Goldberg1990,Costa1992,John1999}, there has been consensus on its superiority by researchers in the field of organisational behaviour upon its performance for two decades. The title is an emphasis, not on the underlying greatness but in the broadness of the select factors in the model. The Big Five personality traits are typically assessed along the constituent five dimensions -- see Figure~\ref{fig:bigfivetraits}.

As per John and Srivastava (1999)~\cite{John1999}, every one of the five dimensions constitutes character at the broadest level of abstraction, and each size summarises a number of distinct, extra and precise personality traits. A study conducted by McCrae in 2002~\cite{McCrae2002}, suggested that personality traits curve does not change over time.

Studies have deduced several characteristics of the Big Five traits \cite{McCrae1992}. First, data suggest that the characteristics can be termed as universal due to their occurrence where there are personality tests involving traits in various languages~\cite{McCrae1997}. Secondly, along with the consideration of the role of genetics or biological factors and environment~\cite{Plomin1990,Digman1989}, research findings show the Big Five traits are highly stable over time~\cite{Gosling2003}. Lastly, the Big Five personality trait model has been accredited as the basic discovery model of personality psychology due to its history of usage, use across different cultures and the empirical evidence from several methods and experimentation~\cite{McCrae1992}. 

According to Costa and McCrae~\cite{Costa1992}, the Big Five personality dimensions can be divided into five factors namely; extraversion, agreeableness, conscientiousness, openness, and neuroticism~\cite{costa+mccrae:1992,McCrae1992,John1999}. Extraversion as the emotional aspects of people characterised by positive feelings and inclination to seek the company of others. People in this dimension are cheerful and sociable while being assertive, optimistic ad naturally talkative. Besides their preference for company results to groups and they relish stimulation. People in this group experience positive effect such as energy, zeal, and excitement~\cite{costa+mccrae:1992,John1999}. On the other hand, Agreeableness is the predisposition to be trusting, compliant, caring, kind, and tender. Such persons have a favourable view of human kind. They are concerned about others and have a yearning to help others; in return they expect others to be helpful. Agreeable individuals are pro-social and have harmonious integration with the public~\cite{costa+mccrae:1992,John1999}.

The third dimension in the Big Five, Conscientiousness, describes the group of people who are very objective in life and show determination in their undertaking. They are responsible for an aspect of self-discipline that is manifested in their ability to work without supervision and surpass the expectations. According to John (1999)~\cite{John1999}, people in this category have a prescribed impulse control that results to goal oriented tasks making them think displays tendency to think before act, strictly follow the rules, regulations, and norms and observe the order of planning, organ sign and giving priority to some tasks over others. 

The fourth group defined by Openness which can be described as an individual's tendency to the imagination while at the time being original in thinking. They have a liking for art and are sensitive to beauty with an attachment to feelings~\cite{costa+mccrae:1992,John1999}. Due to their liberal thought, they are intellectually curious and are willing to entertain new ideas and unique values. Lastly, the neurotic dimension encompasses the individuals who have an array of emotions such as fear, nervousness, sadness, tension, anger, and guilt. These people are subject to emotional adjustment or stability and emotional maladjustment or neuroticism~\cite{costa+mccrae:1992}. 

The aspect of personality that makes its analysis complex is because it is an assortment of behavioural, temperamental, emotional and mental attributes that make a person unique. For instance, communication is a trait that depends on the behaviour, temperament, emotion and psychological status of a person~\cite{Peabody1989,goldberg:1990}. As such, a person's choice of words, the way they say it, their semantic content and their physical prompts creates a significant variation between people. As such, while evaluating a trait, the behaviour, temperament, emotion and mental status of a person should be considered.

By applying factor analysis to the lists of the trait adjectives under the five factors listed above, researchers have obtained the five personality traits~\cite{Norman1963,Peabody1989,goldberg:1990}. The analysis based on the lexical hypothesis~\cite{Allport1936}, which follows the principle that the most consistent personal varieties encoded into the language, and the more significant the difference, the more possible represented as a one word. Although there are some limitations and drawbacks of the Big Five model~\cite{Eysenck1991,paunonen+jackson:2000}, through experiments, the model has succeeded in becoming the standard evaluation technique in psychology. These experiments have shown that personality has a significant impact on task-related individual behaviour. For instance, the personal traits influence how leaders conduct their roles, workplace performance~\cite{Hogan2005}, attitude towards work tools and machinery~\cite{Sigurdsson1991}, sales~\cite{Furnham1999}, the effectiveness of instructors, trainers and observers~\cite{Rushton1987} and academic ability~\cite{Furnham1991,Komarraju2005}. 


\subsection{Linking Online Social Networks and Personality}

The global use of online social networks has increased dramatically over the last fifteen years, across all age demographics~\cite{onssocialmedia2017,persocialmedia2018}. In 2005, a study of social networking websites concluded that an approximate number of users using online social networks totalled 115 million~\cite{golbeck2005computing}. Five years later in 2010, 200 million users were actively using one social network: Twitter. In the process of building social networking profiles, users share a significant amount of detailed, personal, temporal and location-based information about their lives: likes, thoughts, interactions, check-ins and activities. Through self-description, status updates, pictures, videos, group membership and hobbies, much of a user’s emotional state -- and personality -- manifests through their social media profiles. More recently, studies have shown that online social networks have attracted billions of users worldwide, many of whom view online social interactions as a core part of their everyday lives. In a study conducted by Cosenza in 2012~\cite{cosenza2014social}, Facebook had more than two billion active user around the globe and it is the largest online social network, serving 127 countries~\cite{facebook_newsroom}. Further statistics shown by the China Internet Network Information Centre, the total number of users online social network users increased to nearly three billion in early 2013~\cite{Chen2016}.

Psychology researchers have attempted to better understand and explain personality more formally. After extended work to expand and verify commonly-accepted personality models, researchers have revealed relationships between The ``Big Five" personality traits and many types of behaviour. These studies have revealed a strong correlation between personality and psychological disorders~\cite{Saulsman2004}, job production~\cite{BARRICK1991} and well-being~\cite{Judge1999}.

One of the areas that attracted researchers is to Automatic recognition of personality traits based on linguistic analysis, in an attempt conducted by Alam~\cite{Alam2013} to classify the personality using text , provides a very good result of accuracy. Further work has revealed that the user's Facebook profile is a reflection of their real personalities, not self-idealisation~\cite{Back2010}. A 2011 study of Twitter~\cite{quercia2011our} presented an attempt of predicting a user's personality traits using three public parameters from Twitter profiles: \emph{following}, \emph{followers}, and \emph{listed counts}. Identifying these three values will lead to predicting the ``big five" traits of this user~\cite{sumner2012predicting}. In another study supporting Quercia's findings, Golbeck et al. (2011)~\cite{Golbeck2011}, applied the Big Five Personality traits to 279 twitter's users, and extracted their 2000 recent tweets, and were capable to produce a model that can correctly predict each of the ``big five" traits. The capability to predict personality traits has associations in different areas existing research has shown associations between personality traits and success in different levels of social communications (i.e, professional and personal relationships)~\cite{Celli2013}. Another research study, claims it is possible to use Facebook Likes to recognise and potentially predict personality traits, age, gender and sexual orientation~\cite{Kosinski2013}. 

\section{Language Analysis}

Studies based on the relationships between personality traits and language styles have presented a variety of results and outcomes. Certain studies have presented a strong correlation between personality traits and people's writing style~\cite{Harris2004,Jie2006,Dornyei2003}, on the other hand other studies, resulted that there are low significant relationship between personality and language usage~\cite{Cohen1996,Carrell1996}. Pennebaker~\cite{Pennebaker1999} asserts that despite the body of a message being similar, different people will express themselves differently and in unique styles. This is due to the observed defence in the way people speak or write, with varying accents and choice of words in writing. However, researchers have managed to trace word use in an attempt to form ``Linguistics Fingerprinting" for a generation. For instance, analysis of the text and use of words have been used in the 1800s during the wars to the differentiation between soldiers~\cite{Psychology2000} allowing capturing of the way people committed (both verbally and written) to be obtained as a form of a fingerprint. Another text analysis strategy used it the word-based counting system introduced in Stone (1996)~\cite{Pennebaker2003}, Dunphy (1981)~\cite{Broehl1981ContentAI}.

\subsection{Open Vocabulary Approaches}

Open vocabulary is a method of language analysis, which is a popular  approach within computational linguistics and relevant disciplines~\cite{o2011computational}. This approach is a data-driven methodology to the researcher where the dependent class representation typically used in linguistic research. One of the main differences between close and open vocabulary, that the open vocabulary methods use statistical and probabilistic methods to recognise related language patterns or topics. An example of an open-vocabulary method is topic modelling, which handles unsupervised clustering algorithms (i.e., latent Dirichlet allocation or LDA~\cite{blei2003latent}) to find possibly meaningful groups of words in big sample of natural language.

\subsection{Closed Vocabulary Approaches}

The most modern implementation of closed vocabulary analysis in cognitive science is the Linguistic Inquiry and Word Count (LIWC) tool~\cite{Pennebaker2007}, which automatically counts words relating to more than sixty predefined classifications. Such as positive sentiment (e.g., ``happy", ``love", ``nice"), Achievement (e.g., ``make", ``star", ``acquire"), articles (e.g., ``some", ``an") and Tentative words (e.g., ``possibly", ``reasonably", ``maybe").

The closed-vocabulary approach depends on researchers to define categories and psychological labelling~\cite{mulac2006gender}. Define category points to divide the dictionary into groups of words and assign each word to the affiliated group. For instance, a group of first-person singular words (e.g, ``i",``mine") or precautions (e.g, ``in",``at). Psychological labelling, by identify the semantic context if the word belongs to the positive or negative group~\cite{tausczik+pennebaker:2010}.

\subsection{The Linguistic Inquiry and Word Count (LIWC) Tool}\label{lit:liwc}

As mentioned in the previous section, the Linguistic Inquiry and Word Count (LIWC) tool is the most common applications of closed-vocabulary methods, it is broadly used for quantitative text analysis in the social sciences ~\cite{Pennebaker2001}. Although LIWC is able to detect features in text by quantifing them, which allows a text classification and predictions and open the door for further text analysis and modelling, it has been fundamentally used to detect word features that are informative of the underlying psychological states of an individual or group of people~\cite{Friederich2017}. LIWC was first produced to address content analytic issues in experimental psychology~\cite{Pennebaker2003}; nowadays, it has wider applications across different fields such as social science, computational linguistics, shopping, and health care. 

The core of the LIWC program is its dictionary engine, with the most modern version based on the default dictionaries, built up of a total of approximately 4,500 words and word stems. The default dictionaries commonly fall into one of four widespread language dimensions that are grammatical match to the word type, e.g. pronouns, numbers, ,articles social, emotions and cognitive.

The wider literature shows that there has been a focus on critique traits by new researchers in psychology~\cite{giles1990handbook,SANFORD1942,Scherer1979,Weintraub1989} which have utilised different units of analysis on the same~\cite{Pennebaker2003}. We have also identified several ways of computing the Big Five and other personality traits. Pennebaker's approach~\cite{Pennington2014} developed a method that derives the features from textual information, relying on the previous research and corresponds with the traditional procedures. For instance, by using a LIWC psycholinguistic dictionary, meaningful word categories can be obtained, and used to improve the user experience in interacting with computer systems~\cite{Hampson1999}.

The method inferred both values needed to compute the traits. For the values, coefficients between values and LIWC categories was missing from the recent literature. As such, the coefficients from the value scores from surveys with LIWC category scores and texts written by more than 800 individuals which was compared and analysed\cite{Hampson1999}. As for the needs, ground-truth scores used to acquired from a needs survey and text written by more than 200 users. Moreover, the textual features were calculated using a custom dictionary that was constructed from diverse text users. Therefore, using three ground-truth scores associated with the textual elements, the generated a statistical model to come up with the needs~\cite{Hampson1999}.

For the past decades, research into character tendencies increased rapidly to many researchers in different fields, with interest in technology that has the capability to understand people's personality and feelings~\cite{Domahidi2014}. A recent concern has been the relationship between online social behaviour and real-life behaviour; citing examples such as the Facebook and Twitter, researchers assert that the profiles in these accounts have relation to the personality of a person~\cite{quercia2011our}.

Previous studies to establish a correlation between posted interactions on social networks and personality traits suggested mixed evidence of correlations between social network usage and social anxiety. In a study of the behaviour characters conducted by Zahra~\cite{Rizvi2016}, personality is a critical aspect to be examined to better understand the behavioural influences of a sample population. Quercia (2011)~\cite{quercia2011our} argues that there is a strong correlation between personality and online social behaviour on platforms such as Facebook. By using the users' status texts, Zahra~\cite{Rizvi2016} suggested that the findings of the study shows that personality traits and social anxiety did not linked with high usage of the Facebook~\cite{Rizvi2016}. However, according to another study conducted by McCord~\cite{McCord2014}, there are positive correlations between the usage of Facebook and social anxiety, reflecting the emerging nature of this research domain.

A 1986 study conducted by Gottschalk~\cite{Gottschalk1986} shows that there is a relationship between the emotional and cognitive dimensions as observed from the way individuals talk and write. Moreover, empirical evidence over the last four decades shows that a person's choice of words and their utterances are linked to their physical and mental health~\cite{Gottschalk1986}. This has been substantiated by findings from an experiment that showed improvements in psychological and physical health when individuals wrote or talked sincerely about their emotional experiences. These findings can be explained by the text analysis that shows tendency by those who write to use words of positive emotion while avoiding or suing moderate words of negative emotion. Furthermore, the writers tend to benefit from increasing usage of cognitive words which invoke relatively high rates of positive sentiment~\cite{Pennebaker1997}. 

According to a 2004 study by Kendall~\cite{Kendall2004}, in an attempt to create a model that allows learning about the emotional, mental and physical relations between the oral and written texts of individuals, researchers developed the LIWC engine. LIWC has undergone developments since the first version which was developed as part of an exploratory study of language and disclosure. A further innovation, the second version, LIWC2001~\cite{pennebaker2001linguistic} was an upgrade of the dictionary alongside usage of more modern software. The latest versions LIWC2007 offers the merit of a more advanced software and dictionary options. However, despite the advancements, the software retains its initial design objective of analysing one or more language files faster and better while remaining transparent and flexible in its operations.

LIWC is largely applied in the field of social sciences to analyse texts quantitatively. Hence it counts the number of words~\cite{Pennebaker2007}. The software was initially developed to address content analytic issues in experimental psychology and identify word features that can be used to assess the primary hidden psychological states of the writer or speaker.  Moreover, by determining the elements in a text, it can classify it and predict the various associated behaviour outcomes which can after that BE used for psychological applications~\cite{Pennebaker2001}. 


\section{Cognitive Science}\label{cognitiveScience}

\subsection{Emotional Intelligence}

In his work ``What is emotion?"~\cite{Cabanac2002}, Cabanac argues that the world has so far not agreed or come to a consensus regarding a canonical definition of an emotion. However, he identifies a list of motivational states of consciousness that is largely dependant on pleasure. These states include: anger, fear, disgust, joy, sadness, and surprise. However, emotion can be regarded through the lens of events which humans get to experience hedonic content (i.e. pleasure or displeasure). For the majority of humans, emotions make up our everyday life, as very few people are in the position of being devoid of any emotions since they originate from the same part of the brain, the limbic system. Emotions may change over time as people mature, develop and encounter new experiences as they require different triggers, but they are constant. In evolutionary psychology, emotions are deemed necessary for normal day to day living whether they are positive or negative~\cite{Lewis2000}. Behaviour has been defined by the IRIS Center\footnote{The IRIS Center at Vanderbilt University in the USA is a national centre dedicated to improving education outcomes for all children, especially those with disabilities birth through age twenty-one, through the use of effective evidence-based practices and interventions: \url{https://iris.peabody.vanderbilt.edu}} as activities which people engage in and can be visually assessed, quantified and occurs in a repetitive manner. This definition, however, does not bring to the attention of the reader that behaviour is a direct factor of emotion and ultimately manifests through the pleasure principle~\cite{Hampson2012}. 

There are many types of emotions, from the basic emotions identified by Ekman and Friesen (2013) ~\cite{ekman2013emotion}, to a range of secondary emotions. One differentiating factor is the origin of these emotions. An analysis shows that a relationship between developing of the brain and the emotion~\cite{tronick1989emotions}. The basic emotion is the emotions a person has from birth and are less likely to change as one grows up since they already ingrained in the brain at infancy~\cite{Halle2016}. However, some of the basic emits can be modified by saying, for instance, changing the environment although the changes would not develop much over the life time. As for the secondary emotions, these are as a result of the child's experiences from infancy throughout life basing on social and environmental experiences~\cite{Tompkins1963,ekman:1984}. For instance, as one interacts with friends and peers, parents, siblings, caregivers, peers, etc, they will invoke and evoke different forms of emotions depending on the nature of the interaction. Thus, secondary emotions are a factor of how a child views his/her surroundings and how the environment treats the child.

There is one common characteristic of emotions whether basic or secondary: the limited nature of uncontrollably of emotions and that emotions will always result in its disturbance of the initial setup before the occurrence. If emotions were fully controllable them, we would be emotionless (Lewis). As such, Basic emotions such as \emph{fear}, \emph{anger}, \emph{joy}, \emph{sadness}, \emph{disgust}, \emph{interest}, and \emph{surprise} are all shown in early infancy expression and used by all cultures a basic emotions facial expressions are always done quickly without thought~\cite{Izard1992}. Thus, research supports the theorem that basic emotions are universal. Another theory of emotion is by Tomkins (1963)~\cite{Tompkins1963} and Ekman (1984)~\cite{ekman:1984} who are part of the cognitive revolution of the 1960s and 1970s and its states that emotions ``were discrete entities, separate from though interacting with other psychological systems including cognition" The theory had three components with the third being a consequent of the first to indent the subjective experience of emotions. The neurotic behavioural element is the most essential component~\cite{Tompkins1963,ekman:1984}. Supporting their research on Darwin (1872)~\cite{Darwin1872}, Tomkins and Ekman conducted tests on infants in the category of facial expressions across different cultures. The findings showed that the facial expressions in infants did not change regardless of different cultures, a reflection of the basic emotions neural features component. These were the basic emotions also known as the primary emotions~\cite{Duclos1989}. 

Primary emotions and behavioural sciences have widely studied emotions as an essential element of human nature~\cite{Shivhare2012,huysamen1994methodology}. Increasingly, this has been the case across the cognate field of computer science given the frontline applications of computers in the area of human interactions. Through the advanced concept of textual analysis. It has been possible to interpret the emotional aspect of communication into through computational linguistic. This has allowed the researcher to ensure moving detection through sue of the new concepts of textual analysis. However, there has been inadequately minimal efforts into detecting emotion from text. Shivhare (2012) assumes that its word appearance essentially represents emotional reaction of an input sentence~\cite{Shivhare2012}. 

\subsection{Self Assessment of Emotions}\label{selfassessment}

\begin{figure}[!ht]
\centering
\includegraphics[width=\textwidth,height=400px,keepaspectratio]{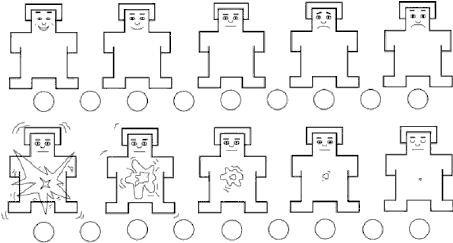}
\caption[The Self-Assessment Manikin (SAM)]{The Self-Assessment Manikin (SAM)~\cite{bergman2008emotion}}
\label{fig:Self-AssessmenSAM}
\end{figure}

Extracting emotions using self assessment is one of the most common approaches in the cognitive science, the Self-Assessment Manikin (SAM) (see Figure~\ref{fig:Self-AssessmenSAM} has been commonly used in the approach of self assessment, it was first introduced by Lang in 1980 \cite{Lang1980} and reported as a fast and simple method for self assessing emotions which can be used in different various contexts, the SAM as methodology have been used for decades~\cite{Bradley1994}, however, with the revolving usage of modern technology with advanced interfaces and methods of communication, it was essential to design a new methodology to support nowadays mindset\cite{MacKenzie2013}. The Affective Slider (AS) first appeared in 2016, and where the design of the slider was used to capture the emotions instead of the SAM methodology and produced more efficient in the usage of the AS~\cite{Betella2016}(see Figure~\ref{fig:SliderAS}).

\begin{figure}[!ht]
\centering
\includegraphics[width=\textwidth,height=400px,keepaspectratio]{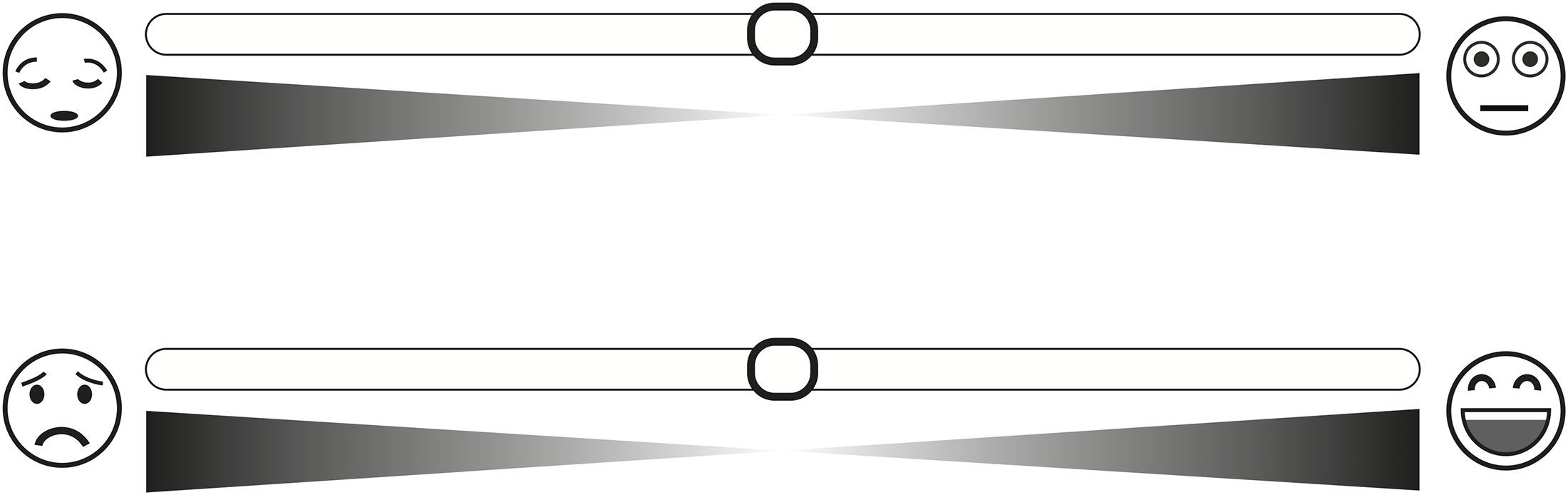}
\caption[The ``Affective Slider" (AS)]{The ``Affective Slider" (AS)~\cite{Betella2016}}
\label{fig:SliderAS}
\end{figure}

\subsection{Temporal Behaviour}

Temporal behaviour is characterised by a change in behaviour patterns over a short time. This is especially true with regards to the digital age which sees trend creation develop and die on the daily. This is elaborated by Roenneberg (2017)~\cite{Roenneberg2017} in his research on studying temporal behaviour on Twitter. Temporal behaviour is especially important in the marketing industry and can be taken advantage of to bring in numbers. Over the years, psychology as a field of study has focused more on scientific modelling and data collection of the field as compared to emotional intelligence importance and role~\cite{Mikolajczak2007}.However, over the past two decades, emotional intelligence has been categorised into ability and trait. The Trait Emotional Intelligence Questionnaire (TEIQue)~\cite{Petrides2001} is an emotional intelligence assessment developed to demonstrate how individuals understand their emotions and how it reflects their communications with others. TEIQue was originally developed by Petrides (2001)~\cite{Petrides2001} as part of his research at the University college of London. It offers variety in the type of audiences it can serve from grown-ups to children. It is widely used in the business industry as a tool of assessing employee's capability adapt and cope at the work place by assessing how intelligently they control their emotions. TEIQue assessment is an emotional intelligence questionnaire that assesses emotional management from a personal perspective to a social aspect which eventually leads to relationship management with self and the rest~\cite{Mikolajczak2007}.

\subsection{Applications of Cognitive Science}

As discussed in Section~\ref{bigfive}, the visual, vernal or written way of communication by an individual can pinpoint their personality traits. For instance, in the social media concept, a status update by a person when communication to the online community can use a wide range of forms of speech depending on factors such as age, gender or occupation. Though these updates, although primarily communicating, such a person can leave evidence of a particular personality trait(s).

In 1992, Costa conducted a research study on the possibilities of acquiring personality models of the Big Five personality traits through observation rather than directly asking the author or user~\cite{Costa1992}. The view is made on the linguistic input of a person by analysing his/her texts and conversation. The study also acknowledges other classification models that use the personality recognition in texts and blog postings. The results reported by Maitresse in 2006 were the first in the field to examine the identification of personality in dialogue, and to apply regression and ranking models that allow us to model personality recognition using the continuous scales traditional in psychology, also systematically examine the use of different feature sets, suggested by psycholinguistic research, and report statistically significant results~\cite{Mairesse2006}.

A 2007 research study conducted by Pennebaker~\cite{Pennebaker2007} determined which language style are significant of specific personality dimensions based on Facebook status updates from {\emph{myPersonality}}\footnote{{\emph{myPersonality}} was a popular Facebook application that allowed users to take real psychometric tests, and allowed recording -- with explicit consent -- their psychological and Facebook profiles.} as a dataset to built an engine that counts words in psychologically classified traits. The result of the research showed a capability to extract a different features, emotions traits, social tones, and personality difference, which was later assessed and verified by Appling in 2013~\cite{Appling2013}.

Another application is in using machine learning approach to make improvements or investigations about the proficiency of play in avenues such as computer gaming~\cite{dix2009human}. The research and development in this direction, however, has not been exhausted with little impact being observed. For instance, while the experience of play is entirely a psychological phenomenon, it is noted that game designs have not widely adopted useful psychological theories, such as the nature of dispositions~\cite{newell1985prospects,Cowley2013}.

Personality traits have been employed in the wider field of artificial intelligence (AI) where current studies have ventured into scrutinising methods to automatically infer other types of logical variations and differences in texts and conversations, such as emotion~\cite{oudeyerprosody2002a,Liscombe}, deception~\cite{Newman2003,Enos2006,Graciarena2006,Hirschberg05distinguishingdeceptive}, speaker charisma~\cite{Rosenberg2005}, mood~\cite{Mishne2005}, dominance in meetings~\cite{Rienks2006}, point of view or subjectivity~\cite{Wiebe2004,Wilson2004,Stoyanov2005,Somasundaran2007}, and sentiment or opinion~\cite{Breck2007,Popescu2005,Pang2005,Turney2002}. However, while the me consideration in AI maybe contextualised or short-lived, the component personality is given a long-term perspective and is viewed as a more stable aspect of individuals~\cite{Scherer2003}.

Further research in this area shows that personality also influences other aspects of linguistic production~\cite{Hampson2012,lambiotte2014tracking}. Strong relations have been observed between the Big Five dimensions with personality interacting and having an effect on them. For example, there is a clear-cut relation between the traits of the extraverts and the conscientious and the extraversion and conscientiousness traits and the constructive outcomes, and amongst neurotic individuals and those with disagreeableness~\cite{Watson1992}. Another case is personality ``vices" -- such as lying -- which contradict the dimension of agreeableness across modes such as visual and acoustic. It is such inconsistencies that create the avenue for human judgement's ability to figure out the cases of deception~\cite{Heinrich1998}. 

For instance, the individuals in the extraversion are outgoing and energetic with higher scores at deception while those in neuroticism are not good at lying. Moreover, the individuals in the agreeableness and openness category are good at identifying potential fraud~\cite{Riggio1988}. Similarly, extraverts have demonstrated excellent public speaking skills~\cite{Enos2006} with a study on those who dominate meetings showing the extrovert features. Thus, the functions used to pinpoint introversion and extraversion automatically can also be used to detect cases of deception automatically~\cite{Newman2003}.

Outgoing and active people -- extraverts -- are more successful at deception, while worried -- neurotic -- people are not as successful in this area~\cite{Riggio1988}, and people that score highly on the agreeableness and openness to experience characteristics are also skilled at identifying deception~\cite{Enos2006}. Newman (2003) and Bono (2004) claimed that features that used to automatically detect introversion and extraversion are also important for automatically identifying deception~\cite{Newman2003,Bono2004}, personality models can be applied to other uses to improve the accuracy of results~\cite{gou2014knowme}. For instance, opinion mining can use personality models to gain more valuable information. Also, the recognition of user personality in computer systems such as social platforms can be applied to other computer applications~\cite{Bono2004}.  Such areas include but not limited to; online dating platforms where character can be matched from analysis of the user text messages resulting to a more prosperous relationships, identifying the right leaders in meetings that require analytical skills through examining the personality dimensions of the candidates, ensuring that tutoring systems are tailored to fit the learner's personality traits and in language systems. Automatically recognising the author's persona in quantity, could also improve language conception, as the differences among people affected in the manner of expressing the concepts~\cite{Rienks2006,Oberlander2006}. 

Another major area of application is in language and conversation where automatically Identification of the author's persona in quantity, could also improve language conception, as the differences among people affected in the manner of expressing the concepts.  As per current research, there are only two studies on programmed identification of user personality~\cite{Hogan1994,Tucker2005,Nunn2005} in addition to our research.  These studies  have shown that users evaluation of conversational agents depends on their personality~\cite{Cassell2003}, which suggests a requirement for such systems to adapt to the user's personality like humans are able to do~\cite{McLarney-Vesotski2006,Funder1993}.

\section{Summary}

This chapter presented a critical review of the core psycholinguistic literature and the state of the art personality theories in the social science domain, summarising the strengths and limitations of personality theories as traits perspectives. A significant advantage of the trait perspectives is their capability to classify apparent behaviours. Several research studies reveal that observing the aggregate actions of people over time and in various situations provides substantial justification for the personality traits and categorised in the traits hypotheses~\cite{Donnellan2004}. Furthermore, the use of objective criteria for classification and identifying behaviours. A rationale for this that many trait theories (e.g. Big Five traits) were developed independently of others and eventually concluded the same classified personality traits~\cite{matthews2003personality, Petrides2001,gosling2003very, Donnellan2004}. On the other hand, other studies argue that trait theories provides a false prediction of the behaviour in some situation where involves other environmental factors, and that combination of trait and situation parameters impact of the behaviour~\cite{boyle2008critique}. For instance, an extroverted person is excited by social communications and tries out social situations, but trait theory does not offer any evidence for why this might happen or why an introvert would avoid such situations~\cite{cahill1992p300}.

Furthermore, this chapter highlights LIWC as a key application to analyse trait theory; LIWC is a critical algorithm used to extract the personality traits and emotions using lexicon-based approach. The broader application of this lexicon-based approach is attractive to researchers, especially with the rise of the social networks; a recent study ~\cite{Pennebaker2007} revealed a strong correlation between Facebook status and personality traits, opening the doors for more application of cognitive science within our daily-life bases and touching the human-computer interactions field. This chapter sets the foundation for the next chapter by highlighting the intersection between the cognitive science and human-computer interaction.

\newpage
\chapter{Human-Computer Interaction}\label{hci}

\section{Introduction}

Human-computer interaction (HCI) is a branch of intellectual science-based practical approaches and is aimed at establishing usage of devices and the components contained in their systems~\cite{CarrollJM2006,dix2009human}. HCI as a field of study is cognitive specific as it includes processing of information in the form of solving problems,making decisions, perceived concepts, alertness and pattern corresponding. HCI is becoming of significant importance as a method of achieving more high-quality, efficient and effective designs that are easier for the consumer to interact with. With the constant shift in consumer needs, HCI's importance cannot be underestimated~\cite{smith2006human,dix2009human}. Technologies are constantly changing and under significant upgrades which may or may not suit consumers' needs, now and in the future. However, our world is shifting towards the concept of what people can do with technology as opposed to what technology can do for them~\cite{smith2006human}. HCI was initially designed to meet and improve five core goals: safety, utility, effectiveness, efficiency, and usability with the latter being the least promoted and advanced. However, these goals have changed over time and established a sense of revolution that encompasses usability as the crucial role of HCI. HCI as a topic is widely research-based and includes some areas that are progressive and some developing progressively. These include; use of multimedia, gesture identification,amplified reality; computer-based cooperative work, natural language processing, simulated reality and gesture recognition~\cite{mcnaughton-et-al:hcis2017,mcnaughton-et-al:cvm2018}. 
As with any other form of technology, HCI is organised in a basic manner that can be simplified as an umbrella of sorts~\cite{Hogendoorn2007}. This is because HCI directly deals with various disciplines such as education, psychology, ergonomics, efficiency, and collaboration. HCI can be subdivided into smaller domains; for example, computer-supported collaborative learning (CSCL), computer-supported collaborative working (CSCW) and lastly computer-supported collaborative research~\cite{Hogendoorn2007,dix2009human}. The domains are organised in such a way that subsequent domains are sub divisions of previous ones.

Usability as a factor of HCI has to be measured and analysed to verify its effectiveness and also as a way of finding out system gaps. There are ten rules which are proposed as basic foundations to guide the HCI usability verification criteria~\cite{Hogendoorn2007}. These are; simple, and natural dialogues should be easily identifiable, using of a user familiar language should be maintained, light memory loads for users to recall, consistency with regards to the system, transparency from developers to the users regarding the system. Others are; ease in exit executions, shortcut availability, easy error relay methods and solutions, error prevention while using the system and ‘help’ and documented information regarding the systems to make clarifications when need be~\cite{Hogendoorn2007}.

\section{Applications}
HCI has found applications in a variety of fields, especially education~\cite{smith2006human,sentance-et-al-wipsce2012,calderon+crick-wipsce2015,mcnaughton-et-al-2017,beauchamp-et-al-bjet2019}. A primary aim of HCI research is to make the interaction with users as pleasurable and effective as possible, allowing researchers to look at technology from the user's point of view. Humans interact with computers using all the channels that they get computer output from in the form of sensory input on their end. These include; visual channels, auditory channels, haptic channels and precise movement. This information has to be stored as various forms long term, short term and sensory. The aim is however not to store the information but rather to create a reaction which can ultimately be measured as the interaction with the computer. These reactions are factors of other external variables giving developers a very critical function of enhancing computer usability regardless of other factors~\cite{tryfona-et-al:hcii2017}. 

\section{User Experience and Usability}

User experience -- commonly referred to as UX -- is a complex concept that can be divided into three camps regarding the relation to usability~\cite{Robert2010}. Namely, UX encompassing usability, UX complementing usability and UX as a factor contained within the spectrum of usability. It is important to note that UX and usability are very different concepts dealing with very changing topics and concepts. While usability deals more with making the users experience easier whilst using computer systems, UX deals with engaging other emotions of the user such as their interest and amusement. This way, computers may be usable but not necessarily engaging, thus failing on the spectrum of UX. 

UX can be categorised into three broad dimensions: the user, the product and the interaction. These dimensions lay forth the foundation of UX development since it is essential first to understand the user’s needs for developers to make products~\cite{JanStage2006}. Product aspects are primarily linked to the human emotions which the product evokes. These include, but are not limited to; memorability of the experience, the ubiquity of the system and the general perception revolving around the system.  All these are considered as outputs from the system whose input include; appropriate product properties; proper consumer needs tackling, the usability of the system, cognitive associations with the system and finally the context in which the user is using the system from.  The interaction in which the user finds themselves engaged in with the product sets forth a new dimension of research and gaps in UX development. Through data generated from the user and product interactions, developers can assess market gaps that lie not addressed and not searched~\cite{Robert2010}. 

It is important to note that user experience and usability are different concepts; this can be shown with the example of UX and usability creation. UX cannot be designed but can instead be planned for. On the other hand, system usability is practical to design as it does not inherently rely wholly on the user of the system. UX is composed of two Meta levels that are exhibited in all dimensions of UX; these two are the sense-making aspect of UX and the aesthetic appeal of UX. UX also addresses functional, physical, perceptual, cognitive, social and holistic dimensions of computing systems. As compared to system usability, UX is very subjective and depends wholly on the attitudes of the user during the moment at which they access the system. It is also not static as it progresses with time as users engage with it. This is a distinct similarity between usability and UX; however, context is king when it comes to UX.

In recent years, we have witnessed significant efforts to raise data and knowledge explaining UX and its relationship with usability~\cite{Robert2010}, as well as within the wider field of software sustainability~\cite{venters-et-al:jss2018}. Of key concern is human emotion during the use of technology as it sets up the foundation on which user experience is laid upon. Based on past research findings, it is difficult to establish definite, specific results by studying conscious human experiences. This is due to the technicalities related to the broad field. Tests techniques are destroyed by incompatible definitions, various assumptions and the ever-changing human state of mind. These constrain thus bring forth human emotion as a gateway for inferences regarding UX and their importance~\cite{JanStage2006}. The emotional facet of UX can be broken down into four main groups; competence frustration models, individual coping differences in human technology interaction, mental contents of emotional experience and non-conscious cognitive process associated with the appraisal process. 

Emotional design is a branch of UX development that deals with the creation of products in the form of systems that elicit appropriate emotions. The main aim of the emotional design is to ‘manipulate’ or evoke positive emotions from the users~\cite{Lockner2014}. Developers thus have to consider the relationships established between users and systems constantly. These relationships can either be negative or positive. Negative relationships curtail the amount of trust that a user gives to systems and becomes rather careful during use. Positive relationships, on the other hand, create bonds with the users in the form of memories which peek at the interest of the users. The context of the type of relationship established influences the user’s final emotional indulgence with the system. For example, horror-themed games elicit fear which is a negative emotion, but at the same time, fear adds to user adaptability by boosting adrenaline from game use. 

Emotional connections act as central connectors between the world and human capacity to learn. Human emotion can be categorised into three levels; visceral, behavioural and reflective~\cite{Robert2010}. Visceral emotions are those elicited when we first encounter the system. This forms the basis for adaptability of the system by the user and aesthetics coupled with the products ability to address consumer needs play a great role. It is therefore essential to first obtain data regarding market opinions on the product before exploring production. The behavioural, emotional design is the usability of the product. Usability in this context is with regards to how effective the system is, how satisfying its use is and how efficient it is. General concept formation clouds this stage of emotional acceptance regarding the product or system and clear opinions have already been established~\cite{Lockner2014}. The final form of emotional design is experience based and makes second purchase/adaptation predictable. It is dependent on the cognitive features of the user and uses objective analysis to affect future decisions. 

Every human interaction made by humans is emotionally aligned in a specific direction~\cite{Lockner2014}. It is thus essential for developers to factor in how users feel while using their products. User emotions are a factor of external and internal factors. Human emotions are of two main streams; dimensional perspective and discrete perspective. The last sets forth human emotion as a sum of categories that can be further divided into other subcategories that come about as a result of human action-reaction interaction, that is, fight or flight coping mechanism~\cite{Lockner2014}. However, this stream of human emotion is limited to a controlled set of external factors that may affect the user thus making human emotion dimensional since it can take various dimensions due to its context. Choosing emotion is crucial as it directly determines the final consumer adaptation of products. 

\section{Usability of Complex Information Systems}

\subsection{System Events}\label{lit:systemstatus}

The computer systems world is organised into clusters and structures that could at first appear as very simple due to the final user interface in which they are represented~\cite{pittphilsci8496}. However, most systems are very complex and require input from various factors for them to be successful. The complexity of systems includes; work complexity, information complexity, technological complexity and topic complexity~\cite{pittphilsci8496}.

All forms of complications have a variety of input which exists in more than one type and works synergistic-ally for the system. The systems in-processing is specific to the system, and so are the outputs. However, outputs appear more straightforward than the processing involved. A good example is the buying action of consumers. While it may seem like a simple action brought about by the need to satisfy a need, buying is a factor of both internal and external consumer-related factors and is affected by other market drivers. With regards to system engineering, the term complexity is very diverse and can range in its spectrum of meaning. Complexity may arise from the presentation of data, non-linearity of data, numerosity of data, data organisation, lack of central control of data, the spontaneity of data and feedback associated with the data~\cite{Bhavnani2000}. Complex information systems have to be developed while paying close attention to consumer usability and user experience. Despite their complexity, ease of use and user emotions have to be well thought of and executed. Thus, complex information systems are a broad set of information technology that deals with processing, evaluating and analysing data. Computer systems can be categorised into complex systems as they involve the incorporation of various aspects of computing to achieve set objectives and functions~\cite{Bhavnani2000}. This is shown by simple computer applications such as word processors that use text formation and visual confirmation which may sometimes be aided by audio endorsement to create desired sets of documented info. Computer systems usability may, however, be phased with challenges in the form of system status. These statuses include system idle, system error, system down and system slow~\cite{Galuten2005}. These status couples the computer during usage and is set by default on the system's framework. 

\subsubsection{Web-Based Applications}

Web applications have been around even before the popularity of the World Wide Web. For instance, Larry Wall (1987)~\cite{wall1999programming}, developed a server-side scripting language named Perl before the mainstream of the internet. In the early 90s, the first web applications were developed to perform simple functions~\cite{o2005web}. Nowadays, current web applications classified as complex applications as it is used across the globe to perform complex functions (e.g. taxes, online banking, socialising and more)~\cite{ginige2001web}.The web application uses the Hypertext Transfer Protocol (HTTP) protocols~\cite{Fielding1999}. The web application consists of two main components: client and server. The ``client" is the application used to enter the information, and the ``server" is the application used to store the information~\cite{berson1992client}. Web applications generally use a mixture of server-side script (e.g. PHP, JSP)~\cite{heninger2002server} and client-side script (e.g. Javascript, HTML5) to produce an application~\cite{flanagan2006javascript}. The client-side script handles the visualisation of the data while the server-side script deals with the back-end functions~\cite{berson1992client}; for instance, handling the server connections, storing the information in the database, etc. 

Hypertext Transport Protocol (HTTP) is a stateless application-level protocol that is used for distributed, collaborative and hypermedia information systems~\cite{Fielding1999}.HTTP integrates servers as the key information transfer mechanism between the user and itself. Users send requests to the server which creates a response back to the consumer in the form of an HTTP protocol. The chain of command is straightforward but may include other parties designed to send out information within the request-response chain. An excellent example of such an intermediary is the existence of a proxy that is used to send out information and acts as a gateway or forwarding agent~\cite{Fielding1999}. 

HTTP includes the following elements: text marked up using HTML and CSS, scripts, and hypermedia~\cite{Fielding1999}. Hypertext is first media in the form of text that can be viewed easily by uses and can contain connected networks called hyperlinks to other hypertext. Hypermedia is the representation of hypertext in modes that represent pre-determined sets of logical extensions. Hyperlinks, as the name suggests, are network structures formed over and around the web to create links to related information from one source to the other. Scripts are instructions that can be made and put into action by the user from their end. HTTP as a concept is primarily built on its simplicity which enhances the ease of use coupled with faster response based platforms that interact with the user quickly. Client requests are independent and co-occur due to the amount of capacity that is built on the framework of the HTTP system use~\cite{Fielding1999}. 

\subsubsection{Statuses of Web Applications}

web applications are commonly to experience a variety of system status that includes but are not limited to, \emph{system errors}, hardware malfunction (\emph{Server Down}) Runtime/Application errors~\cite{yen2002system} and \emph{Server Slow} Stop Error (Termination)~\cite{gray1986computers}. Server errors are more severe in effect as they affect a chain of user computers which rely on them for functionality. It is thus important to constantly upgrade, maintain and assess data provided by servers after they have decoded system information~\cite{jones1995patterns}. 

\emph{System idle} is commonly displayed by computers on the task manager and comes with clear declarations of the username it bears on the system, the memory it occupied within the system, and a general descriptor used to describe it~\cite{JanStage2006}. During system idle, the process runs in the background with the primary aim being continuous processing of instructions in the computing unit. The idle system status allows one to know how much of the system is not under usage from other processes. System idle is constant and appears in the task bar every time. The values of system idle range on a wide spectrum between individual computers due to internal and external conditions. 

\emph{System error} is a default setting that acts as a warning mechanism to alert users of incompatibility of systems running on the computer and the inherently built system~\cite{Galuten2005}. System error is a temporary status for most computer systems and requires the click of a button for it to disappear.  Most errors as indicated above show incompatibility and are a result of safety concerns within the system. Fatal system errors lead to system stop or crash~\cite{pittphilsci8496}. In modern web systems each type of error has a standard HTTP error code, in 1999, Internet Engineering Task Force (IETF) \footnote{The Internet Engineering Task Force develops and promotes voluntary Internet standards, in particular the standards that comprise the Internet protocol suite} divided the error codes as following indicating any 5xx server error is an indication of error cause from the server side. \cite{Fielding1999}:

\emph{System down} is another form of computer status. This is mainly a concept of computer software and applications. During system down, the system undergoes a crushing event in which it loses its intended functionality and exits from the system. System down is characterised by information ‘hangs’ and in the worst case scenario~\cite{gray1986computers}, fatal system errors which break down the whole computing system. This status is mainly as a result of inputting the wrong set of information into the system primarily in the form of key instructions. The instructions may overlap and have a ripple effect on the system thus causing failure~\cite{Galuten2005}. 

\emph{System slow} as the name suggests is a system status primarily based on processing speed. It is associated with the continuous flow of information that may serve as inadequate in executing specific functions of the system. System slow is performance-based and evaluated. System errors are either handled automatically by built in systems within computers or manually by the users. If manually done, the system provides guidelines which are in the form of displayed text for the user to follow. The instructions can range from simple to complex actions with regards to the type of error~\cite{Galuten2005}. It occurs due to a variety of dynamics such as: internal computer factors such as space contained, number of processes being run on the computer, type of computer and capacity. External factors can include; data being input into the system which may be faulty, software introduced into the system which could induce the lag, system hardware elements. 

\subsection{Response Times and Human Perceptions}

For decades researchers investigated the relationship between the response time of a system that satisfies the user~\cite{Dannenbring1983}. According to Miller~\cite{Miller1968} and Myers~\cite{Myers1985}, the server response time can be clustered into three main points \emph{Discontinuity of waiting time at 15 seconds} and \emph{Time recovery from errors and failures} and linked to the performance of the user after recovering from those two errors~\cite{Nielsen1993Article,Nielsen1993} \emph{0.1 seconds}, \emph{1.0 seconds} and \emph{10 seconds}. The \emph{0.1 second}, indicate that the system is working idle and displayed the requested output to the user. while in case of, \emph{2 to 10 seconds}, the user stars to lose that the system is slow and that there is not operating as expected, furthermore, \emph{more than 10, 15 seconds}, clearly indicates to the user that the tasks has occurs an error and the system is not expected to return any feedback to the users.

\subsubsection{Server Response Analytics}
Technology as a field has brought with it tools of trade that make everyday living experiences much faster and simpler~\cite{Plaza2011}. This is especially true in the analytics world. The diversity in types of internet based media has generated tones of data that needs to be analysed as it represents the consumer directly. One of Google’s most useful tools in analysis is the Google Analytics which is an analytics tool that includes tracking websites and accessing reports to view data that is collected from the websites~\cite{Plaza2011}.  The tool can thus be used to estimate the status of data servers as the reports suggest direct relationships with website status and ultimately server status, part of the reports provided by Google analytic is \emph{Average server response time}, the time for the server to respond to a user action or event.

\section{Summary}
This chapter has presented and critically reviewed the key domain literature of the wider field of human-computer interaction (HCI), recognising the importance and impact of the HCI domain, providing the theoretical and practical foundation for design and developments that make it easier for users to interact with digital devices~\cite{MacKenzie2013}. The user usability is a critical factor of the HCI domain to measure the effectiveness and systems and interfaces; however, the user experience interested more of the cognitive science, emotions and user's behaviours~\cite{smith2006human}. The emerging of the cognitive science with user experience leads to the development of a new branch called emotional design that deals with the user's emotions and behaviours opening a new topic to attract researchers from a different discipline (e.g. system design, marketing, business development)~\cite{nass2007emotion}. The chapter sum up the system status and human-interactions from two perspectives technical and human perception. With broadly identify four system status from a technical perspective that includes \emph{System errors}, hardware malfunction (\emph{System Down}) runtime/application errors~\cite{yen2002system} and \emph{System Slow} and running idle mode \emph{System idle}. The technical classification aligns with the three response time limit theory introduced by Nielsen~\cite{Nielsen1993Article}, a concept based on the human perception of server waiting time. Nielsen argues that even a few seconds' delays are enough to create an uNLPeasant user experience, triggering various emotions that affect their interactions~\cite{Nielsen1993}. In the following chapter, we will focus on the use of emerging -- and increasingly impactful -- domain of artificial intelligence and machine learning in the context of linguistic analysis and HCI.

\newpage
\chapter{Artificial Intelligence}\label{ComputationalIntelligence}

\section{Introduction}
The ``digital" era has become almost ubiquitous -- especially in the developed world -- with technology finding its way into the home, businesses, government, military, environment, education and healthcare~\cite{russell2016artificial,Luger2005}. There is an ever-increasing market demand for intelligent machines and intelligent approaches to a variety of domain problems. This is mainly because of demand and competitive nature -- higher capacity and ease of access, processing and analysing of data whether input or output~\cite{Luger2005}. This need has thus lead to the awareness, growth, development and application of artificial intelligence (AI). AI can be defined as the science (and art) that aims to create human-like intelligence capabilities in machines. While existing traditional computational paradigms are fundamentally limited by physical constraints, they are not limited by the biological constraints as for human intelligence. The application areas of AI and users are diverse (and widening), and certain AI applications have already surpassed their expectations falling into a class of technology referred to as part of the singularity event phenomenon; this category of AI has developed at a near-exponential rate~\cite{Singh2010}. With AI finding use in applications such as autonomous vehicles, statistical analysis (especially through the application of natural language processing approaches) and medical diagnostics, its superiority over humans has already been established~\cite{martinez2005emotions}.

AI as a broader field of study is founded on the basis that intelligent thoughts are a way through which computation is established; that is, one that can be formalised and ultimately mechanised~\cite{Singh2010}. This means that knowledge has to be firstly represented and then manipulated. Knowledge representation is solely based on human imagination and perception of specific items. Knowledge manipulation, on the other hand, is solution based offering solutions to problems designed by the human imagination process~\cite{Singh2010}.

The development and application of AI comes with both positives and negatives; the advantages include: development of autonomous vehicles, lack of cognitive bias, high flexibility and adaptability, and variety regarding applications~\cite{Singh2010}. Furthermore, AI systems are very flexible since they can be easily re-purposed to suit specific needs as compared to human experience which may be aligned in specific fields of interest and savvy~\cite{nilsson2014principles}. Its flexibility also comes from the variety which it exists; this means that the scope of application is very wide and easy to integrate~\cite{negnevitsky2005artificial}. Autonomous vehicles are of a wide range including cars, industrial systems, analytic systems, cognitive science~\cite{tveter1997pattern} and medical systems to state a few. The introduction of the automated system not only introduces more capacity in what can be done at a go but also creates a higher chance of accuracy~\cite{cohen2014handbook}. The disadvantages are primarily represented as risks associated with AI as a field, as well as its potential applications -- AI systems lack cognitive bias in that they operate based on statistical data pre-fed into the systems and thus limit chances of error as compared to systems that are human operated. AI has not been able to function as full-brain activities such as self-consciousness, self-control, self-control and self-motivation. Furthermore, other limitations of the AI, is lack of original creativity as any AI engine base it is choices and decision based on pre-defined data~\cite{garnham2017artificial}.

The wide applicability and potential of AI use come with great risks that are directly translated into challenges that are already taking shape. Risks as a challenge from AI adaptation can be looked at as either being a positive or negative risk~\cite{Singh2010,wenger2014artificial}. Positive risk is risks initiated by developers as a vision of the potential opportunity or failure of the system. Negative risk, on the other hand, is the type of risk associated with loss of the chance~\cite{Singh2010} -- AI as a field faces human misuse as a major risk in its implementation. AI developers have to be extremely careful when carrying out a risk assessment to reduce human misuse of their developments, with the emerging field of ``ethical AI" taking shape. Humans tend to be selfish and could manipulate AI to carry out agendas of self-interest which fall back on the developers~\cite{kephart2004artificial}.

\section{Computational Intelligence}

Computational intelligence (CI) is the ability of a computer to learn a specific task from data or experimental observation~\cite{Eberhart200717,engelbrecht2006fundamentals,eberhart1996computational}; while there is no commonly accepted definition of computational intelligence, it is recognised as a sub-branch of AI and commonly considered a synonym of soft computing. According to the IEEE Computational Intelligence Society,  CI is ``a domain focusing on the natural intelligence and behavior". The main objective for researchers in this field to link the Nature with artificial methodologies to replicate the Nature activities inhuman to a computer intelligent activities in an attempt to improve the efficiency and benefit from the computer advantages. The three main areas of CI are neural networks, fuzzy systems and evolutionary computation~\cite{Eberhart200717,engelbrecht2006fundamentals,eberhart1996computational}. This section focuses on the core parts of CI that are used in this study as following neural networks, with in-depth discussion to wide range of neural network applied (see Section~\ref{nn}), natural language processing (NLP) (see Section~\ref{sec:nlp} and an application using NLP with IBM Watson (see Section~\ref{IBMWatson}). 

\subsection{Neural Networks}\label{nn}

Neural networks are generated and developed as a way of increasing the effectiveness of machine pattern classification. As with biological neural networks, machine neural networks are specific and bound to limitation s of their specificity and can be simple or complex and highly integrated. This means that they only give an output of what they are programmed to provide as output. Neural networks have anticipatory capacities and are considered as new age mathematical, computational methods which are used to solve unanticipated dynamic problems in developed behavioural systems during a specific time or period~\cite{Adya1998}. These connections thus have the ability to unravel based on pre-learned patterns by use of vast survey models that can anticipate various variables. Neural networks can process information at increasingly faster speeds thus rationalising the time that would have been used if other methods were used.

Neural networks are grouped into two broad paradigms: unsupervised and supervised. Unsupervised neural networks are best suited for the purposes of clustering patterns and can be approached in three ways: self-organising feature maps, competitive learning, and adaptive resonance theory (ART) networks~\cite{Singh2007}. Supervised paradigms are established to be universal approximates of continuous/discontinuous functions and are thus suitable for applications where approximations are made regarding the input and the output data. In this case, the network is first trained using specific input and output that is approximated. Based on this, inputs are then fed into the network to counter check that the output map reflects the original map from the comparative data. In software engineering, neural networks have found practical use especially as test oracles, effort estimators, and cost estimators. 

\subsubsection{Multilayer Perceptron}

A multilayer perceptron (MLP) of neural networks is a variation of the novel perception model proposed by Rosenblatt in 1957~\cite{Rosenblatt1958}, in his research work ``The perceptron: A theory of statistical separability in cognitive systems", and is a class of feedforward artificial neural network. Rosenblatt explains that future electric and electro-mechanical systems would be able to learn and thus recognise similarities or identities between patterns of optical, electrical or tonal information in a manner which may be closely analogue to the human brain~\cite{Block1961}. The systems would be dependent on a probabilistic model of operation as compared to the use of a deterministic principle-based approach for its operation. Further, the system would be able to work well with large populations of diversified elements. Once the system had incorporated all the above, it would be referred to as a perception~\cite{Block1961}.  

The modern model for multilayer perceptron is not dissimilar from this model and includes the use of backpropagation training algorithm~\cite{Ramchoun2016}. The network is designed to bear one or more hidden layers between its input and output layers. The neurons are also arranged in layers, with connections arising from the lower end all the way to the upper layers. However, neurons located in the same layer are not interconnected~\cite{Ramchoun2016}. With regards to numbers, the network design is such that the number in the input layer is equal to the number of the measurement for the pattern problem and the neurons number in the output layer.

\subsubsection{Random Forest Trees}\label{RFTree}

Random Forests is data classification approach that creates random trees and primarily used in fields that classify data by creating random trees~\cite{Breiman2004}.  The main advantage of random forests as a method of classification is the ease of interpretation which it offers to the end users. The data obtained is easy to explain and interpret. Random forests can be as a factor of various variables as the trees formed different support variables. These trees can term as simple algorithms that show relationships in a tree-based approach~\cite{Breiman2004}. This approach solves problems from a bagging approach in which different variables clustered in different ways and the final result obtained by getting the average across the different trees. When creating random forests, it is crucial to consider homogeneity of the data as this enables clusters to established on the right principle background~\cite{Breiman2004}. Random forests, in essence, work the same way as deeply connected neural networks by creating simple multi-layers of information in the form of input or output. 

Random forests application is broad due to its ease of operation and interpretation~\cite{Breiman2004}.  They are however mainly applied in the analysis sector as they form the broad range of algorithms, meaning that they can be used in regression and individually as a tool in machine learning. Their efficiency in this specific field based on pre-training and input data previously provided. In classification, random forests used for internet traffic interception, video and other media classification, image classification, and voice classification. All these uses can be termed as media classification as they are different forms of media. 

Random trees are classification tools that can also be used to solve regression problems as with random forests mentioned above~\cite{Breiman2004}. Random tree classification works by the use of input information referred to as the vector which classified into a tress classified in a forest. The outputs of the resulting class become the final classes. Trees within forests undergo the same type of training but have specific training sets for individual tree allocation. The vectors occur in random subsets that will be present or absent in a random manner or subspace. Errors are unavoidable when using random trees. However, by establishing reference error limits during the training,  it reduces the errors~\cite{Breiman2004}. 
Random trees have advantages over other types of data in that they are easy to read, can manage both statistical and categorical data, and perform well on large datasets, extremely fast, easy to understand. However, they also require algorithms which demand allocation of various optical choices making them a centralised option for classification~\cite{Breiman2004}. The other disadvantage is that they are inherently prone to over-fitting of items due to the amount of specificity required while making them.  According to Breiman and Cutler (2004)~\cite{Breiman2004} random trees grow based on the following; firstly, if the amount of instances in the training set is \emph{N}, sample \emph{N} cases at random - but with replacement, from the primary data. This training set will be consisted of the samples and used as the training set for developing the tree. Secondly, if there are \emph{M} input variables, a number 
\begin{equation}
m<<M 
\end{equation} is specified such that at each node, \emph{m} variables are selected at random out of the \emph{M}, and the best split on this \emph{m} is used to split the node. During the forest growing the value of \emph{m} is held constant. Lastly, each tree grows to the most significant extent possible. There is no pruning of the trees.

\subsubsection{J48 Decision Tree Classifier}

The ID3 (Induction of Decision Tree) method is used generate a decision trees from the dataset and were introduced by Quinlan (1979)~\cite{quinlan1986induction}. In early 90s, Quinlan (1993)~\cite{quinlan2014c4} produced the C4.5 method is an extension of the ID3 methods, and has been used for classifications purposes~\cite{HSSINA2014}. Furthermore, Quinlan (1996)~\cite{Quinlan1996} developed the J48 classifier is an improvement classifier based on the C4.5 method. The decision tree generated by the J48 classifier is referred to as a \emph{statistical classifier} as it creates trees based on datasets~\cite{Patil2013}. J48 is part of a larger group of classifiers mentioned above; supervised classifiers. It works by deciding the target value that is the dependent variable of a sample based on a variety of characteristics. In between the branches of` different classes exists nodes that dictate the outcomes that the characteristics can be seen in the sample population. The predicted variable is referred to as the dependent variable~\cite{Patil2013}.

\subsection{Natural Language Processing}\label{sec:nlp}

Natural language processing represents a programmed method of carrying out analysis on writing founded on models and technological expertise~\cite{ElizabethD.Liddy2001}. The semantic is very theory motivated, allowing very large variety of automated methods for carrying out analysis and expressing normally. Output strings occur naturally relate to various languages, modes and genres which the text can take~\cite{ElizabethD.Liddy2001}. Natural language processing originated from various disciplines; linguistics, computer science, cognitive psychology, electrical and electronics engineering, robotics, mathematics and artificial intelligence~\cite{Chowdhury2005}. Linguistics forms the formal structures of language while technology from computer science developed in-house illustrations of information and proficient handling of the models and finally, intellectual psychology.  offers a way into the human cognitive process~\cite{ElizabethD.Liddy2001}. The need for a linguistic way of analysing texts has been essential over time due to constant technological innovations that are moving at an exponential rate. Technology thus has to be used to facilitate change~\cite{tryfonas+crick:petra2018}. Natural language processing is a form of technology consequentially referred to as human-like language processing, this alludes to the fact that it is a discipline within the wider AI domain~\cite{ElizabethD.Liddy2001}.

The main aim of natural language processing is to build database or generate summaries in a manner that was human-like. Artificial intelligence was denoted as natural language understanding. It was mainly composed of: paraphrased input text, translated texts, answered questions regarding the context and drawn up inferences. On the contrary, natural language processing has its main goal as achieving natural language understanding~\cite{ElizabethD.Liddy2001}. The systems work by maintaining dialogue with the user as part of database retrieval~\cite{Allen2003}. This means that other than information retrieval, natural language processing is also used for machine translation, expert systems, speech recognition, artificial intelligence and question and answering as part of the analysis~\cite{Chowdhury2005}.

One of the common applications of NLP is with chatbots~\cite{shawar2003machine}, which works as an intelligent agent to respond to customer's inquiry grounding on NLP to understand the context of the conversation, which is able of managing any situation of dialogue with people (for example, api.ai, Microsoft Language Understanding Intelligent Service (LUIS))~\cite{mcneal2013introducing}.

\subsection{IBM Watson Tone Analyzer}\label{IBMWatson}

The IBM Watson ``brand" is as a well-known term for a broad range of various intelligent applications including emotions recognition, expression recognition, NLP and sentiment analysis~\cite{ibm_cloud_docs}. The IBM Watson's Tone Analyzer basic fundamental working principle is rooted in human behaviour and thus factors in how humans interact with the world. This can be simplified through the Big Five personality traits and emotion; fear, disgust, anger, joy, and sadness. These five are both as a result of needs from people and values imparted to them by the society. The tone analyser factors in these two while generating output from language input made into its system as it generates answers of questions depicted in the form of natural language, IBM Watson follows the LIWC approach as discussed in Section~\ref{lit:liwc}, furthermore, in the emotion extraction, IBM Watson based their research on \emph{emotion lexicon}~\cite{Wang2015,Kim2010}. Concerning the Big Five, the system describes them in three major ways; facets, the range of characteristics and primary and secondary dimensions~\cite{Lerner2000}. Characteristics have descriptors that lead to single term eluding personal traits. This kind of behavioural forecasting is essential, especially in emotional mining and marketing. According to a 2014 technical report from IBM~\cite{ibm_cloud_docs}, for the human analysis, IBM used a well-known annotation crowdsourcing platform called \emph{CrowdFlower}, which filters the participant to choose the top-rated annotators for the task, five participant had to confirm the classification of the sentence to ensure the highest quality of the annotation process and eventually comparing the \emph{F1} score for each analytic tone the overall difference were acceptable which indicate that the tool is working in a good performance. The Tone Analyzer looked at five different emotions (see Table~\ref{tbl: emotionscodes})~\cite{ibm_cloud_docs}.

\begin{table}[]
\centering
\begin{tabular}{|p{2cm}|p{8cm}|}
\toprule
Emotion & Description                                                                                                                             \\ \midrule
Joy     & Joy or happiness has suggestions of pleasure and satisfaction. It is a feeling of safety and comfort~\cite{griffiths2008emotions}.                                   \\
Fear    & A reaction to threatening dangers. It is a survival approach that is a response to any negative motive~\cite{ohman1993fear}.                                 \\
Sadness & Shows a perception of losing. Sometimes sadness is noticeable when a person is seen to be calm less active and isolated~\cite{karp2016speaking}.                \\
Disgust & An emotional response of disgust to something considered offensive or uNLPeasant. It is a sensation that refers to something revolting~\cite{cisler2009attentional}. \\
Anger   & Triggered due to abuse, conflict, embarrassment, carelessness or dishonesty~\cite{torestad1990anger}.                                                            \\ \bottomrule
\end{tabular}
\caption{Emotion codes for IBM Tone Analyzer}
\label{tbl: emotionscodes}
\end{table}

Earlier released of the engine used the LIWC with its machine-learning approach. However, the open-vocabulary engine just performed better than the LIWC-based approach.

\section{Machine Learning}

Machine learning is one of the most popular and rapidly growing fields in the wider AI/computer science domain. The increase in the digital technology and the social networks played a major role in increasing the integration in our daily life and widen our digital footprints and rate of data generation which require a quicker processing to relay specific information~\cite{alpaydin2014introduction}. This leeway from data has not only challenged machine learning but also caused a push and diversity in the way we look and engage with at machine learning as a concept~\cite{Smola2008}. Machine learning takes various forms of our everyday technology use and is mainly used to re-arrange and re-organise data. For the last decade, statistical data has been flooded by estimation models of actual situations. This is mainly due to the bulky nature of data. This means that the data classified was not categorical, lacked data points and had spread out data points. Machine learning is the science that deals with understanding the ways through which machines improve how knowledge is acquired~\cite{Singh2007}. Machine learning allows systems to learn directly from examples, data, and expertise. To increase expert performance, knowledge has to be smart and very specific to enhance the process of knowledge engineering.

Machine learning can be categorised into two types: inductive and deductive machine learning~\cite{Singh2007}. As the name suggests, deductive learning works by deducing information. It uses past knowledge and facts to infer outcomes or new knowledge. The system works by clustering information from large data into simple easy to understand knowledge. On the other hand, inductive education forms establish computer programs/knowledge by mining rules and designs from large data clusters~\cite{Singh2007}. Machine learning largely overlaps with statistics and finds its basis in statistics algorithms.  There are a variety of uses of machine learning including. These are search engines, natural language processing, medical diagnosis, bioinformatics, cheminformatics, and stock market analysis. Additionally, speech and handwriting recognition, genetic sequencing; game playing classifying DNA sequences, object recognition in computer vision, robot locomotion, and banking; credit card fraud recognition~\cite{Singh2007}. 

The future of machine learning is of great importance to its users. This is mainly because of the role which it plays in knowledge generation. It is therefore of key importance for developers and societies, in general, to think critically and carefully regarding the role of machine learning in the society~\cite{Smola2008}. Research has to change its focus and aim at taming the various benefits that can be brought forth by machine learning. These benefits also have to be shared across the society for them to be of great impact. Some areas of machine learning require public acceptance for them to be impactful in the community at large. Researchers can engage themselves in issues surrounding algorithm interpretability, robustness, privacy, and fairness, the inference of causality, human-machine interactions and security with regards to the future. 

\section{Classifiers and Regressions}\label{sec:classifiers}

Data mining is the process through which patterns are derived though analysis of information presented as data which later becomes a source of knowledge~\cite{Gupta2010}. Data mining uses co-relations and establishes various relationships among a given set of stored datasets. Through data mining, businesses are able to predict the future trends in consumerism and business approaches. Classifiers and classification analysis as the name implies classifies data into various clusters and is used as a data mining technique. It involves a dual-phased process, that is, model construction which precedes model usage. In model construction, the classes are pre-determined and classified under sets of rules, decision trees and mathematical formulas~\cite{Gupta2010}. Model usage on the other hand classifies unknown objects. It is very useful in estimating the accuracy of the model since it uses a pre-determined label of  experimental specimens which are related to the categorised data from tested models . There are a variety of algorithms used for classification, namely; regression trees, decision tree induction, rule-based classifiers, Bayesian classifiers, nearest neighbour classifiers, support vector machine, ensemble classifier, artificial neural network, rule based classifiers, decision tree induction, nearest neighbour classifiers, Bayesian classifiers, artificial neural network, support vector machine, ensemble classifier, regression trees~\cite{Gupta2010}. 

Regression analysis is a form of predictive modelling that is frequently used to determine the relationship between dependent targets and independent variables otherwise called predictors~\cite{Jadhav2015}. This form of analysis is thus primarily used to estimate or predict relationships between various variables.  The model established through this form of analysis indicates the importance of established relationships and the effect which more than one variable may induce on the dependent variable. Factors are pre-determined using the foundation that one factor is defined as the dependent/explained while the other factor is referred to as the independent/predicting variable. While working with regression, the variables show elasticity in nature with regards to their economies of scale or existence; regression thus helps find out the average correlation amongst a group of associated examinations that are expressed through the use of a regression equation~\cite{Jadhav2015}. There are various forms of regression with the specific types being situational and variable specific. Namely, they include: polynomial regression, step-wise regression, logistic regression, lasso regression, elastic net regression, linear regression and ridge regression.  

\subsection{Linear Regression}
Linear regression is an approach to statistical analysis that assumes that the relationship established between a variable and an independent factor fit in a linear scale. The variables can be one or more. There are various forms of linear relationships established between different factors. In the case of a single variable, the relationship is termed as a simple linear regression while in case the variable is more than one, the relationship becomes a multiple linear regression. Multivariate linear regression on the other hand involves multiple co-related dependent variables. Linear regression adopts the use of predictor functions expressed in a linear manner making use of unfamiliar paradigm limits that process set of data~\cite{Seal1967}. Linear regression is the most widely read on and oldest form of regression known~\cite{Su2012}.  In simple linear regression, the predictor variable x is represented with a scalar response variable y. The model however makes various assumptions; that a weak erogenous exists between the predictors, linearity in the relationship between the variables, constancy in the variance in errors, independence of errors, and lack of perfect multicollinearity in the predictors. During data forecast and with the use of linear regression, past information is used to forecast trends that vary in usability depending on the person undertaking the analysis. This form of analysis is also used widely in the business field to pre-determine occurrences, manage product quality and assess the differences established in data types as input in decision making~\cite{Jadhav2015}.

\subsection{Multiple Linear Regression}

As indicated previously, multiple linear regressions include the existence of more than one variable with relation to a single independent factor. The relationship works to fit the information obtained from data into a single linear equation and as with linear regression, the relationship between $x$ and $y$ has to be established. The variables are expressed as: 

\begin{equation}
x_1, x_2, x_3,...,x_p 
\end{equation} 

\noindent where:

\begin{equation}
y =  b_0 +  b_1x_1 + b_2x_2 + \ldots + b_px_p 
\end{equation}

This line shows how the mean response  y changes with the explanatory variables~\cite{Winship1984}. The observed values for $y$ vary about their means $y$ and are assumed to have the same standard deviation . The fitted values $b_0$, $b_1$, $\ldots$, $b_p$ estimate the parameters  0,  1, ...,  p of the population regression line. Multiple regression is great model for establishing variable linear relationships. 

Multiple linear regressions include the existence of more than one variable with relation to a single independent factor. The relationship works to fit the information obtained from data into a single linear equation, and as with linear regression, the relationship between x and y has to be established. The relationship between independent and dependent variables is established by using regression standard multiple regression where all of the independent factors are combined within the regression equation at the same time~\cite{Jadhav2015} R and R² are used to determine the power of relations amongst the dependent variables~\cite{Jadhav2015}. Multiple regressions are great model for establishing variable linear relationships. This form of regression analysis best befits the forecast of a continuous dependent variable from a variety of independent factors. t Multiple linear regression can be further categorised into either hierarchical or step-wise multiple regression. In hierarchical or sequential regression, independent factors are submitted in binary steps, and the statistical change in R² is applied to determine the significance of the variables introduced in the second stage~\cite{Jadhav2015}. Step-wise or statistical regression, on the other hand, is useful in identifying the subclass of independent factors which display the strongest relations to the dependent factors more economically inclined with regards to regression analysis~\cite{Jadhav2015}.

\subsection{Ordinal Regression}

Ordinal regression can also be referred to as ordinal classification. This model of regression is widely used for the classification of ordinal variables~\cite{Winship1984}. These are variables which have values on an arbitrary scale thus creating a situation whereby only where the relative ordering between different values is of importance. There are two main examples of ordinal regression; ordered logit and ordered probit. The main use for ordinal regression is in social sciences where sociological literature is a factor of importance~\cite{Winship1984}. 

\subsection{Multinomial Logistics Regression}

This is another model of regression that is commonly used to provide estimations regarding nominal dependent variable given one or more independent variables~\cite{Cohen2003}. Through constant updates and research, this model has been consequentially referred to as being somewhat an annex of binomial logistic regression that offers a platform for the dependent factor to have more than two classes~\cite{Bayaga2001}. This means that it its use is paramount if the dependent factor contains more than two nominal or ordered categories. Multinomial logistic regression results to dependent variables that exist in the form of dual ship and independent variables that are incessant and or specific~\cite{Bayaga2001}. When using multinomial logistic regression analysis, dummy codding is frequently used as a tool of the trade. A close relationship with other types of regression is that multinomial logistic regression also exhibits the ability for the nominal and continuous independent variables to interact. It is widely used in risk analysis as it offers an advantage over other types of regression analysis methods~\cite{Bayaga2001} 

\subsection{Binomial Logistic Regression}

This model of regression is similar to the multinomial logistic regression model. It is commonly referred to as logistic regression and predicts probability on the basis that tendencies, as noted above are very binary with regards to the number of dependent factors founded on more than one independent variable. If the groups are more than one, we have a multinomial logistic regression~\cite{Menard2010}. 

\subsection{Mahalanobis Distance}
The Mahalanobis distance is a method of measuring distance between a point and a distribution centre~\cite{McLachlan2004}. The points were defined as P while the distribution was defined as D. This theory is a way of establishing proponents that could cause deviations from the point P to the mean dimension of the distribution point D. It is commonly used in classifying data and involves consideration of variances that may occur within different data clusters~\cite{McLachlan2004}.

\subsection{Naive Bayes Classifier}

Naive Bayes classifier is a cluster/classification technique that users the Bayes theorem which asserts a level of independence between variables~\cite{Pang+Lee:05a} and features that are to be classified. This means that the theory assumes the existence or non-existence of very feature is not connected to other features of the objects to be classified~\cite{Mitchell1997}. As demonstrated by Gabriele et al. (2016)~\cite{Trovato2016}, a Naive Bayes classifier is best suited for dataset that is heterogeneous, incomplete, of small set sizes and categorical variables. Different datasets include data is very categorical and involves variables such as gender, nationality or race. The incomplete dataset is data that may have other attached variables to it. A good example is education level which can be a factor of nationality and gender~\cite{Trovato2016}.  This type of classifier is most widely used in machine learning that requires the use of heterogeneous sources for data such as social sciences. Robotics integration into the human world is increasing at a swift rate, and thus the need to study their cognitive behaviour is important since it forms a basis of knowledge on how they adapt to various social environments. 

\section{Sentiment Analysis}

Sentiment Analysis is a form of information analysis primarily employing a range of techniques to obtain individual clusters of information from test data provided~\cite{Pang+Lee:05a}. Sentiment analysis uses include, though they are not constrained to; business intelligence, politics, and sociology. Data used for sentiment analysis is not limited to what users say but can also be analysed from what they prefer when they are on internet-based platforms~\cite{Beigi2016}. This includes the videos they choose to watch, the sites they visit, the kind of information they look for and the items they upload. These leads to their opinions and sentiments since they are first-hand choices made by the user. Sentiment analysis is readily available for data mining since the social media industry has been on an upward growth curve. The ability to link different platforms also makes it easy to obtain more specific and accurate data. 

According to Ghazaleh et al.~\cite{Beigi2016} sentiment analysis is multidisciplinary and examines attitudes, emotions, opinions analysis regarding  people oriented organisations, services and with other people. Other are events, topics and includes multiple fields such as natural language processing, computational linguistics, information retrieval, machine learning and artificial intelligence. However, sentiment classification approaches can be classified into three main methods: lexicon based, hybrid approach and machine learning~\cite{DAndrea2015}. In the machine learning approach, sentiment analysis is used to predict how polarised sentiments are based on trained as well as test datasets. The lexicon approach does not take into account previous training or induction but instead uses a pre-existing list of words that have been previously defined to assert certain sentiments. The last approach; the hybrid approach is a combination of machine learning and lexicon-based approach. This approach is the most diverse and has a better potential of improving sentiment classification performance~\cite{DAndrea2015}.

The use of sentiment analysis dates back to the spread of the Web in the early 2000s. It has undergone development into various morphological states but can be categorically classified into five major steps of analysis; sentiment classification data collection sentiment detection, text preparation, and lastly presentation output~\cite{DAndrea2015}. As the name suggests, data collection involves gathering information from the source. Natural language processing and test analytics are used to extract the data since it is bulky and disorganised making it impossible to collect manually. Text preparation is a data clean-up process before analysis and involves removing all the unwanted content. Sentiment detection is an examination phase where the subjectivity and objectivity of the data are verified. Objective data is discarded while subjective data is stored. The last step; presentation of output is the end goal of sentiment analysis as it converts various personal information to specific and meaningful data~\cite{DAndrea2015}.

However, despite its importance, sentiment analysis can pose challenges when using it and often end up providing incorrect results. Opinions are shaped by trends, social status, economic status, geographical factors and even political factors. This means that opinions may not be as independent as they should be. The bias created by factors affecting the opinion of people consequentially means incorrect results when it comes to sentiment analysis. Sentiment analysis also faces the challenge that emanates from ambiguity that can couple social media posts~\cite{Nasukawa2003}. Posts being analysed contain may contain forms of opinions that are ironical and sarcastic which is very difficult for analysing tools to detect. 

\section{Summary}

This chapter summarises keys elements of the intersections between social networks, artificial intelligence, natural language processing and sentiment analysis. With the rise of research on large, complex networks and their characteristics, a substantial number of studies have investigated social networks in an attempt to Understand its structures whose nodes represent persons in the social context, and whose edges represent communication, collaboration, or connections between nodes and entities. With the increased usage of the social networks, increasing the availability of big, complex datasets leading to the stimulated extensive study of their fundamental properties ~\cite{Adamic2003,Castro1999,Newman2003,Newman2006,Watts1998}. Attracting scientists from a different discipline, to contribute to revealing more properties of the social network. Artificial intelligence and linguistics analysis play a vital role in such process~\cite{Singh2010}. AI attempts to simulate the human intelligence by the computer, Language/text data is one of the primary sources of interactions and expressions for human-intelligence~\cite{nilsson2014principles}. Computational intelligent is a significant part of the field of AI, especially with the growth of the Linguistics field and natural process area~\cite{eberhart1996computational}. The core issue of computational intelligence is the modelling of the primary linguistics process -- ``learning" the languages and context. Overlapping with the broader problem of AI, to learn perception, interaction, planning, decision making based on reasoning~\cite{Allen2003}. A combination of sentiment analysis and NLP is frequently used to help machine understanding the text~\cite{Nasukawa2003}, using machine learning algorithms to enable the program to learn from the previous dataset, popular applications include but not limited to chatbots, detecting spam filters and opinion mining. This chapter also highlights cutting-edge classifiers and statistics analysis used in the study, emphasises the usage of the sentiment analysis. Furthermore, describing state of the art technology presented by IBM in the field of extracting personality and emotions using state-of-art languages analysis and AI~\cite{shawar2003machine}.

\newpage
\chapter{Methodology}\label{methodology}
\section{Introduction}
Building on from the wider domain context and critical review of the literature from the previous chapters, this chapter provides an detailed overview  of the research methods that were developed and followed in the study. It provides information on how the data have been extracted from the complex computer system used as part of the stud. Explain different types of data sources and how they were sampled. The researcher describes the research design that was chosen for the purpose of this study and the reasons for this choice.

\section{System Overview}

\begin{figure}[!ht]
\centering
\includegraphics[width=\textwidth,height=400px,keepaspectratio]{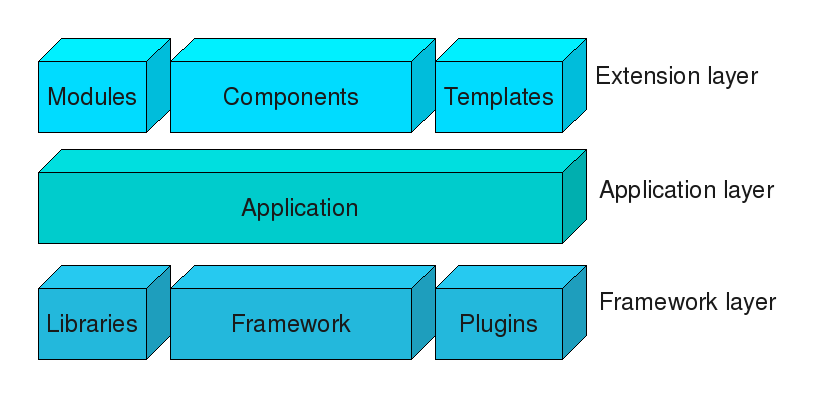}
\caption[Joomla! Framework Architecture]{Joomla! Framework Architecture~\cite{DocumentationJoomlaArchitecture}}
\label{fig:JoomlaFramework}
\end{figure}
The dataset used in this study was collected and extracted from the online portal for a European Union (EU) international scholarship mobility hosted at a UK university, which we will further explain in Section~\ref{SourceData}; however, this section gives an overview about the architecture of the system from technical perspective. The primary framework for the web portal was, Joomla!~\footnote{Joomla! Is a free and open-source content management system (CMS) for publishing web content.} and it is divided into three layers as presented in Figure~\ref{fig:JoomlaFramework}~\cite{Tiggeler2012}.

\begin{description}
\item[Extension layer] with responsibility of handling the \emph{modules} displaying information on the interface, \emph{components}, producing a complete functionality and interactive dynamic like mini application. \emph{template}, handling the development and design of front-end and back-end template.
\item[Application Layer] to allow developers to run other and integrate other application into Joomla core functionality.
\item[Framework Layer] is for writing pure web and command line applications in PHP.
\end{description}
 
\begin{figure}[!ht]
\centering
\includegraphics[width=\textwidth,height=400px,keepaspectratio]{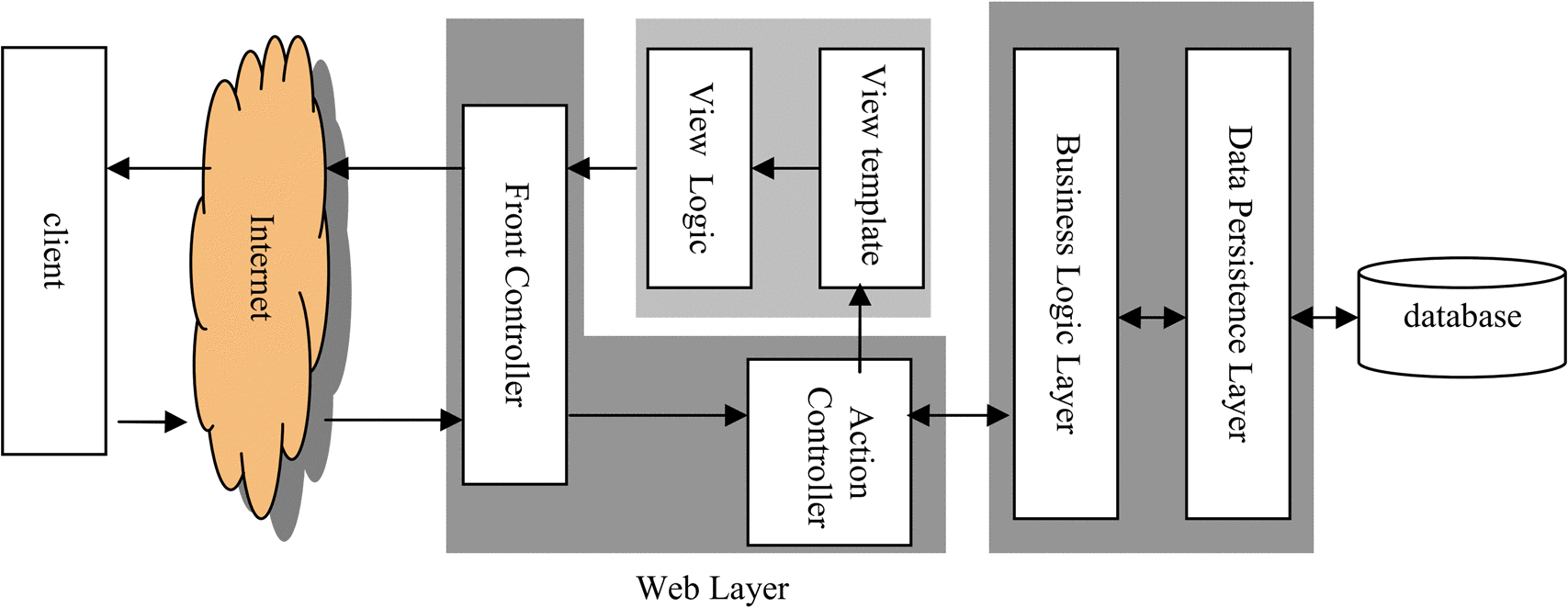}
\caption[MVC Framework Architecture]{MVC Framework Architecture~\cite{CuiThePattern}}
\label{fig:MVCFramework}
\end{figure} 
The underpinning software design architecture used in Joomla is the Model/View/Controller (MVC) paradigm, a software design model commonly used for developing user interfaces that separate an application into three interconnected parts (see Figure~\ref{fig:MVCFramework}). The MVC design pattern is a frequently used methodology for the developing of well-structured modular applications~\cite{Hofmeister2009}; it is the main design pattern using in Joomla's Components. The primary function is to split an application into three layers to give the application flexibility of debugging and investigation process for any performance issues. By separate models and views as shown in fig \ref{fig:MVCFramework}, MVC helps to improve the ease of the complexity in architectural design and to increase the refactoring and reuse of programming code~\cite{CuiThePattern}.

The primary programming language in the web-based system is PHP\footnote{PHP: Hypertext Preprocessor is a commonly-used open source multi-purpose scripting language that is particularly suited for web development and can be embedded into HTML.} and in the backend database, are MySQL and the server running Linux on an Apache web server.

\section{Data Sources and Workflows}\label{SourceData}

The digital footprint dataset used as part of this study is extracted from the behaviour of the user on a web-based educational application portal: ``{\emph{Online Portal for Scholarship Mobility}}''. We make use of textual data, analysing with these same psycholinguistic techniques, and employ standard statistical methods on non-textual data. The textual data also includes interaction with a dedicated Facebook Page for resolving problems with the applications (of which there were many); and, the actual documents submitted, including a free-text application motivation letter. The non-textual data includes the final scoring of the individual for the grant they applied for (e.g. success, reserved, failure) and the individual's behaviour on the site (when they uploaded their documents, how close to the deadline and so on).

The data comes from an online portal for a European Union (EU) international scholarship mobility hosted at a UK university. The mobility programme aims to enhance quality in higher education through scholarships and academic cooperation between Europe and the rest of the world. It provides mobility grants for students at different educational levels (Undergraduate, Masters, PhD, Post-Doctoral, Faculty) and has numerous courses available from a wide range of institutions across the EU.

The features of the call were as follows: there were 2,706 applications submitted by 1,170 candidates, applying to 10 EU universities and ten non-EU universities. The system allows an applicant to use to up to three courses from all courses offered by the ten universities, and the applicant is required to assign a priority for each module. This priority field is the primary source of final selection status in the selection stage, for instance, if the applicant \emph{Accepted} at Course A as a priority (1) and Course B as a priority (2), then the 1st priority will be offered.

Each mobility call has an opening date/time and closing date/time, with occasional extensions given for specific reasons (for instance due to administrative reasons or technical issues with the portal). Applicants are required to submit for their application specific necessary files, such as motivation letter, passport/identification, curriculum vitae), as well as optional data (supporting documents). The primary modes of communication between candidates and the project team are via emails, telephone and the dedicated Facebook Page. The selection process divided into three stages: Eligibility, Evaluation and Final Selection.

Based on which rank the applicant assigned to the host university, the final selection is the top {\emph{n}} of applicants. {\emph{n}} is calculated based on the host capacity and budget of the project. This process results in the final classification of the applicant as either:

\begin{itemize}\label{lbl:finalselection}
\item {\emph{Accepted}} (ranked highest);
\item {\emph{Reserved}} (passed but not selected);
\item {\emph{Rejected}} (below passing grade); 
\item {\emph{Ineligible}} (missing documents or out-dated documents).
\end{itemize}

While the call is running there were three approaches used to provide the user with a platform to communicate with the technical support team and the administrator's team in case of project coordinator team:

\begin{itemize}
\item Facebook Page;
\item Ticketing (help desk platform);
\item Emails.
\end{itemize}

\subsection{Motivation Letters}

All users asked to submit a motivation letter (personal statement) as part of the application process; the uploading process applies certain technical restrictions to the file type, in which only limits to image files, not PDF files. For this study, all motivation-letter image files converted using an OCR Java SDK\footnote{ABBYY FineReader v11.0.102.583 OCR Corporate Edition JAVA SDK} and Google Cloud Vision API for OCR \footnote{https://cloud.google.com/vision/docs/ocr}, to save all image files into Text files.

\subsection{Role of the Facebook Page}

Facebook Pages play a vital role in improving the communication between users and program coordinators, and the significant positive influence of ``Friends like", ``online activities" in promoting to business~\cite{Richard2014}. In the current system, a Facebook created and managed by project coordinators to interact and communicate with users,  During the scholarship calls as described previously, the Facebook page is used to receive an administrator enquires and technical reports. Facebook provides an SDK to allow developed to retrieve and to interact with their database via the API~\cite{Guide2008}. A PHP with Joomla Framework script was developed to retrieve all posts and store it in a local MySQL DB for further analysis (see Section~\ref{lst:retrieveFacebook}).

\begin{lstlisting}[language=PHP,caption=Retrieve Facebook Posts and Store to local MySQL for further analysis,label={lst:retrieveFacebook}]
$db=JFactory::getDBO();
$graphEdge=$response->getGraphEdge();
do {
foreach ($graphEdge as $graph)
{
	if (isset($graph['message']))
	{
		$message=$graph['message'];
		$msg_id=$graph['id'];
		try {
		  $response_comments = $fb->get('/'.$msg_id.'/comments/?limit='.$maxPages, $accessToken);
		} catch(Facebook\Exceptions\FacebookResponseException $e) {
		  echo 'Graph returned an error: ' . $e->getMessage();
		  exit;
		} catch(Facebook\Exceptions\FacebookSDKException $e) {
		  echo 'Facebook SDK returned an error: ' . $e->getMessage();
		  exit;
		}
		$commentNext=0;
		$graphComments=$response_comments->getGraphEdge();
		do {
		//comments
		foreach($graphComments as $graphcomment)
		{
			$filter = JFilterInput::getInstance();
			$msg=$filter->clean($graphcomment['message'], 'filter');
			$msg = preg_replace('/[^A-Za-z0-9]/', ' ', $msg);

			$sql='INSERT INTO `allCommentsData` (`id`, `message_id`, `fb_id`, `date_time`, `name`, `message`, `system`) VALUES ("NULL", "'.$graphcomment['id'].'", "'.$graphcomment['from']['id'].'", "'.$graphcomment['created_time']->format('Y-m-d h:i:s').'", "'.$graphcomment['from']['name'].'", "'.$msg.'", "'.$system.'");';
			$db->setQuery($sql);
			if (!$db->Query())
			{
				exit();
			}

			$commentNext++;
		}
		}while ($commentNext < $maxPages && $graphComments = $fb->next($graphComments));
			
	}
	//print_r($graph);
}
	$pageCount++;
} while ($pageCount < $maxPages && $graphEdge = $fb->next($graphEdge));

}
\end{lstlisting}

\subsection{Help Desk Platform and Ticketing System}

The ticketing help desk deployed and installed as part of the methods to communicate with the uses in case they have any questions or problem using the system. The help desk in our system~\footnote{HESK -- a free PHP help desk -- https://www.hesk.com/} used MySQL as the backend database. The stored data has been collected with all DateTime stamp and merged with the data gathering from Facebook posts.

\section{Identifying Computer System Status and Events}\label{serverStatus}

As presented in Section~\ref{lit:systemstatus}, we have a range of computer errors, system errors and applications and that map on the HTTP/Web application to HTTP status codes for server errors for server slow and both can be triggered by a different computer error. To identify the system events to be able to investigate the personality and emotion. For this analysis started by dividing the source of data into three types:

To investigate the Google Analytics data and specifically speed-time loading of the page to monitor the page impression and detect the number of users on the system and this point. And investigate the Apache log server to track the triggered errors at the different time to understand more about the system behaviour.  The four categories reflect the detailed system status as an output from this process:

\begin{itemize}
\item
\emph{Idle}: The system reported as working fine and server response time is fast.
\item
\emph{Slow}: The system reported as slow in response by the users and the server response time is below acceptable average.
\item
\emph{Down}: The system reported as not accessible by the users and the server response time is zero. 
\item
\emph{Error}: The system reported as accessible by the users but not working as expected with a cretin error code or unexpected behaviour
\end{itemize}

\subsection{Identifying System Status}

In this section we present the steps undertaken to identify the system status from the dataset collected:

\subsubsection{Step 1}

\begin{table}[!ht]
\centering
\begin{tabular}{@{}ll@{}}
\toprule
Server Status & Keywords                         \\ \midrule
Idle          & Working fine, thanks                     \\
Error         & Error, FTP, SQL, code            \\
Down          & Down, not working, cannot access \\
Slow          & Cannot upload, slow, upload      \\ \bottomrule
\end{tabular}
\caption{Keywords used to identify each server status}
\label{tbl: keywordsServerStatusTable}
\end{table}

For the dataset -- collected from the system as explained in Section~\ref{SourceData} -- the users had two platforms to submit their technical problem: a Facebook Page and the HESK helpdesk platform. All text collected and search by keywords as following to identify the status of the system as shown in Table~\ref{tbl: keywordsServerStatusTable}.

\subsubsection{Step 2}

From the start of the project, Google Analytics is integrated into the web-application system to monitor the system behaviour and to detect and verify of the reported system status, the average server response (seconds) were all stored on MySQL database and investigated in the next step. Figure~\ref{fig:averageServerResponse} shows an indicative sample of the server response from 8 January 2012.

\begin{figure}[!ht]
\centering
\includegraphics[width=\textwidth,height=400px,keepaspectratio]{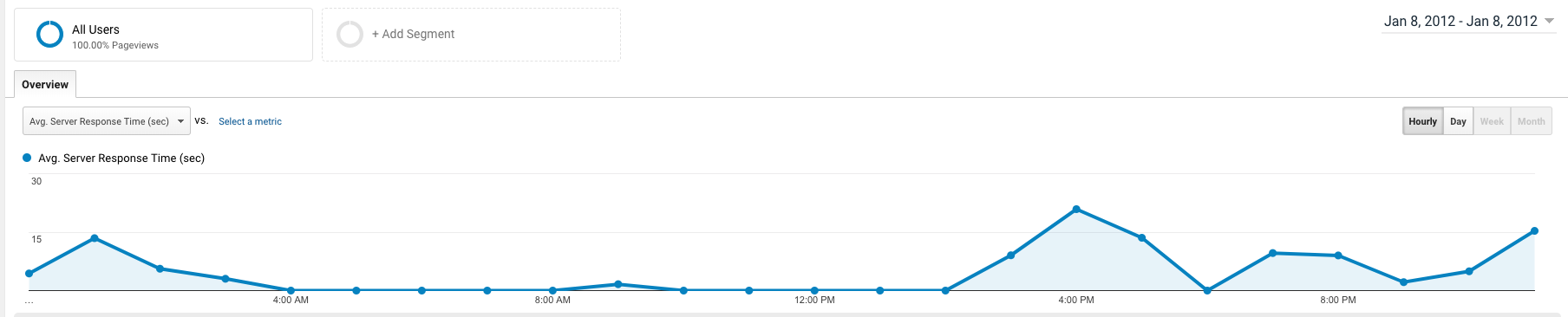}
\caption[Average server response (in seconds) on 8 January 2012]{Average server response (in seconds) from Google Analytics on 8 January 2012}
\label{fig:averageServerResponse}
\end{figure}

\subsubsection{Step 3}
Searching through all posts posted by users during the usage of the system with a sample of keywords as per Table~\ref{tbl: keywordsServerStatusTable}. Investigating the Google Analytics, in the exact date and time of each post and check if the server response time confirms the status or not. 

\begin{table}[!ht]
\centering
\resizebox{\textwidth}{!}{%
\begin{tabular}{|p{2cm}|p{6cm}|p{3cm}|p{2cm}|}
\toprule
\textbf{User ID} & \textbf{Facebook Post}                                                                                  & \textbf{Date time}  & \textbf{Server Response (Seconds)} \\ \midrule
1466             & {\emph{i have this msg Database Error Unable to connect to the database Could not connect to MySQL}}             & 2012-01-08 21:26:50 & 2.16                               \\
2449             & {\emph{Database Error Unable to connect to the database Could not connect to MySQL this message error show now}} & 2012-01-08 21:42:51 & 4.95                               \\
2304             & {\emph{please advice about the below:An Error has occurred! Unable to open JFTP connection}}                     & 2012-01-08 22:57:58 & 15.33                              \\ \bottomrule
\end{tabular}
}
\caption{Facebook posts with keyword \emph{error} and  server response }
\label{tbl: fb error server response}
\end{table}

\begin{table}[!ht]
\centering
\resizebox{\textwidth}{!}{%
\begin{tabular}{|p{2cm}|p{6cm}|p{3cm}|p{2cm}|}
\toprule
\textbf{User ID} & \textbf{Facebook Message}                                                                                                                                                                                                                                                                                                                                             & \textbf{Datetime}   & \textbf{Server Response (Seconds)} \\ \midrule
4669             & {\emph{the site is down i just need to upload few doc only}}                                                                                                                                                                                                                                                                                                                   & 2012-01-22 05:32:38 & 0                                  \\
163              & {\emph{i didnt complete the upload of my papers and i am trying from friday and the site give me always error and always down please if you can help me Youssif Rady please send me the solution}}                                                                                                                                                                             & 2012-01-23 11:02:26 & 0                                  \\
3650             & {\emph{i know time is over now but i am one of many people who wanted to just click submit the application but the site was down due to high traffic and that s not just today yesterday and 2 days before Justice is needed Admin}}                                                                                                                                           & 2012-01-22 22:10:53 & 0                                  \\
4895             & {\emph{i just want to click my submission button and the system is down}}                                                                                                                                                                                                                                                                                                      & 2012-01-21 20:57:36 & 0                                  \\
4895             & {\emph{everything is already uploaded since yesterday i just need to click on the submission button and the system have bee down for the past 24 hours and i dont know what to do}}                                                                                                                                                                                            & 2012-01-21 21:03:16 & 0                                  \\
3396             & {\emph{Thanks it wasn t necessary to extend it 13 days more 6 or 7 days would have been enough to downsize the traffic load but anyway thanks a lot}}                                                                                                                                                                                                                          & 2012-01-09 19:45:50 & 0                                  \\
3339             & {\emph{The site is down Element Scholarships Program}}                                                                                                                                                                                                                                                                                                                         & 2012-01-20 19:15:19 & 0                                  \\
1841             & {\emph{Please check the website it is down to complete the uploading}}                                                                                                                                                                                                                                                                                                         & 2012-01-21 12:58:43 & 0                                  \\ \bottomrule
\end{tabular}
}
\caption{Facebook posts with keyword ``down" and  server response  }
\label{tbl: fb down server response}
\end{table}

\begin{table}[!ht]
\centering
\resizebox{\textwidth}{!}{%
\begin{tabular}{|p{2cm}|p{6cm}|p{3cm}|p{2cm}|}
\toprule
\textbf{User ID} & Facebook posts                                                                                                                                                                                                                                & \textbf{Datetime}   & \textbf{Server Response (Seconds)} \\ \midrule
2016             & {\emph{Thanks for considering me I think the problem is not only the traffic because I haven t wait to the last minutes I ve tried to upload my documents since lots of days}}                                                                         & 2012-01-22 16:32:05 & 0.03                               \\
1918             & {\emph{thanks I received am to login and submit but I already logged in}}                                                                                                                                                                              & 2012-01-22 18:01:28 & 0.04                               \\
1918             & {\emph{Dear Element Scholarships Program admin I received an email but I can t to submit can you help please thanks}}                                                                                                                                  & 2012-01-22 18:19:09 & 0.04                               \\
2498             & {\emph{i am the only student who applied from the faculty of science ain shams university only only only i need to submit my application to graz university could you do that please username ahmed mounir email bashkora yahoo com thanks very much}} & 2012-01-22 18:28:25 & 0.04                               \\ \bottomrule
\end{tabular}
}
\caption{Facebook posts with keyword ``idle" and  server response  }
\label{tbl: fb idle server response}
\end{table}

\begin{table}[!ht]
\centering
\resizebox{\textwidth}{!}{%
\begin{tabular}{|p{2cm}|p{6cm}|p{3cm}|p{2cm}|}
\toprule
\textbf{User ID} & \textbf{Facebook Message}                                                                                                                                                                         & \textbf{Date time}  & \textbf{Server Response (Seconds)} \\ \midrule
1361             & {\emph{when i upload any word file not accpted what is the type of file you need to submit}}                                                                                                               & 2012-01-07 14:23:47 & 14.72                              \\
339              & {\emph{i have a question this error message means that the paper is not uploaded although this massege appeared i saw correct sign in the front of the required paper so that i submitted my application}} & 2012-01-09 00:31:27 & 28.89                              \\
2022             & {\emph{am tryn 2 upload only 2 files for7 hours still didn t succeed}}                                                                                                                                     & 2012-01-09 02:49:27 & 16.42                              \\
3598             & {\emph{i have one document left plzzz i struggled to get my documents fixed i need to upload the last one}}                                                                                                & 2012-01-09 14:39:44 & 22.77                              \\
3937             & {\emph{Same here I cannot upload all my documents due to lack of access to the programme s website}}                                                                                                       & 2012-01-09 16:10:44 & 19.51                              \\
3339             & {\emph{The site is not working at all please help I wanna upload the invitation letter}}                                                                                                                   & 2012-01-19 11:46:49 & 54.15                              \\ \bottomrule
\end{tabular}
}
\caption{Facebook posts with keywords related to uploading/slow performance, alongside server response}
\label{tbl: fb slow server response}
\end{table}

\begin{itemize}
\item 
Table~\ref{tbl: fb down server response} shows a sample of the posts posted on Facebook page with the keyword \emph{down}, and the server response extracted from Google Analytics; as shown, the server response reported were zero, which confirms the server status as \emph{down}.

\item 
Table~\ref{tbl: fb idle server response} shows a sample of the posts posted on the Facebook page with the keyword \emph{thanks} or \emph{working fine}, alongside with the server response rate reported by Google Analytics in same time and date. The server response rate stated was below 0.1 and above 0 and according to Nielsen (1993)~\cite{Nielsen1993} that considered as acceptable idle system behaviour for the users, which confirms the server status as \emph{Idle}.

\item 
Table~\ref{tbl: fb error server response} shows sample of the posts posted on Facebook page with the keyword \emph{error}, and figure ~\ref{fig:averageServerResponse} shows the Server response time reported in Google Analytics. At the time of the posts reported in the ~\ref{tbl: fb error server response} the server response reported were above the idle system response reported by Nielsen ~\cite{Nielsen1993} which confirms the server status as \emph{error}

\item 
Table~\ref{tbl: fb slow server response} shows sample of the posts posted on Facebook page with the keyword \emph{slow} or \emph{uploading}. Using Google analytics as the main source of extracting the server response a shown in the Table~\ref{tbl: fb slow server response}.  The average seconds reported is above 10 seconds which confirms the status as \emph{slow} assumed by Nielsen (1993)~\cite{Nielsen1993}.
\end{itemize}

\paragraph{Server response time}, as part of Google Analytics, there are behaviour reports, to indicate how system behaviour over time, as Reduce back-end processing time or place a server closer to users. According to Nielse (1993)n~\cite{Nielsen1993}, to determine an excellent acceptable server response time from the server is measured as following with the three Important Limits, there are three primary time limits (which are defined by human perceptual abilities) to keep in mind when optimising web and application performance. The fundamental knowledge regarding response times has been about the same for thirty years as it is human perceptual abilities.
\begin{itemize}
\item
0.1 second is approximately the limit for having the user sense that the system is reacting immediately, meaning that no particular feedback is required except to display the result.
\item
1.0 second is approximately the boundary for the user's stream of thought to stay constant, even though the user will notice the lag. Typically, no feedback expected during stoppages of longer than 0.1 but less than 1.0 second, but this may cause the user to lose interest in continue working on the data directly.
\item
10 seconds is approximately the limit for retaining the user's concentration on the system interface. For continued delays, users will want to accomplish other jobs while waiting for the system to finish, in such cases it is suggested to give the user an indicates about the system's progress. In many cases, the user will lose interest and might interrupt or cancel the task.

\end{itemize}

\section{Extracting Personality Traits}\label{section:ExtractBIG5}

As part of the feedback process, and for propose of this research. A Big Five Questionnaire sent to users used this system to for further analysis as part of this study. The number of responders if 80 users filled out the questionnaire. For extracting the personality traits, a Java engine developed to check if the user filled out the survey it will be the primary source of Big Five Traits. However, in case the user did not fill in the questionnaire, then the motivation letter combined with all Facebook and Help Desk correspondents will be used as the primary source to extract Big Five Personality Traits (Using either Mairesse tool or IBM Watson Tool).

\subsection{Using the Mairesse Approach}\label{Javatool}

Mairesse (2007)~\cite{Mairesse2007}, research study, presented a model to extract the personality and posted the source code publicly to be used.\footnote{Online Java code based on Mairesse model}. Suggested model built based on querying the Medical Research Council (MRC) Psycholinguistic Database \footnote{Psycholinguistic information about more than 150,000 words over 14 linguistic features.} and LIWC, also, the Mairesse's tool were validated and assessed ~\cite{Wilson1988,pennebaker+king:1999,tausczik+pennebaker:2010}. Mairesse's tool used as one of the leading approaches to extract the personality traits until IBM released a new tool to allow us to retrieve the personality traits more efficiently (See Section~\ref{releaseIBM} for further information}).

\subsection{Rationale of Using the Big Five Personality Theory}

According to the literature we reviewed in Section~\ref{bigfive}, the Big Five personality traits is state of the art in classifying the personality, furthermore, the classification based on the \emph{lexical hypothesis}, which was first produced in 1884, by Sir Francis Galton ~\cite{Galton1907}. For last decades psychology researchers have used the adjectives to describe personality and classify the traits according to the \emph{adjective} in English dictionary till it is now limited to five factors~\cite{Bagby2005}\cite{Tupes1992} ~\cite{Norman1963}. As the Big Five personality traits original based in the \emph{lexical hypothesis}, it was decided the best fit in this study, as the main dataset stream (see Section~\ref{SourceData}) to extract social networks interactions as text posted or motivation letters uploaded to the system.

\section{Extracting Emotions from Text}\label{section:extractEmotions}

As the literature revealed a powerfull produced by IBM (See section - ~\ref{IBMWatson}) Using emotion tone the methodology is output from their emotion analysis research, which is an ensemble framework to infer emotions from a given text. 

Generalisation-based ensemble framework is applied To derive emotion scores from the text. A stacked generalization is a general method of using a high-level model to combine lower-level models to achieve more significant predictive accuracy.

Features such as n-grams (unigrams, bigrams and trigrams), punctuation, emoticons, curse words, greeting words (such as ``hello", ``hi", and ``thanks"), and sentiment polarity are fed into state-of-the machine learning algorithms to classify emotion categories.

Emotion categories are the benchmark against standard emotion datasets such as ISEAR\footnote{ISEAR Databank: Over a period of several years during the 1990s, a large group of psychologists all over the world gathered data in the ISEAR project~\cite{isear_computing_2018}.} and SEMEVAL\footnote{Sentiment Analysis in Twitter~\cite{task4_2018}.}. The emotion tone engines outcomes reveal that the average performance of the model (macro-average F1\footnote{F1 score is a means of a test's accuracy} score is approximately 41\% and 68\%, sequentially). The output is stated to be statistically better than the top efficiency by the state-of-the-art models (F1 are approximately 37\% and 63\% sequentially)~\cite{ibm_cloud_docs}.
 
\section{Mapping Facebook User Profiles}\label{matchingAlgorithm}


As explained previously, the primary dataset of this study is coming from Facebook as social interactions and professional interaction with the complex computer system used to accept an application for the particular program. The dataset collected from web-application and Facebook, however, one of the challenges were to match Facebook User and their account on the system, since the \emph{EU Scholarship System} did not deploy the \emph{Login using Facebook} OAuth. OAuth is a login protocol introduced in late 2007 to support login by social networks\cite{ko2010social}.

\begin{figure}[!ht]
\centering
\includegraphics[width=\textwidth,height=400px,keepaspectratio]{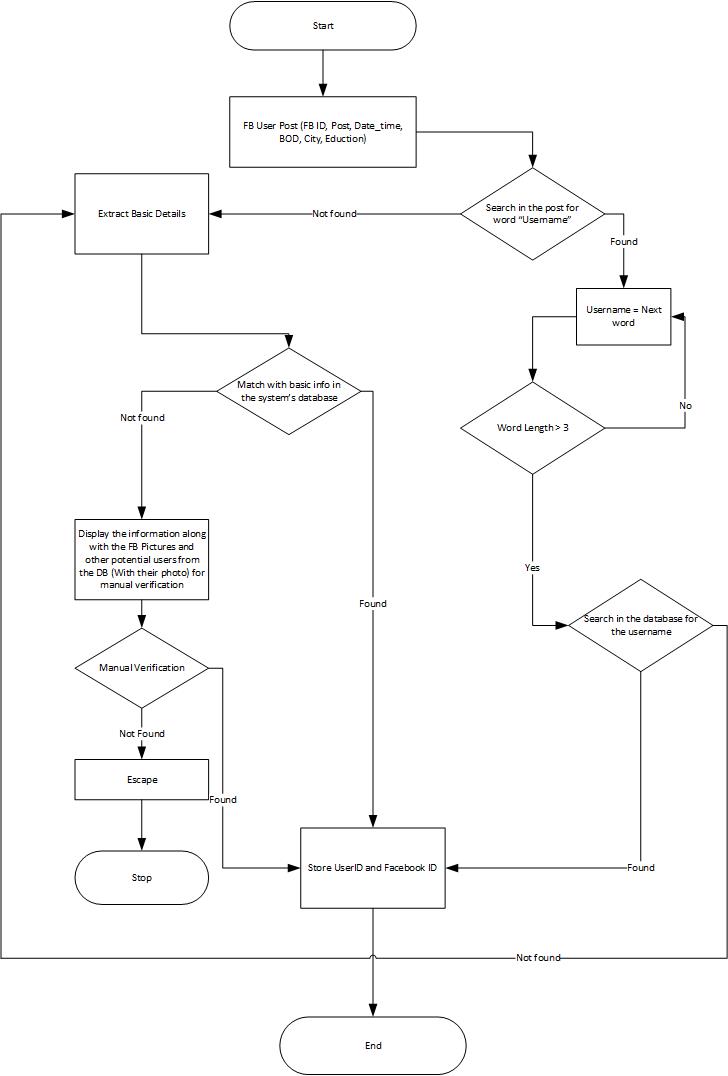}
\caption{Algorithm to match users from Facebook with users on our system}
\label{fig:FlowChart1}
\end{figure}

Figure~\ref{fig:FlowChart1} shows the flowchart of the matching algorithm to achieve this task as well as manual detection is included in different stages and a random sample of the output of the algorithm to ensure the accuracy.

The primary objective of the matching algorithm is to match between users from the Facebook page detected to the system and with the users from the system. Furthermore, the number of Facebook posts collected is 2,681. Therefore, an automation algorithm suggested producing to obtain all text from the social network (Facebook) and start to classify the interactions text by \textit{FacebookID} 871 user. However, although the automation model proofed an excellent accuracy the nature of the Facebook privacy restrictions and the use of Nicknames rather than the real names, impacted on the outcome of the automation process.

\begin{figure}[!ht]
\centering
\includegraphics[height=60px,keepaspectratio]{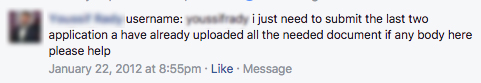}
\caption[Usage of username in a post]{Example 1: Usage of username in a post}
\label{fig:fbpostex1}
\end{figure}

\begin{figure}[!ht]
\centering
\includegraphics[height=60px,keepaspectratio]{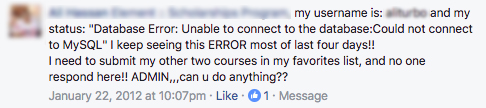}
\caption[Usage of username in a post following \textit{is} after username]{Example 2: Usage of username in a post following \textit{is} after username}
\label{fig:fbpostex2}
\end{figure}

The usage of the Facebook Page was not only for social or announcement purposes. The project coordinators decided to lunch the page as for technical errors support. Furthermore, the administrators of the project meant to ask the Facebook users to post their username on their correspondences on social media to allow the administrator team to located the user's quickly especially when it is related to technical support. Therefore, part of the automation process is to search for the word \textit{Username} and return the next word as the username as shown in Figure~\ref{fig:fbpostex1}, however, as shown in Figures~\ref{fig:fbpostex2} some other cases the next word was \textit{is}. Therefore, the algorithm as shown in Figure~\ref{fig:FlowChart1} and see Listing~\ref{lst:findusername} check if the following word number of characters is less than three (Minimum number accepted by Joomla!) then escape this word and pass the word after to the \textit{matchUserName} function to return the username is found.

The function \textit{matchUserName} verifies if the username matches the name of the user on Facebook and returns the User ID on the system; if not then Return False, and move to the other process to match the user details from Facebook to the User's Basic Information.
 
The outcome of the first phase \textit{find username in the post} successful matched 9\% of the dataset. The next phase in the automation process is to match the user's details from the Facebook with the essential information from the system, the primary limitation for this process was the privacy applied by each user is different from each. The algorithm managed to successfully match 39\% of the dataset by matching \textit{Name, Gender, City, University}. The final output of the automation process is 58\% successful matches.

As shown in Table~\ref{tbl: MatchingPerfo} the semi-manual verification phase, is to match the rest of 42\% of the dataset that failed to be matched by the previous automated system, by manually match the profile picture of the user with \textbf{proposed} users from the system. The reason it is called semi-manual, the proposed list of users is generated by the algorithm to get the closest users from the system to the Facebook profile with the available data although, it is not totally match, it is still suggested to the manual verification due to the fact the users on Facebook uses a nicknames instead of the real names. The outcome of the semi-manual verification is 15\%. And the overall failed to match is 36\%.

\begin{lstlisting}[language=PHP,caption=Matching Algorithm - PHP SnapShot Code: Searching for username,label={lst:findusername}]
/**
* Function getUsername
* @param: $message String,$name String,$system
* @param: Return false if not found, user_id if found
* @throws: none
*/
function getUsername($message,$name,$fb_id,$sys)
{
	//Param Sys refer to which system since there was three systems running on same time.
	$message=strtolower($message);
	$msg=explode(" ",$message);
	for ($i=0;$i<count($msg);$i++)
	{
		if ($msg[$i]=="username" || ($msg[$i]=="name" && $msg[$i-1]=="user"))
		{
			$next=$msg[$i+1];
			if (strlen($next)<=3)
			{
				$check=$this->matchUserName($msg[$i+2],$name,$fb_id,strtolower($sys));
				if ($check)
				{
					return $check;				
				}
			}
			else
			{
				$check=$this->matchUserName($next,$name,$fb_id,strtolower($sys));
				if ($check)
				{
					return $check;
				}
			}
		}			
	}
	return false;
}
\end{lstlisting}

\begin{table}[!ht]
\centering

\begin{tabular}{@{}cc@{}}
\toprule
\textbf{Process}        & \textbf{Progress of matching} \\ \midrule
Find username in post   & 9\%                           \\
Match Basic Information & 39\%                          \\
Semi-Manual Verification     & 15\%                          \\
Failed to match         & 36\%                          \\ \bottomrule
\end{tabular}
\caption{The performance of the matching algorithm}
\label{tbl: MatchingPerfo}
\end{table}

\section{Verifying Accuracy using the IBM Watson Tone Analyzer}\label{releaseIBM}

\subsection{Introduction}



In 2013, IBM announced the release of its Watson Tone Analyzer service to allow the identification of personality traits based on how people write. The tool is based on uses linguistic analysis to read and demonstrate emotions, personality traits, and language usage found in text. Emotions extracted based on basic emotion methodology as discussed in Section~\ref{IBMWatson}.

The {\emph{Personality Insights}} service suggests personality traits from textual data based on an open-vocabulary method. This tool represents the latest research developments in inferring personality~\cite{10.1371/journal.pone.0073791, Plank2015}. This section will highlight a specific service inside IBM Watson tool, which used mainly to extract personality traits. The service first segment the input text to develop the design in a \emph{n-dimensional space}. The service uses an open-source word-embedding method called GloVe\footnote{GloVe is an unsupervised training algorithm for getting vector representations for words~\cite{pennington_2018}.} to obtain a vector representation for the words in the input document \cite{Pennington2014}. It then provides this output to a machine learning algorithm that predicts a personality profile with the ``Big Five". The tool uses scores collected from questionnaires carried between thousands of users along with data from their Twitter supplies to train the model.

This experiment is to verify the accuracy of the IBM Watson Engine before using it for further analysis. Although, IBM conducted a validation study to understand the accuracy of the service's approach to understanding a personality profile. IBM collected questionnaire responses, and Twitter feeds for more than 2000 active users for all features and languages~\cite{10.1371/journal.pone.0073791, Plank2015}. It is thus essential to verify the tool for the dataset used in this thesis (as presented in Section~\ref{SourceData}).


The dataset of all experiments extracted from a complex web-based application designed to accept scholarship applications from users. After the result is announced the program administration posted certain questions in order to get a feedback about the scholarship and services in general for propose of improving user experience and system quality. Approaching administration team to post Big Five questionnaire as part of the follow up stages in order to improve user experience by understand more about the user's personality type.

The Big Five Questionnaire were filled out by 87 participant as it was posted the Facebook page of the scholarship program, and were open for anyone to fill in the Questionnaire, however, not all of the 87 participant were an existed user of the previous existed dataset. Using the matching algorithm in Section~\ref{matchingAlgorithm}, to match users from Facebook with users from the system 67 users have been found. Only 43 users from the 67 dataset reported to have filled in the motivation letter which will be used as Text Source for the this experiment. The final dataset consist of 43 user filled out the Big Five questionnaire and had records of the motivation letter and Facebook interactions on the database.

\subsection{Comparing Statistical Differences Between Traits}

Independent-samples t-test, the independent-samples t-test is used to determine if a difference exists between the means of two independent groups on a continuous dependent variable. More specifically, it will let you determine whether the difference between these two groups is statistically significant. Before applying the t-test to the dataset, a check to determine if the dataset is normally distributed or not is part of the assumption before using t-test. Table~\ref{tbl:TeNormality} shows the Tests of Normality, independent variables are Big Five traits and groups divided into \emph{Questionnaire result} and \emph{IBM Watson Personality insight}. 

\begin{table}[!ht]
\centering
\begin{tabular}{@{}lllllll@{}}
\toprule
                  & \textbf{Kolmogorov-Smirnova} & \textbf{}   & \textbf{}     & \textbf{Shapiro-Wilk} & \textbf{}   & \textbf{}     \\ 
                  & \textbf{Statistic}           & \textbf{df} & \textbf{Sig.} & \textbf{Statistic}    & \textbf{df} & \textbf{Sig.} \\ \midrule
Extraversion      & .114                         & 86          & .008          & .960                  & 86          & .009          \\
Agreeableness     & .118                         & 86          & .005          & .922                  & 86          & .000          \\
Conscientiousness & .145                         & 86          & .000          & .931                  & 86          & .000          \\
Neuroticism       & .094                         & 86          & .057          & .966                  & 86          & .023          \\
Openness          & .159                         & 86          & .000          & .922                  & 86          & .000          \\ \bottomrule
\end{tabular}
\caption{Independent samples t-test - Tests of Normality}
\label{tbl:TeNormality}
\end{table}

The Big Five traits \emph{(Extraversion, Agreeableness, Conscientiousness, Neuroticism and Openness)} were usually not distributed, as assessed by Shapiro-Wilk's test (${p < .05}$). The assumption is not met, however, the decision is to run the test regardless because the independent-samples t-test is relatively robust to deviations from normality and in the run the Mann-Whitney U test to confirm the result.

\begin{table}[]
\centering
\resizebox{\textwidth}{!}{%
\begin{tabular}{@{}lllllllllll@{}}
\toprule
\multicolumn{11}{c}{\textbf{Independent Samples Test}}                                                                                                                                                                                                       \\ \midrule
                  &                             & \multicolumn{2}{l}{Levene's Test for Equality of Variances} & \multicolumn{7}{c}{\textbf{t-test for Equality of Means}}                                                                                    \\
                  &                             & F                             & Sig.                        & t      & df     & Sig. (2-tailed) & Mean Difference & Std. Error Difference & \multicolumn{2}{l}{95\% Confidence Interval of the Difference} \\
                  &                             &                               &                             &        &        &                 &                 &                       & Lower                          & Upper                         \\
Extraversion      & Equal variances assumed     & 2.505                         & .117                        & 1.836  & 84     & .070            & .09279          & .05054                & -.00771                        & .19329                        \\
                  & Equal variances not assumed &                               &                             & 1.836  & 80.811 & .070            & .09279          & .05054                & -.00777                        & .19335                        \\
Agreeableness     & Equal variances assumed     & .004                          & .951                        & 1.065  & 84     & .290            & .05837          & .05483                & -.05066                        & .16741                        \\
                  & Equal variances not assumed &                               &                             & 1.065  & 83.990 & .290            & .05837          & .05483                & -.05066                        & .16741                        \\
Conscientiousness & Equal variances assumed     & 2.981                         & .088                        & 1.814  & 84     & .073            & .07674          & .04230                & -.00737                        & .16085                        \\
                  & Equal variances not assumed &                               &                             & 1.814  & 80.459 & .073            & .07674          & .04230                & -.00742                        & .16091                        \\
Neuroticism       & Equal variances assumed     & 19.769                        & .000                        & .972   & 84     & .334            & .03907          & .04018                & -.04083                        & .11897                        \\
                  & Equal variances not assumed &                               &                             & .972   & 68.676 & .334            & .03907          & .04018                & -.04109                        & .11923                        \\
Openness          & Equal variances assumed     & 3.754                         & .056                        & -1.258 & 84     & .212            & -.06674         & .05307                & -.17227                        & .03878                        \\
                  & Equal variances not assumed &                               &                             & -1.258 & 82.017 & .212            & -.06674         & .05307                & -.17231                        & .03882                        \\ \bottomrule
\end{tabular}%
}
\caption{Independent Samples Test}
\label{WatsonQSamplesTest}
\end{table}

There was a statistically significant difference in mean traits score between \emph{Big Five Questionnaire} and \emph{IBM Watson Personality Insight} across all traits as reported in the Table~\ref{WatsonQSamplesTest}.

\subsection{The Mann-Whitney U Test}

\begin{table}[!ht]
\centering
\resizebox{\textwidth}{!}{%
\begin{tabular}{@{}llll@{}}
\toprule
\textbf{Null Hypothesis}                                                     & \textbf{Test}                               & \textbf{Sig.} & \textbf{Decision}          \\ \midrule
The distribution of Extraversion is the same across categories of Group      & Independent C5- Samples Mann-Whitney U Test & 0.056         & Retain the null hypothesis \\
The distribution of Agreeableness is the same across categories of Group     & Independent C5- Samples Mann-Whitney U Test & 0.188         & Retain the null hypothesis \\
The distribution of Conscientiousness is the same across categories of Group & Independent C5- Samples Mann-Whitney U Test & 0.89          & Retain the null hypothesis \\
The distribution of Neuroticism is the same across categories of Group       & Independent C5- Samples Mann-Whitney U Test & 0.344         & Retain the null hypothesis \\
The distribution of Openness is the same across categories of Group          & Independent C5- Samples Mann-Whitney U Test & 0.327         & Retain the null hypothesis \\ \bottomrule
\end{tabular}%
}
\caption{Hypothesis Test Summary -  Mann-Whitney U Test}
\label{ Mann-WhitneyUTest}
\end{table}

The Mann-Whitney U test is a non-parametric method to verify the output of the previous independent-samples t-test, used to confirm the t-test result.

\begin{figure}[!ht]
\centering
\includegraphics[scale=0.60]{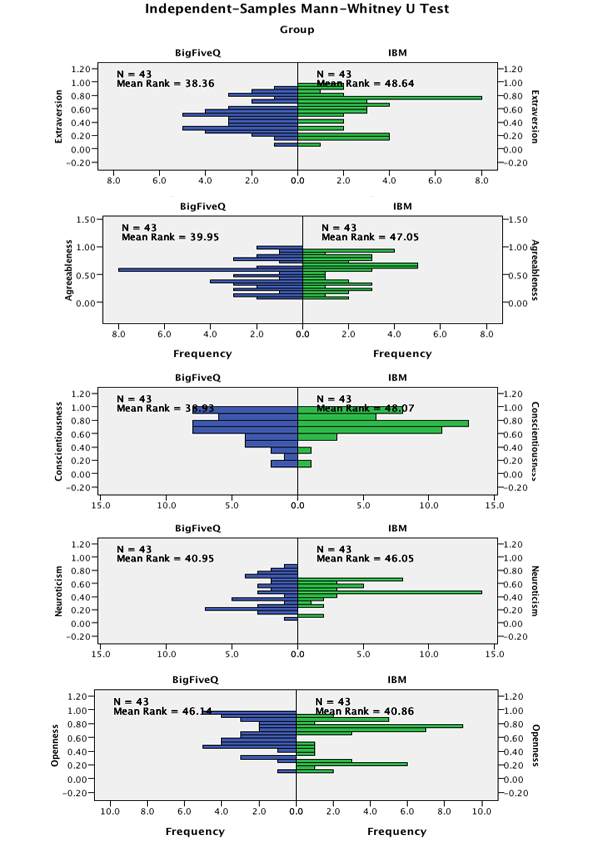}
\caption{Population pyramid representing personality traits}
\label{fig:populationPyramid} 
\end{figure}

The Mann-Whitney U test was run to determine if there were differences in traits score between personality traits produced from IBM Watson and the questionnaire. Distributions of the all traits for both IBM Watson and the Questionnaire were similar, as assessed by visual inspection~\ref{fig:populationPyramid}. \emph{Extraversion} for IBM Watson (48.64) and Questionnaire (38.36) was not statistically significantly different, U = 43, p = 0.056, \emph{Agreeableness} for IBM Watson (47.05) and Questionnaire (39.95) was not statistically significantly different, U = 43, p = 0.188,  \emph{Conscientiousness} for IBM Watson (48.07) and Questionnaire (38.93) was not statistically significantly different, U = 43, p = 0.89,  \emph{Neuroticism} for IBM Watson (46.05) and Questionnaire (40.95) was not statistically significantly different, U = 43, p = 0.344 and \emph{Openness} for IBM Watson (40.86) and Questionnaire (46.14) was not statistically significantly different, U = 43, p = 0.327. 

\subsection{Summary}
The output from The Mann-Whitney U tests shows no significantly statistically different between both groups. Furthermore, it agrees with the evaluation report produced by IBM discussed previously in Chapter~\ref{IBMWatson} and confirms the conclusion from Independent-samples t-test.

\section{Summary}
Following on from the discussion and analysis in this chapter, the following chapter outlines the flow of the experiments and how each experiment builds upon the previous one, underpinned by the literature review findings.

\newpage
\chapter{Empirical Grounding for the PMsys Engine}\label{pmsys}

\section{Introduction}

This chapter presents the experiments conducted as part of this study towards extracting the key features, to allow the development of the conceptual model. It starts by exploring the structure of the available variables of the dataset and demonstrates the flow the feature process. Furthermore, this chapter examines the underlying association between personality and emotions, and its impact on users’ behaviour in a digital domain, identifying the main features and elements of our proposed model. 

\section{Profiling Complex Online Interactions}\label{subs:ProfilingComplex}

\subsection{What Behaviour Can You Infer From a Digital Footprint?}

Understanding the software development process is essential to facilities it's effective and efficient to use as a core part of the broad field of human-computer interaction. Different users from different conceptual models about their interactions and have different ways of obtaining and developing knowledge and skills; cultural and national differences may also play a significant role. Another consideration in human-computer interaction is that technology -- and in particular, user interface technology -- changes rapidly, offering new interaction possibilities to which previous research findings may not necessarily apply. Alongside this, user preferences (and the way in which they interact with the software) change as they gradually master new interfaces and environments. Personality and behaviour is determined from digital data~\cite{Pennebaker2001,vazire+gosling:2004,Iacobelli2011,oatley+crick:2014,oatley-et-al-soccogcomp2015}. Previously, the textual information consisted of the container of the blogs, status posts and photo comments~\cite{blamey-et-al-2012,blamey-et-al-2013}, there is also a wealth of information in the other ways of interacting with digital artefacts; for instance, it is possible to observe the ordering (and frequency) of button clicks for a user\cite{Mairesse2007}. Demonstrating the use of features from the psycholinguistic databases LIWC~\cite{Pennebaker2001} and MRC~\cite{wilson:1988} to create a range of statistical models for each of the Big Five personality traits~\cite{Norman1963,Peabody1989,goldberg:1990}. As discussed previously, these five traits are: {\emph{Extraversion}}, {\emph{Emotional Stability}}, {\emph{Agreeableness}}, {\emph{Conscientiousness}} and {\emph{Openness to Experience}}. Equation~\ref{eqn:extraversion} describes {\emph{Extraversion}}, where each feature is prefixed by the containing database.

\begin{equation}\label{eqn:extraversion}
\small
\begin{aligned}
\verb!Extraversion =!\\
&\verb!-0.0379 * MRC.K_F_NSAMP +!\\
&\verb!-0.0803 * LIWC.UNIQUE +!\\
&\verb!-0.6074 * LIWC.ABBREVIATIONS +!\\
&\verb! 0.1445 * LIWC.PRONOUN +!\\
&\verb!-0.3941 * LIWC.HEARING +!\\  
&\verb!17.1407;!
\end{aligned}
\end{equation}

Initially, in this analysis it is divided into three different types of experiments: 

\begin{description}
    \item[Experiment 1] comparing the motivation letters against the Facebook interactions;
    \item[Experiment 2] examining the interaction footprints against the motivation letters;
    \item[Experiment 3] validating the raw data using multiple regression.
\end{description}

\subsection{Parameters and Feature Extraction}

Applicants are required to upload a description of why they are applying for this particular mobility grant, the motivation letter. Applicants also communicated with the project team through the project Facebook page~\cite{oatley-et-al-soccogcomp2015}.

This experiment is part of the feature extraction process, one of the objectives it to the determination whether to include \emph{final selection} as part of the features or not (see Section~\ref{lbl:finalselection}). The text was extracted from all motivation letters and Facebook interactions and analysed both blocks of text according to the Five Factor personality traits as discussed previously. To examine the strength of the relationship between the two extracted five big personality traits list. \emph{Kendall’s Tau} and \emph{Spearman’s rank} correlation coefficient assess statistical associations based on the ranks of the data. Kendall's tau! Roger News (1990)~\cite{kendall_gibbons_1990} argues that the distribution of Kendall’s tau has better statistical properties that Spearman's rank and the interpretation of Kendall’s tau regarding the probabilities of observing the agreeable (concordant) and non-agreeable (discordant) pairs are very direct. Kendall rank is used to investigate the relative position of each and compare both lists, after extracting the five factors for each applicant. Kendall rank correlation statistic~\cite{kendall:1938}. For these groups, the average Kendall's tau coefficient value is reported, for each of the Five Factor features. By considering rank position and not absolute value, we mitigate against explaining values without baseline experimentation~\cite{oatley-et-al-soccogcomp2015}.

\paragraph{Key Findings}

\emph{Experiment 1}, Table~\ref{tbl:avrankcorr} shows the results of the Kendall's tau
coefficient, specifically the variant that makes adjustments for ties
({\emph{Tau-b}}). Values of {\emph{Tau-b}} range from -1 (100\%
negative association, or perfect inversion) to +1 (100\% positive
association, or perfect agreement). A value of zero indicates the
absence of association.

\begin{table}[!ht]
\centering
\begin{tabular}{llllll}
\hline
Group                 & E        & ES                   & A        & C     & O                       \\ \hline
\emph{All}      & $-0.094$ & 0.099                & 0.145    & 0.025 & \textbf{$-0.379$} \\
\emph{Accepted} & 0.000    & 0.000                & 0.000    & 0.000 & \textbf{0.800}    \\
\emph{Rejected} & $-0.244$ & \textbf{0.333} & $-0.067$ & 0.022 & $-0.244$                \\
\emph{Reserved} & 0.010    & 0.010                & 0.162    & 0.153 & \textbf{$-0.6$}   \\ \hline
\end{tabular}
\caption[Average rank correlation for applicant group versus personality traits]{Average rank correlation for applicant group versus personality
traits (E: \emph{Extraversion}; ES: \emph{Emotional Stability}; A: \emph{Agreeableness}; C:
\emph{Conscientiousness}; O: \emph{Openness to Experience})}
\label{tbl:avrankcorr}
\end{table}

The most significant positive relationship is between those applicants \emph{Accepted} and the feature \emph{Openness to Experience} (\emph{Tau-b} = 0.8). A strong negative relationship exists between those applicants \emph{Reserved} and the feature \emph{Openness to Experience} (\emph{Tau-b} = -0.6).

\subsection{Relating a User's Digital Behaviour and Personality Traits}

\begin{figure}[!ht]
\centering
\includegraphics[height=250px,keepaspectratio]{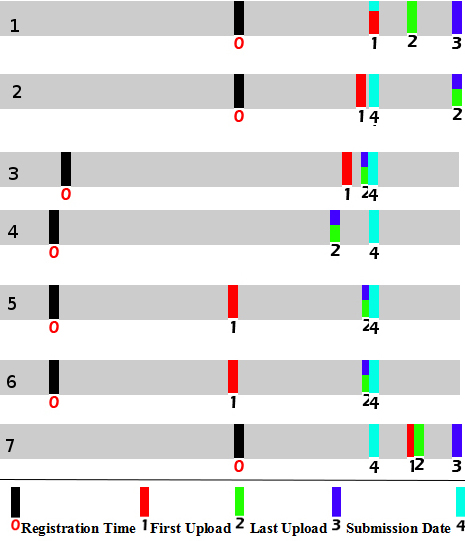}
\caption[Seven example user time-lines]{Seven example user time-lines. {\emph{T0}} (black bar) is when the applicant first
registered with the call. {\emph{T1}} (red bar) represents when the applicant
uploaded their first document, or First Action. {\emph{T2}} (green bar)
represents an applicants' Last Action. {\emph{T3}} (blue bar) represents the
applicants' Submission. {\emph{T4}} (aquamarine bar) represents the first
deadline (certain calls had initial deadlines extended)}
\label{fig:timelines}
\end{figure}

Classification the applicant's timeline by simplified an applicant's interaction, or timeline, with the portal to include the following milestones: {\emph{T0}} Registration Time; {\emph{T1}} First Action; {\emph{T2}} Last Action; and, {\emph{T3}} Submission. Additionally, representing the extension to the submission deadline as {\emph{T4}} Extension. In this way it is possible to represent an applicant's interaction as shown in Figure~\ref{fig:timelines}, which shows seven example timelines.

Using these milestones it is possible to identify interesting behaviours that compare and contract with personality traits and other sources of information. Behaviours such as: how long it was before an applicant became aware of the call, and when they registered; how long after registration did the applicant carry out their first action with the system; how long did they take to complete their application; and, how close to the deadline did they submit their application.

The timeline of the call was divided into five segments as presented in the following Table~\ref{tbl:tlperiods}. The complete timeline from opening to final close was 125 days. There was an extension from day 112 until day 125. The presentation of the segments or timeline periods is as percentage chunks of the entire timeline, for instance, segment {\emph{S0}} is the first 20\% of the timeline, and so ranges from day one until day 25, segment {\emph{S1}} ranges from day 26 until day 50, and so on~\cite{oatley-et-al-soccogcomp2015}.

\begin{table}[!ht]
\centering
\begin{tabular}{c c c} 
\hline
Segment & Start & Finish  \\ 
\hline
{\emph{S0}} & 0 & 20\\
{\emph{S1}} & 20 & 40\\
{\emph{S2}} & 40 & 60\\
{\emph{S3}} & 60 & 90\\ 
{\emph{S4}} & 90 & 100\\ 
\hline
\end{tabular}
\caption{Timeline periods as percentages of total timeline}
\label{tbl:tlperiods}
\end{table}

Using these segments it is possible to assign the various applicant actions ({\emph{T0}} Registration, {\emph{T1}} First Upload, {\emph{T2}} Last Upload, {\emph{T3}} Submission) to various time periods. The segmentation allowed us to assign applicants to statistically significant categories, and also to add in a few categories from observations. These shown in Table~\ref{tbl:apptlseg}; a small number of applicants ({\emph{n}}=4) registered within the segment S1 (20-40\% of the timeline), and then uploaded all of their documents and submitted within the segment S3 (60-90\% of the timeline). \emph{Class A} represent this segment. The rest of rows applies the same classification.

\begin{table}[!ht]
\centering
\begin{tabular}{c c c c c c } 
\hline
{Class} & {\emph{n}} & {\emph{T0}} & {\emph{T1}} & {\emph{T2}} & {\emph{T3}} \\ 
\hline
{\emph{A}} & 4 & S1 & S3 & S3 & S3\\
{\emph{B}} & 14 & S2 & S2 & S2 & S2\\
{\emph{C}} & 128 & S2 & S3 & S3 & S3\\
{\emph{D}} & 29 & S2 & S3 & S4 & S4\\
{\emph{E}} & 678 & S3 & S3 & S3 & S3\\
{\emph{F}} & 202 & S3 & S3 & S4 & S4\\
{\emph{G}} & 9 & S3 & S4 & S4 & S4\\
{\emph{H}} & 54 & S4 & S4 & S4 & S4\\
\hline
\end{tabular}
\caption{Applicants' time-line actions assigned to segments}
\label{tbl:apptlseg}
\end{table}

\begin{table}[!ht]
\centering
\resizebox{\textwidth}{!}{\begin{tabular}{|p{2cm}|p{6cm}|p{4cm}|}
\hline
Class & Description & Potential Alias  \\ 
\hline
{\emph{A}} & Register early, and take some time to upload documents, but submit with plenty of time before deadline & EverythingEarly\\
{\emph{B}} & Register reasonably early, but then upload documents and submit straight after with plenty of time before deadline, making no amendments & QuiteEarlyAndQuick\\
{\emph{C}} & Similar to Class B, but submitting more slowly & Cautious\\
{\emph{D}} & Registers reasonably early, and then takes time to upload, and only submits at the last days & VeryCautious\\
{\emph{E}} & Latecomer to registration, but then uploads and submits
quickly thereafter & Cautious\\
{\emph{F}} & Latecomer to registration, but then uploads and submits
slowly & Cautious\\
{\emph{G}} & Latecomer to registration, but delays uploading and
submission to last days & Cautious\\
{\emph{H}} & Does everything at the last days, from registration to submission & EverythingLastMinute\\
\hline
\end{tabular}
}
\caption{Description of each class}
\label{tbl:classdesc}
\end{table}

We do not want to be too quick to ascribe an alias to the behaviours, as we recognise that there are several possible interpretations; nevertheless, we have used the `Potential Alias' column in Table~\ref{tbl:classdesc} to indicate some initial thoughts.

\paragraph{Key Findings}
The following Figures~\ref{fig:extraversion}--\ref{fig:openexp} show box and whisper plots for each of the five factors, with the y-axis of each figure displaying the range for that particular feature. For example, Figure~\ref{fig:extraversion} displays the Extraversion feature, and the y-axis displays these values accordingly. The x-axis is comprised of the various classes from Figure~\ref{fig:extraversion}, combined with the status of the
application (1. {\emph{Accepted}}, 2. {\emph{Rejected}}, 3. {\emph{Reserved}}, 4. {\emph{Ineligible}}). Therefore, A1 are the Class A applicants who were {\emph{Accepted}}, distinguished from A2, who were the same class (i.e. same activity based on timeline/milestones), but who were {\emph{Rejected}}.

\begin{figure}[!ht]
\centering
\includegraphics[width=\textwidth]{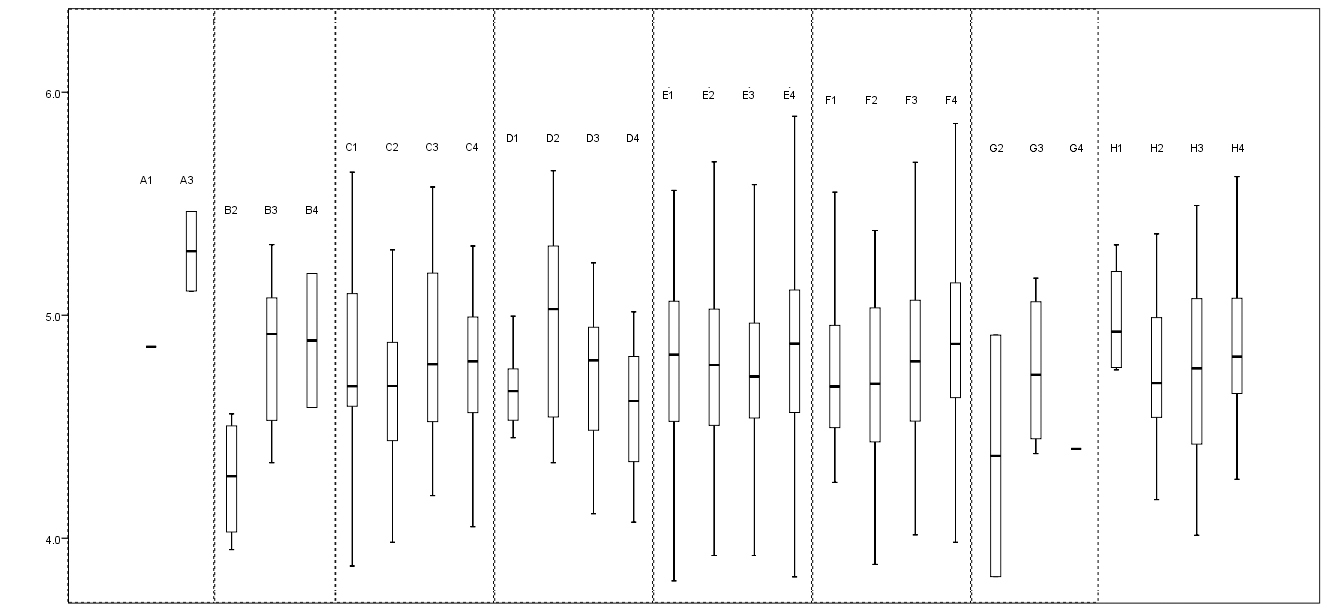}
\caption[Key findings: Extraversion]{{\emph{Extraversion}}. All features are hard to distinguish between, excepting that B2 is
significantly smaller than B3 and B4}
\label{fig:extraversion}
\end{figure}

\begin{figure}[!ht]
\centering
\includegraphics[width=\textwidth]{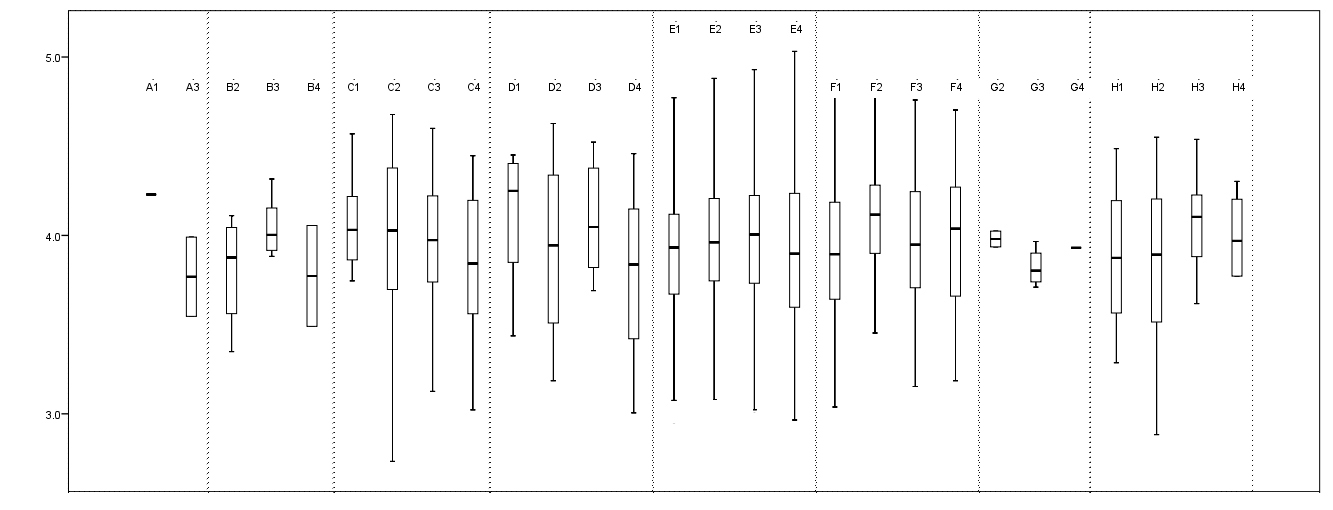}
\caption[Key findings: Emotional Stability]{{\emph{Emotional Stability}}. No real features larger or smaller, although the range on all of the E
features seems much greater than the other features}
\label{fig:emotstab}
\end{figure}

\begin{figure}[!ht]
\centering
\includegraphics[width=\textwidth]{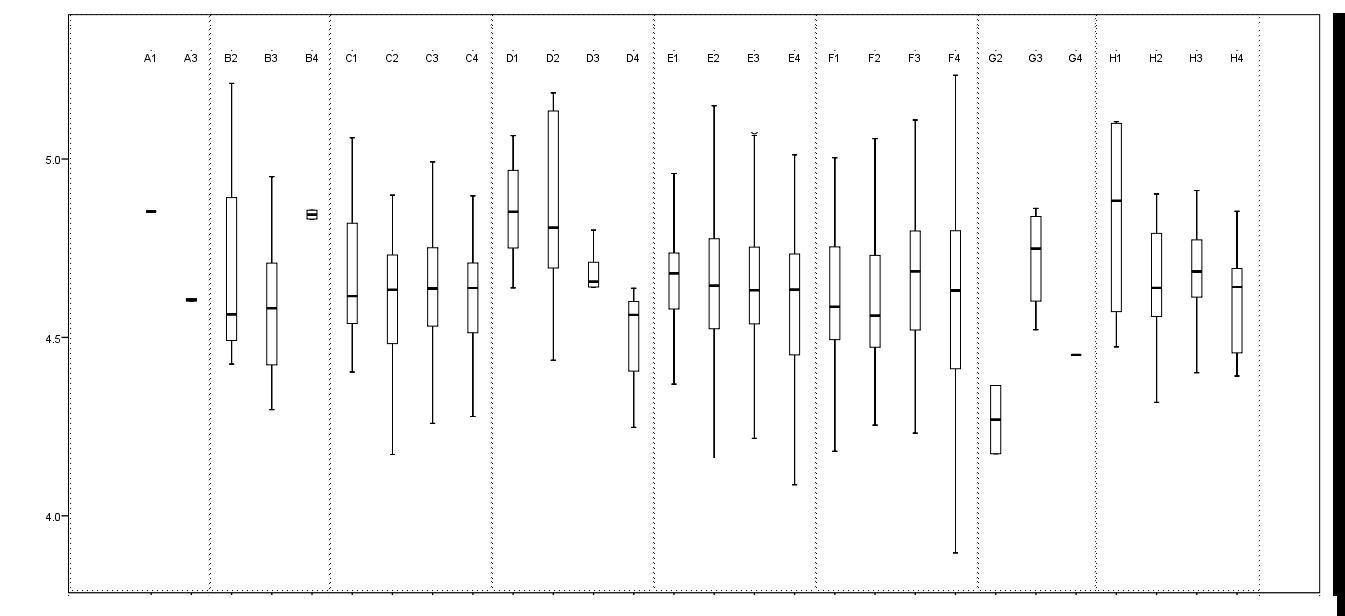}
\caption[Key findings: Agreeableness]{{\emph{Agreeableness}}. D4 is significantly smaller than D1, D2, and D3. G2 appears
significantly less conscientious than G3}
\label{fig:agreeableness}
\end{figure}

\begin{figure}[!ht]
\centering
\includegraphics[width=\textwidth]{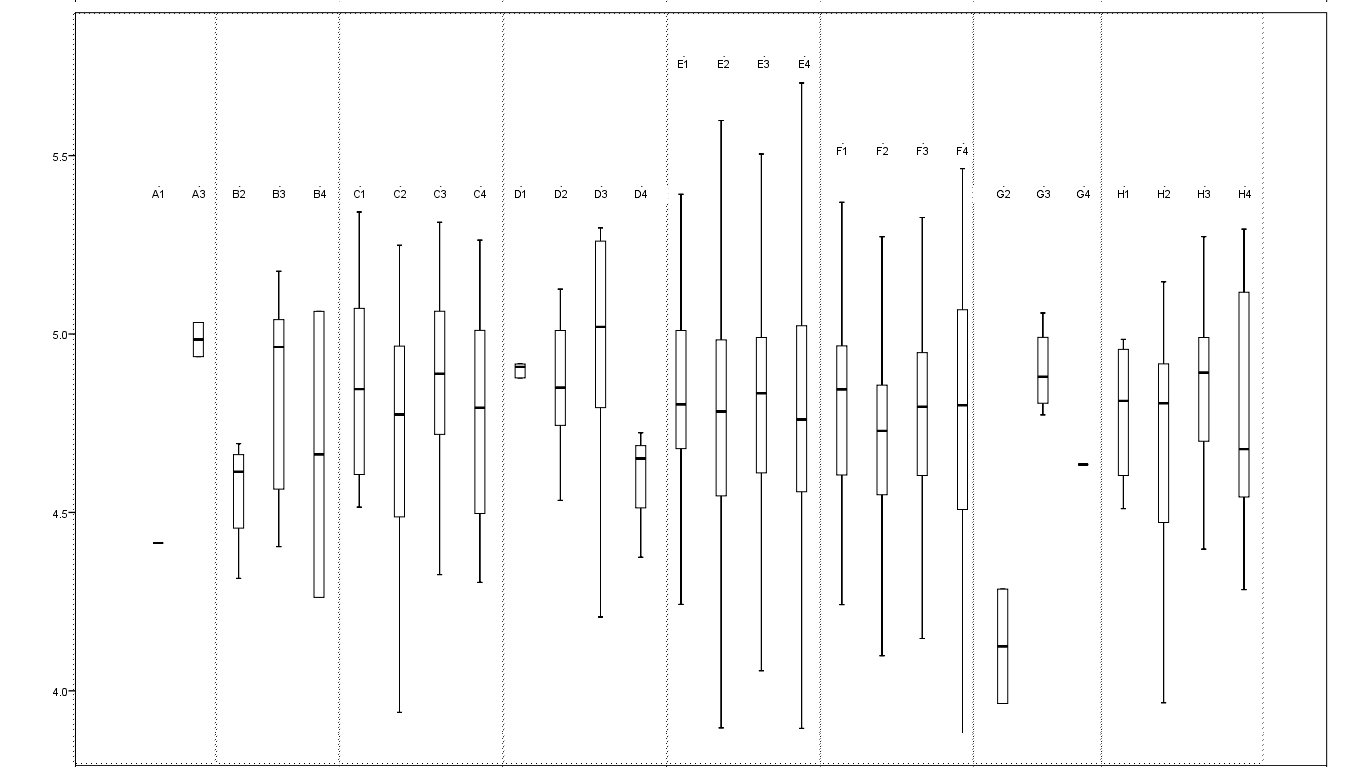}
\caption[Key findings: Conscientiousness]{{\emph{Conscientiousness}}. G2 appears significantly less conscientious than G3. To a lesser
degree D4 is smaller than D1, D2, and D3}
\label{fig:conscientiousness}
\end{figure}

\begin{figure}[!ht]
\centering
\includegraphics[width=\textwidth,height=300px]{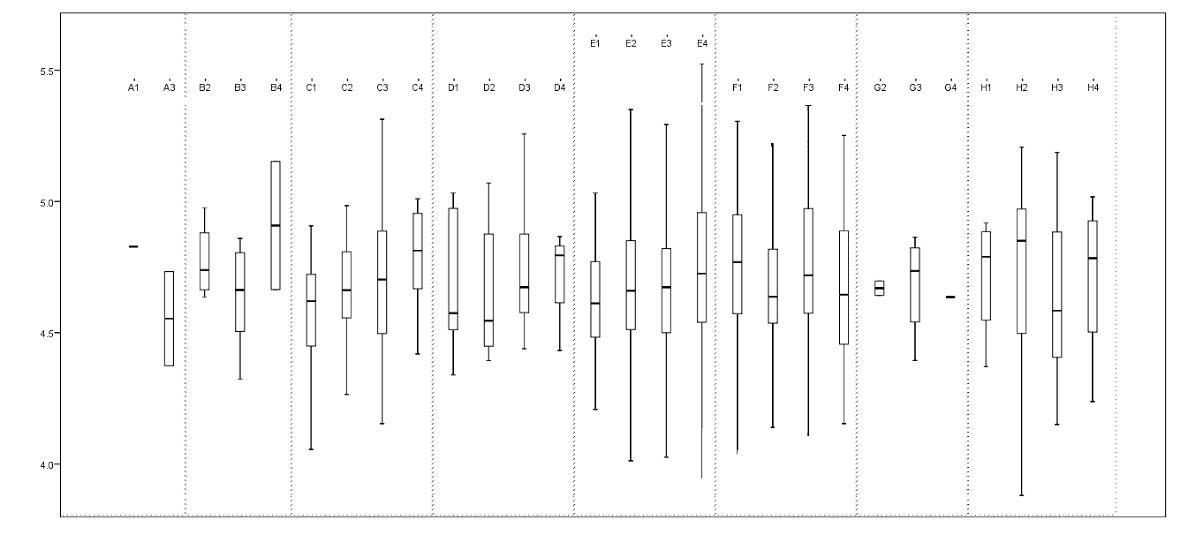}
\caption[Key findings: Openness to Experience]{{\emph{Openness to Experience}}. As with Emotional Stability, there are no exceptional features,
although the range on all of the E features seems much greater than
the other features. The Class E were the applicants that were relative
late comers to registration, but who then uploaded and submitted
quickly thereafter. Openness to Experience would seem to have very
little relationship with this class of applicant}
\label{fig:openexp}
\end{figure}

\paragraph{Mahalanobis Distance}

Checking the Mahalanobis distance, we found that 102 records exceed the
critical values, so we have removed these records since it is more
than 2\% of the total number. 
\begin{figure}[!ht]
\centering
\includegraphics[height=150px,keepaspectratio]{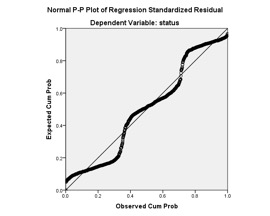}
\caption{Normal P-P Plot}
\label{fig:normalpp}
\end{figure}

\begin{figure}[!ht]
\centering
\includegraphics[height=150px,keepaspectratio]{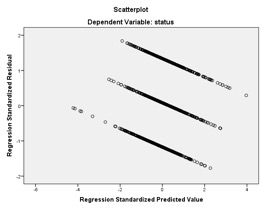}
\caption{Scatterplot}
\label{fig:scatterplot}
\end{figure}

\subsection{Extracting LIWC Data Features}
While the compound features of the five factors are an interesting
perspective, we also needed to check the raw data underneath this, in
the form of the psycholinguistic features LIWC and MRC. For this
investigation we chose multiple regression.

We extracted the LIWC row data features (87 features) from the
motivation letters and analysed the input dataset against the
`status' of the application. The method used was multiple regression,
the dependent variable being `status' and the independent variables
are the LIWC features. We proceeded as follows:

\begin{itemize}
\item Our first assumption is multicollinearity, which refers to the
  relationship when two independents variables are highly correlated;
\item Removing the above features, carry out regression;
\item Detecting outliners using Mahalanobis distance, and since we
  have 60 features remaining after the multicollinearity elimination,
  our critical value is: {\textbf{99.607}} (see Table~\ref{tbl:sampledata});
\item Screening for outliners, since multiple regression is very
  sensitive regarding outliners;
\item Making sure that we have linear relationship between the
  independent variables and the outcome.
\end{itemize}

\paragraph{Key Findings}

We extracted the LIWC row data features (87 features) from the motivation letters and analysed the input dataset against the \emph{final selection} of the application. The method used was multiple regression, the dependent variable being `status' and the independent variables are the LIWC features (see Table~\ref{tbl:sampledata}).

\begin{table}[!ht]
\centering
\begin{tabular}{llll}
\hline
\multicolumn{2}{c}{\multirow{2}{*}{Model}} & \multicolumn{2}{l}{Collinearity Statistics} \\
\multicolumn{2}{c}{}                       & Tolerance             & VIF                 \\
\hline
1           & (Constant)                   &                       &                     \\
            & REFERENCE\_PEOPLE            & 0.003                 & 290.011             \\
            & LEISURE\_ACTIVITY            & 0.004                 & 266.827             \\
            & AFFECTIVE\_PROCESS           & 0.006                 & 181.557             \\
            & PHYSICAL\_STATES             & 0.006                 & 181.512             \\
            & POSITIVE\_EMOTION            & 0.006                 & 176.569             \\
            & SPORTS                       & 0.007                 & 139.147             \\
            & OTHER                        & 0.007                 & 138.914             \\
            & BODY\_STATES                 & 0.008                 & 122.386             \\
            & YOU                          & 0.009                 & 108.779             \\
            & SENSORY\_PROCESS             & 0.011                 & 89.824              \\
            & HOME                         & 0.015                 & 67.555              \\
            & PRONOUN                      & 0.022                 & 46.061              \\
            & DIC                          & 0.023                 & 43.246              \\
            & SEEING                       & 0.025                 & 39.69               \\
            & SELF                         & 0.029                 & 34.103              \\
            & WE                           & 0.033                 & 30.494              \\
            & MUSIC                        & 0.037                 & 27.254              \\
            & TV\_OR\_MOVIE                & 0.041                 & 24.26               \\
            & FEELING                      & 0.043                 & 23.21               \\
            & SLEEPING                     & 0.044                 & 22.841              \\
            & HEARING                      & 0.047                 & 21.488              \\
            & SEXUALITY                    & 0.049                 & 20.536              \\
            & OCCUPATION                   & 0.05                  & 19.857              \\
            & SOCIAL\_PROCESS              & 0.07                  & 14.252              \\
            & COGNITIVE\_PROCESS           & 0.075                 & 13.407              \\
            & NEGATIVE\_EMOTION            & 0.089                 & 11.213             \\
\hline
\end{tabular}
\caption{Coefficients of multicollinearity variance influence factor}
\label{tbl:coefficmulti}
\end{table}
\begin{table}[!ht]
\centering
\begin{tabular}{c c c c c} 
\hline
UID & WC & WPS & UNIQUE & SIXLTR\\ 
\hline
1003 & 364 & 24.2667 & 41.4835 & 37.9121\\
1008 & 275 & 22.9167 & 61.4545 & 22.5455\\
1010 & 197 & 8.20833 & 68.0203 & 37.0558\\
1014 & 577 & 19.2333 & 53.7262 & 28.9428\\
1016 & 348 & 19.3333 & 55.4598 & 29.5977\\
1023 & 538 & 16.8125 & 53.9033 & 26.9517\\
1033 & 517 & 23.5 & 54.352 & 35.9768\\
1035 & 165 & 23.5714 & 62.4242 & 27.8788\\
1039 & 388 & 16.1667 & 56.1856 & 31.701\\
1040 & 491 & 14.8788 & 58.2485 & 33.4012\\
1049 & 462 & 25.6667 & 55.8442 & 33.1169\\
1058 & 293 & 32.5556 & 55.2901 & 26.9625\\
1069 & 436 & 29.0667 & 52.5229 & 26.1468\\
1073 & 162 & 27 & 61.1111 & 25.9259\\
1078 & 334 & 17.5789 & 55.988 & 34.4311\\
\hline
\end{tabular}
\caption[Dataset sample]{Sample of the dataset: 87 LIWC features and more than
  1000 candidates; UID represent the user and rest of the columns represent the LIWC features}
\label{tbl:sampledata}
\end{table}
\begin{table}[!ht]
\centering
\begin{tabular}{llllll}
\hline
         & Sum of Sq. & df  & Mean Sq. & F     & Sig.      \\ \hline
Regress. & 41.602     & 56  & 0.743    & 1.172 & 0.187$^b$ \\
Residual & 563.465    & 889 & 0.634    & -     & -         \\
Total    & 605.067    & 945 & -        & -     & -         \\ \hline
\end{tabular}
\caption{Evaluation of the model and ability to predicate the status values}
\label{tbl:anova}
\end{table}

\begin{table}[!ht]
\centering
\begin{tabular}{ c c c c c}
\hline 
\multicolumn{5}{c}{Model Summary}\\
\hline 
Model & R        & R Square & Adjusted R Square & Std. Error of the Estimate \\ 
\hline

1     & .262$^a$ & 0.069    & 0.01              & 0.796                      \\ 
\hline
\end{tabular}
\caption{Model summary after removing the multicollinearity features
  and above critical value of Mahalanobis distance}
\label{tbl:modelsum}
\end{table}

\begin{table}[!ht]
\centering
\begin{tabular}{llll}
\hline
\multicolumn{2}{l}{Model} & Stnd. Coeff. & Sig.  \\
1     & (Constant)        &              & 0.000 \\
      & NEGATIONS         & 0.109        & 0.004 \\
      & QMARK             & 0.107        & 0.199 \\
      & SPACE             & 0.098        & 0.044 \\
      & ABBREVIATIONS     & 0.076        & 0.038 \\
      & CAUSATION         & 0.073        & 0.051 \\
      & DASH              & 0.068        & 0.093 \\
      & UNIQUE            & 0.068        & 0.247 \\
      & INHIBITION        & 0.062        & 0.072 \\
      & JOB\_OR\_WORK     & 0.051        & 0.188 \\
      & DISCREPANCY       & 0.045        & 0.289 \\
      & SCHOOL            & 0.044        & 0.279 \\ \hline
\end{tabular}
\caption{Top effective coefficient LIWC features over the model}
\label{tbl:topliwc}
\end{table}


The first result set shows the correlation between a dependent variable (status) and independents variables (LIWC features). In our first assumption multicollinearity, we use an R-value of $0.7$ or higher to say two predictable values have multicollinearity. Based on Table~\ref{tbl:sampledata}, if the tolerance is smaller than 0.1 then we have the probability of multicollinearity, while VIF is the inverse of the tolerance, and so in the case of VIF greater than 10 then we have a case of multicollinearity. In this way, we found that the below LIWC features are related through multicollinearity.

\subsection{Discussion}

There seems to be a strong relationship between the Five Factor feature {\emph{Openness to Experience}} with a strong correlation with the Accepted group. The exploration of timeline behaviour is dependent on our representation used for interactions, and the classes derived. The same feature Openness to Experience has no group/class combinations that are significantly different than others. The outcome suggests to ignore the \emph{Final selection} as a feature in any further analysis as it only correlates with one personality trait and also, has different other parameter affecting it.

\section{Mapping User Behaviour to System Stages}\label{subs:sentimentStages}

\subsection{Introduction}






Previous experiment encourage to progress forward in more investigation specially in relationship between user's personality and different stages or events in the system. In this experiment the data collected from the main dataset and consist of $322$ record. The experiment objective is to explore the relationship between personality, stages, and sentiment of the users.

The typical approaches investigate extracting the personality traits and emotion from text using linguistic analysis \cite{Pennebaker2001} with the recent growth of human-computer interaction on daily bases the need to understand how personality and emotions incorporate in different stages. In the current dataset the web-application involved different stages before submission of the application form. Therefore, for this analysis the system have been divided into stages as shown on Table~\ref{stagestablel}.

\begin{table}[!ht]
\centering
\resizebox{\textwidth}{!}{%
\begin{tabular}{@{}ll|p{12cm}p{3cm}|@{}}
\toprule
StageID & Phase Name             & Description                                        \\ \midrule
1	&	Start Stage            & From start of the call till the first engagement from the user with the system  \\
2	&	Uploading Stage        & During the uploading process of the documents.   \\
3	&	Submission Stage       & The period after the uploading and submission \\
4	&	After Submission Stage & Stage after submission and before end of the call \\
5	&	Extension Stage        & Stage where an extension have been granted for users to continue applying (After deadline - Extension date) \\ \bottomrule
\end{tabular}
}
\caption{System Stages}
\label{stagestablel}
\end{table}

\begin{figure}[!ht]
\centering
\includegraphics[height=350px,keepaspectratio]{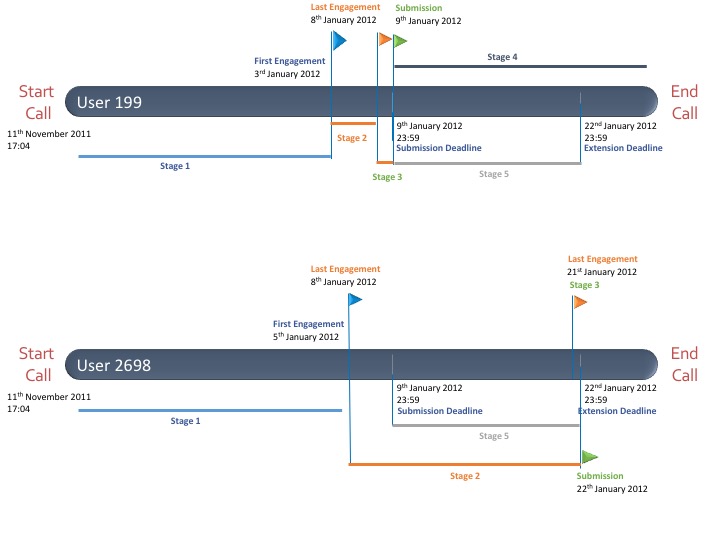}
\caption[Timestamp of two users with respect to proposed stages]{Timestamp of two users (\textit{Users 199 and 2698}) with respect to proposed stages}
\label{fig:stagesTimeline}
\end{figure}

\paragraph{Retrieve user's timeline in the proposed stages} each user has his own time stamp on the system, which represents the user's interactions during the life cycle of the call~\cite{oatley-et-al-soccogcomp2015}. In this study, user's timeline will be segmented with respect to the proposed stages as shows on Table~\ref{stagestablel}. To give an illustration of such classification. Figure~\ref{fig:stagesTimeline}, shows a timeline of one of the users on the system and how the proposed stages are cross overlapped. Code A used to pick the text associated with the user at each stage and saved it with the StageID based on Start-End date explained on Table~\ref{stagestablel}.

An algorithm responsible to detect the stage of each user by identifying the dates of each stage from the system interaction timestamps. Start and end of each stage have been identified as per the Figure~\ref{fig:stagesTimeline} the algorithm automatically detect the start/end of each stage and identify which stage and extract all text associated with the user. The Table~\ref{tbl:StagesText} shows a sample of the data after preparation, notice for each user there are different text associated with different stageID, which associated with Table~\ref{stagestablel}. 

\begin{table}[!ht]
\centering
\begin{tabular}{@{}ll|p{12cm}p{3cm}|@{}}
\toprule
user\_id & stageid & text                                                                                                                                                                                                                                                                                                                                                                                                                                                                                                                                                                                                                                                                                                                                                                                                                                                                                                                                                                                                                                                                                                                                                                                                                                                                                                                                                                                                                                                                                                                                      \\ \midrule
3254     & 1       & {\emph{I have 3 questions 1-for the required selected documents  there are two requirements MASTER TRANSCRIPT and TRANSCRIPTS  What is meant by the Transcripts because there is a slot already to upload the master certificate . 2- you say it is needed to I have 3 questions}}  \\
3254     & 2       & {\emph{i have only two documents to upload the site sometimes open but when it does i cant reach the upload page   my username amrnawar what about me i only have two documents left because the research proposal needed to be revised by college staff and they just did revising yesterday   i need to upload it and submit     please admin please thanks alot thats very kind of you thak you very very much}}                                                                                                                                                                                                                                                                                                                                                                                                                                                                                                                                                                                                                                                                                                                                                                                                                                                                                                                                                                                                                                                                                                                                \\
3254     & 5       & {\emph{admin thankkkkkkkkkkkkkks to you i think i uploaded necessary documents but i cant submit now please help me to submit user amrnawar i have only two documents to upload the site sometimes open but when it does i cant reach the upload page   my username amrnawar what about me i only have two documents left because the research proposal needed to be revised by college staff and they just did revising yesterday   i need to upload it and submit     please admin please thanks alot thats very kind of you please urgently help    istarted uploading everything only 2 pages of research proposal left when site crashed    all my work will go in vain}}                                                                                                                                                                                                                                                                                                                                                                                                                                                                                                                                                                                                                                                                                                                                \\
2878     & 5       &  {\emph{username  Luis Olano  tried to apply for Mass Communication program Ain Shams University several times but it s impossible to upload any document  please tell me if i can send by email so we spent two weeks fully dedicated to obtaining every document needed to apply for this scholarship and in the end all the effort made was not worth at all    it is very sad}}  \\
2766     & 5       & {\emph{My username is mo3taz elsawy  i already uploaded my papers and only want to submit my application  it still isn t working for me up tell now i can log in but i can not do any thing after that the site is very slow mo3taz elsawy is my user name}}                                                                                                                                                                                                                                                                                                                                                                                                                                                                                                                                                                                                                                                                                                                                                                                                                                                                                                                                                                                                                                                                                                                                                                                                                                                                                       \\
6223     & 5       & {\emph{Thaaaaaaaaaaaaaaaaaaaaaanks    Reassure for my Documents}}                                                                                                                                                                                                                                                                                                                                                                                                                                                                                                                                                                                                                                                                                                                                                                                                                                                                                                                                                                                                                                                                                                                                                                                                                                                                                                                                                                                                                                                                                  \\
2937     & 5       & {\emph{my username is  Hatem Hassan     plz just one document to complete my submission}}                                                                                                                                                                                                                                                                                                                                                                                                                                                                                                                                                                                                                                                                                                                                                                                                                                                                                                                                                                                                                                                                                                                                                                                                                                                                                                                                                                                                                                                          \\
3094     & 2       & {\emph{whenever i upload any file with it's required extension (JPG )website says message appears BAD REQUEST Your browser sent a request that this server could not understand}}  \\ \bottomrule
\end{tabular}
\caption{Sample of Stages Text collected per user~\cite{oatley-et-al-soccogcomp2015}}
\label{tbl:StagesText}
\end{table}

Applying sentiment analysis allowed us to label each text as positive or negative with respect to each stages, with the expected output for each text being {positive, negative}. The API Text-Processing used \footnote{http://www.text-processing.com} where it expect the text input and the output consist of 3 attributes (\textit{Pos},\textit{Neg} and \textit{Natural}) as shown in Table~\ref{tbl:StagesSentiment}, the first column \emph{StageTextID} represent a record on the Table~\ref{tbl:StagesText} to match the text, stageID with the output of this text. \emph{Negative}, \emph{Neutral} and \emph{Positive} represent the probability of each label. \emph{negative} and \emph{positive} adding up to 1, however, \emph{neutral} is standalone. If \emph{neutral} is greater than 0.5 then the label marked as \emph{neutral}. For this experiment this rule will be ignored the study focus in the probability of \emph{positive} and \emph{negative} only. Therefore, the \emph{neutral} label replaced with other higher probability. For example in the Table~\ref{tbl:StagesSentiment} in the case of record number $12$ of \emph{StageTextID} the \emph{positive} = $0.492328951$ , \emph{Neutral} = $0.670260584$ and \emph{negative} = $0.507671049$ while it is labelled \emph{neutral} the \emph{negative} probability is higher than \emph{positive} therefore, the label changed to \emph{neg}

\begin{table}[!ht]
\centering
\begin{tabular}{@{}lllll@{}}
\toprule
StageTextID & Negative    & Neutral     & Positive    & Label   \\ \midrule
1           & 0.518402657 & 0.366022022 & 0.481597343 & neg     \\
2           & 0.733416831 & 0.167587171 & 0.266583169 & neg     \\
9           & 0.606781287 & 0.165370162 & 0.393218713 & neg     \\
10          & 0.823030133 & 0.176314817 & 0.176969867 & neg     \\
11          & 0.780114198 & 0.303985047 & 0.219885802 & neg     \\
12          & 0.803123608 & 0.229718524 & 0.196876392 & neg     \\
13          & 0.810792691 & 0.168979215 & 0.189207309 & neg     \\
14          & 0.740935943 & 0.235929961 & 0.259064057 & neg     \\
15          & 0.762828024 & 0.768115205 & 0.237171976 & neg     \\
16          & 0.762828024 & 0.768115205 & 0.237171976 & neutral \\
17          & 0.38153628  & 0.420256869 & 0.61846372  & pos     \\
18          & 0.332127272 & 0.3738164   & 0.667872728 & pos     \\
19          & 0.492328951 & 0.670260584 & 0.507671049 & neutral \\
20          & 0.715223337 & 0.310831638 & 0.284776663 & neg     \\
21          & 0.549909425 & 0.661750668 & 0.450090575 & neutral \\ \bottomrule
\end{tabular}
\caption{Sample of the sentiment analysis output}
\label{tbl:StagesSentiment}
\end{table}

The text processing uses the NLTK Naive Bayes Classification in the methodology with a dictionary of movies review labelled as following a movie reviews corpus has 1000 positive files and 1000 negative files. 75\% of the dataset use as  the training set, and the rest as the test set. Training and Testing the Naive Bayes Classifier. The outcome of the modelling is 73\% accuracy. 

There are two sources for extracting the personality traits \emph{motivation letter} and \emph{Big Five Questionnaire} as part of the user's experience the program coordinator encourage users to fill in the \emph{Big Five Questionnaire}. Therefore, before an attempt to extract the personality traits from the motivation letter \footnote{Personal statement document submitted as part of the application process}, first checking the if the User filled out the \emph{Big Five Questionnaire} if not then move to the personality trait extraction methodology stated on Section~\ref{section:ExtractBIG5}.

\subsection{Binomial Logistic Regression (Logistic Regression)}

The dataset consists of the Big Five traits, stages of the system and the sentiment positive/negative output. Since the ``Big Five" traits are continues values and sentiment output are either positive or negative (dichotomous dependent variable ), and the \emph{stageID} is a \emph{ordinal variables}. Therefore,  the mix of data types suggests to use Binomial logistic regression, to calculate the probability sentiment behaviour of a user based on personality trait and stage. To observe the possibility of modelling the user's behaviour based on specific stages.

For a logistic regression to be accurate, the ``big five" traits values need to be linearly related to the logit transformation of the \emph{Sentiment Output}. This hypothesis can be examined using the Box-Tidwell (1962) method.  The Box-Tidwell is a method to examine this assumption~\cite{Box1962}.To achieve this, we follow the below steps:

\begin{itemize}
    \item Using IBM SPSS, the transformation of all ``big five" traits to its natural logs.
    \item Create interactive term for each of all ``big five" traits to its original values and respective natural logs.
    \item It is well-known practice to use Bonferroni correction based on all terms in the model when evaluating this linearity hypothesis~\cite{Box1962,Tabachnick2001}.
\end{itemize}

\begin{table}[!ht]
\centering

\resizebox{\textwidth}{!}{

\begin{tabular}{lllllll}
\hline
\multicolumn{1}{|l|}{}                               & \multicolumn{1}{l|}{B}      & \multicolumn{1}{l|}{S.E.}  & \multicolumn{1}{l|}{Wald}   & \multicolumn{1}{l|}{df} & \multicolumn{1}{l|}{Sig.}  & \multicolumn{1}{l|}{Exp(B)} \\ \hline
\multicolumn{1}{|l|}{stageid}                        & \multicolumn{1}{l|}{}       & \multicolumn{1}{l|}{}      & \multicolumn{1}{l|}{21.862} & \multicolumn{1}{l|}{4}  & \multicolumn{1}{l|}{0}     & \multicolumn{1}{l|}{}       \\ \hline
\multicolumn{1}{|l|}{stageid(1)}                     & \multicolumn{1}{l|}{-1.575} & \multicolumn{1}{l|}{0.435} & \multicolumn{1}{l|}{13.084} & \multicolumn{1}{l|}{1}  & \multicolumn{1}{l|}{0}     & \multicolumn{1}{l|}{0.207}  \\ \hline
\multicolumn{1}{|l|}{stageid(2)}                     & \multicolumn{1}{l|}{-0.472} & \multicolumn{1}{l|}{0.458} & \multicolumn{1}{l|}{1.065}  & \multicolumn{1}{l|}{1}  & \multicolumn{1}{l|}{0.302} & \multicolumn{1}{l|}{0.624}  \\ \hline
\multicolumn{1}{|l|}{stageid(3)}                     & \multicolumn{1}{l|}{0.031}  & \multicolumn{1}{l|}{0.705} & \multicolumn{1}{l|}{0.002}  & \multicolumn{1}{l|}{1}  & \multicolumn{1}{l|}{0.965} & \multicolumn{1}{l|}{1.032}  \\ \hline
\multicolumn{1}{|l|}{stageid(4)}                     & \multicolumn{1}{l|}{-1.554} & \multicolumn{1}{l|}{0.459} & \multicolumn{1}{l|}{11.481} & \multicolumn{1}{l|}{1}  & \multicolumn{1}{l|}{0.001} & \multicolumn{1}{l|}{0.211}  \\ \hline
\multicolumn{1}{|l|}{Openness}                       & \multicolumn{1}{l|}{-0.017} & \multicolumn{1}{l|}{1.306} & \multicolumn{1}{l|}{0}      & \multicolumn{1}{l|}{1}  & \multicolumn{1}{l|}{0.989} & \multicolumn{1}{l|}{0.983}  \\ \hline
\multicolumn{1}{|l|}{Conscientiousness}              & \multicolumn{1}{l|}{-0.296} & \multicolumn{1}{l|}{1.296} & \multicolumn{1}{l|}{0.052}  & \multicolumn{1}{l|}{1}  & \multicolumn{1}{l|}{0.819} & \multicolumn{1}{l|}{0.744}  \\ \hline
\multicolumn{1}{|l|}{Extraversion}                   & \multicolumn{1}{l|}{-1.305} & \multicolumn{1}{l|}{0.835} & \multicolumn{1}{l|}{2.443}  & \multicolumn{1}{l|}{1}  & \multicolumn{1}{l|}{0.118} & \multicolumn{1}{l|}{0.271}  \\ \hline
\multicolumn{1}{|l|}{Agreeableness}                  & \multicolumn{1}{l|}{2.031}  & \multicolumn{1}{l|}{0.798} & \multicolumn{1}{l|}{6.475}  & \multicolumn{1}{l|}{1}  & \multicolumn{1}{l|}{0.011} & \multicolumn{1}{l|}{7.624}  \\ \hline
\multicolumn{1}{|l|}{Neuroticism}                    & \multicolumn{1}{l|}{-0.272} & \multicolumn{1}{l|}{0.929} & \multicolumn{1}{l|}{0.086}  & \multicolumn{1}{l|}{1}  & \multicolumn{1}{l|}{0.77}  & \multicolumn{1}{l|}{0.762}  \\ \hline
\multicolumn{1}{|l|}{Log\_OP by Openness}            & \multicolumn{1}{l|}{3.508}  & \multicolumn{1}{l|}{3.293} & \multicolumn{1}{l|}{1.135}  & \multicolumn{1}{l|}{1}  & \multicolumn{1}{l|}{0.287} & \multicolumn{1}{l|}{33.365} \\ \hline
\multicolumn{1}{|l|}{Conscientiousness by Log\_Cons} & \multicolumn{1}{l|}{0.843}  & \multicolumn{1}{l|}{3.154} & \multicolumn{1}{l|}{0.071}  & \multicolumn{1}{l|}{1}  & \multicolumn{1}{l|}{0.789} & \multicolumn{1}{l|}{2.324}  \\ \hline
\multicolumn{1}{|l|}{Extraversion by Log\_ext}       & \multicolumn{1}{l|}{0.141}  & \multicolumn{1}{l|}{2.595} & \multicolumn{1}{l|}{0.003}  & \multicolumn{1}{l|}{1}  & \multicolumn{1}{l|}{0.957} & \multicolumn{1}{l|}{1.151}  \\ \hline
\multicolumn{1}{|l|}{Agreeableness by Log\_Agree}    & \multicolumn{1}{l|}{-2.017} & \multicolumn{1}{l|}{2.288} & \multicolumn{1}{l|}{0.778}  & \multicolumn{1}{l|}{1}  & \multicolumn{1}{l|}{0.378} & \multicolumn{1}{l|}{0.133}  \\ \hline
\multicolumn{1}{|l|}{Log\_neuro by Neuroticism}      & \multicolumn{1}{l|}{1.965}  & \multicolumn{1}{l|}{2.538} & \multicolumn{1}{l|}{0.599}  & \multicolumn{1}{l|}{1}  & \multicolumn{1}{l|}{0.439} & \multicolumn{1}{l|}{7.133}  \\ \hline
\multicolumn{1}{|l|}{Constant}                       & \multicolumn{1}{l|}{3.505}  & \multicolumn{1}{l|}{2.205} & \multicolumn{1}{l|}{2.527}  & \multicolumn{1}{l|}{1}  & \multicolumn{1}{l|}{0.112} & \multicolumn{1}{l|}{33.297} \\ \hline
                                                     &                             &                            &                             &                         &                            &                            
\end{tabular}
}
\caption{Logistic Regression -- variables in the equation table}
\label{BinaryRegressionVarl}
\end{table}

According to the Logistic Regression result, Table~\ref{BinaryRegressionVarl} Linearity of the ``big five" traits concerning the logit transformation of the ``Sentiment Analysis" was evaluated via the Box-Tidwell method \cite{Box1962}. A Bonferroni correction was applied using all twelve (Big five, Log Big Five, StageID and Sentiment Output) terms in the model occurring in a valid statistical significance, since ${p < .00416}$ \cite{Tabachnick2001}. Based on this evaluation, all the ``big five" traits were observed to be linearly correlated to the logit of the sentiment output.

\subsection{Key Findings and Discussion}

There was one studentised residual with a value of ${-5.817511} standard deviations, which retained in the investigation. The logistic regression model was statistically significant, ${χ2(9) = 31.853}, ${p < .0005}$. The model explained 9\% (Nagelkerke R2) of the change in sentiment raised in different stages and accurately classified 80.2\% of cases as shown in Table~\ref{BinomialLogisticModelTable}. Sensitivity was 99.2\%, specificity was 7.4\%, the positive predictive value was 19.5\%, and the negative predictive value was 80.4\%, of the nine predictor variables, there were statistically significant: Stages (1,4), \emph{Extraversion}, \emph{Agreeableness}, \emph{Conscientiousness}  Table~\ref{BinaryRegressionVarl}. It is notable that there is a big gap between the positive predictive value and the negative value.

\begin{table}[!ht]
\centering
\begin{tabular}{@{}c|llll@{}}
\toprule
\multirow{3}{*}{Observed} &                       & \multicolumn{3}{c}{Predicted}                                      \\ \cmidrule(l){2-5} 
                          &                       & \multicolumn{2}{l}{Sentiment}                 & Percentage Correct \\ \cmidrule(l){2-5} 
                          &                       & Pos                   & Neg                   &                    \\ \midrule
sentiment\_label          & Pos                   & 5                     & 63                    & 7.4                \\
\multicolumn{1}{l|}{}     & Neg                   & 2                     & 259                   & 99.2               \\ \midrule
Overall Percentage        & \multicolumn{1}{l|}{} & \multicolumn{1}{l|}{} & \multicolumn{1}{l|}{} & 80.2               \\ \bottomrule
\end{tabular}
\caption{Binomial Logistic regression Classification}
\label{BinomialLogisticModelTable}
\end{table}

\section{Relationship Between Personality Traits and Emotion}\label{personalityTraitsTemporal}

\subsection{Introduction}

The increase usage of online platform which involve the daily bases life has challenge us to develop a model for the type of user’s using these platform, by understanding the personality and emotion raised while using the system, that would lead to signification improvement in the architecture of the complex computer system not only the design, it is delivering the information to the users.

Research into personality traits have been challenge for many researchers in different fields , for past 100 years interest in developing technology that has the ability to recognise people’s personality and emotions~\cite{Taylor2005} has grown rapidly . Recently researchers have started to investigate the relation between the social on-line behaviour and the real life behaviour. This research interested in capturing the emotions used in the test, and according to the literature as discussed in Section~\ref{cognitiveScience}, different approach has been introduced to capture the emotions, one of them is using IBM Watson (see Section~\ref{IBMWatson}) and other is using TEIQue assessment~\cite{Petrides2001}, therefore, it was vital to investigate the association between Big Five Personality Traits and the EI traits using TEIQue Assessment. 

This experiment focus on investigating the relation between the personality traits (Big Five) and the Emotional Intelligent behaviour (TEIQue), for the Facebook dataset, in order to decide either to use TEIQue or Lexicon Basic Emotion approach.

\subsection{Personality Traits and Temporal Behaviour}\label{bigfiveEI}

The dataset using in this experiment is retrieved from the web-based scholarship system (see Section~\ref{SourceData}), 72 participants completed a Big Five Personality Traits personality questionnaire and TEIQue assessment questionnaire after submitting their application, as part of the user experience improvement suggested by Scholarship Administrator team. 

The type variables in the dataset suggested a use of Pearson's correlation o investigate the association between the Big Five Personality traits and the EI traits.

\subsubsection{Pearson Correlations}

\begin{table}[!ht]
\centering
\resizebox{\textwidth}{!}{

\begin{tabular}{@{}lllllll@{}}
\toprule
\multicolumn{7}{c}{\textbf{Tests of Normality}}                                           \\ \midrule
                  & Kolmogorov-Smirnova & Shapiro-Wilk &           &       &      &       \\
Statistic         & df                  & Sig.         & Statistic & df    & Sig. &       \\
Extraversion      & 0.088               & 28           & .200*     & 0.98  & 28   & 0.857 \\
Agreeableness     & 0.135               & 28           & .200*     & 0.938 & 28   & 0.096 \\
Conscientiousness & 0.148               & 28           & 0.122     & 0.948 & 28   & 0.178 \\
Neuroticism       & 0.176               & 28           & 0.026     & 0.93  & 28   & 0.06  \\
Openness          & 0.131               & 28           & .200*     & 0.966 & 28   & 0.477 \\
Wellbeing         & 0.192               & 28           & 0.01      & 0.81  & 28   & 0     \\
Selfcontrol       & 0.15                & 28           & 0.106     & 0.942 & 28   & 0.121 \\
Emotionality      & 0.083               & 28           & .200*     & 0.986 & 28   & 0.965 \\
Sociability       & 0.118               & 28           & .200*     & 0.932 & 28   & 0.068 \\ \bottomrule
\end{tabular}
}
\caption{Shapiro-Wilk's normality check for Big Five traits and EI traits.}
\label{tbl:normalityCheckBigFiveEITraits}
\end{table}

To assess the statistical significance of Pearson's correlation coefficient, a normality assumption need to be verified first and test the level of normality for all variables involved before proceeding to Pearson Correlation. Table~\ref{tbl:normalityCheckBigFiveEITraits} shows that none of the variables were normally distributed, as assessed by Shapiro-Wilk's test (${p < 0.05}$), Therefore, Spearman's rank-order correlation suggested to be used instead as it can be used to measure the strength and direction of the association between either two continuous variables. Furthermore, it is still possible to run Pearson's Correlation Coefficient as the test is somewhat robust to deviations from normality.

\begin{table}[]
\centering
\resizebox{\textwidth}{!}{

\begin{tabular}{@{}llllll@{}}
\toprule
\multicolumn{6}{c}{\textbf{Correlations}}                                                                 \\ \midrule
                             &                     & Wellbeing & Selfcontrol & Emotionality & Sociability \\
\textbf{Extraversion}      & Pearson Correlation & .255      & .026        & .222         & .350        \\
                             & Sig. (2-tailed)     & .190      & .895        & .256         & .067        \\
\textbf{Agreeableness}     & Pearson Correlation & .452*     & .219        & .297         & .175        \\
                             & Sig. (2-tailed)     & .016      & .263        & .125         & .373        \\
\textbf{Conscientiousness} & Pearson Correlation & .465*     & .076        & .173         & .143        \\
                             & Sig. (2-tailed)     & .013      & .702        & .379         & .468        \\
\textbf{Neuroticism}       & Pearson Correlation & -.550**   & -.506**     & -.235        & -.492**     \\
                             & Sig. (2-tailed)     & .002      & .006        & .229         & .008        \\
\textbf{Openness}          & Pearson Correlation & .393*     & .296        & .263         & .323        \\
                             & Sig. (2-tailed)     & .039      & .127        & .177         & .094        \\ \bottomrule
\end{tabular}
}
\caption{Pearson correlation coefficient, Big Five and EI traits}
\label{PearsonBigFiveEi}
\end{table}

According to Table~\ref{PearsonBigFiveEi}, there was a moderate positive correlation between \emph{Well being}, \emph{Agreeableness}, \emph{Conscientiousness} and \emph{Openness}, r = .452, r=0.465 and r=0.393 and negative moderate correlation, between \emph{Neuroticism} and \emph{Well being}, \emph{Self control} and \emph{Sociability}, r=-.550, r=-.506 and r=-.492.

\subsubsection{Spearman's Rank-Order Correlation}

\begin{table}[]
\centering
\resizebox{\textwidth}{!}{

\begin{tabular}{@{}llllll@{}}
\toprule
\multicolumn{6}{c}{\textbf{Correlation}}                                                                      \\ \midrule
                             &                         & Wellbeing & Selfcontrol & Emotionality & Sociability \\
\textbf{Extraversion}      & Correlation Coefficient & .158      & -.023       & .278         & .288        \\
                             & Sig. (2-tailed)         & .423      & .907        & .153         & .137        \\
\textbf{Agreeableness}     & Correlation Coefficient & .413*     & .193        & .255         & .066        \\
                             & Sig. (2-tailed)         & .029      & .324        & .190         & .737        \\
\textbf{Conscientiousness} & Correlation Coefficient & .514**    & -.035       & .178         & .114        \\
                             & Sig. (2-tailed)         & .005      & .859        & .364         & .563        \\
\textbf{Neuroticism}       & Correlation Coefficient & -.383*    & -.484**     & -.212        & -.357       \\
                             & Sig. (2-tailed)         & .044      & .009        & .280         & .062        \\
\textbf{Openness}          & Correlation Coefficient & .329      & .214        & .260         & .236        \\
                             & Sig. (2-tailed)         & .087      & .274        & .182         & .227        \\ \bottomrule
\end{tabular}
}
\caption{Spearman's Rank-Order correlation output}
\label{SpearmanBigFiveEITraits}
\end{table}

As the normality check on Table~\ref{tbl:normalityCheckBigFiveEITraits} reported that none of the variables were normally disrupted and to verify the output of the Pearson Correlation Coefficient reported on Table~\ref{PearsonBigFiveEi}. Spearman's rank-order correlation has been suggested to confirm the association. Table~\ref{SpearmanBigFiveEITraits}, shows that there was a positive correlation between \emph{Conscientiousness} and \emph{Well being}, rs = .514, and Negative correlation between \emph{Neuroticism} and \emph{self control}. rs=-.484.

\subsection{Association between Personality Traits and Six Basic Emotions}\label{bigfiveBasicEmotions}

This experiment is a replication of the previous experiment to investigate the association between Big Personality Traits and basic emotions (as initially presented in Section~\ref{cognitiveScience}). The dataset used in this experiment is consisted of 477 interaction extracted from scholarship system, through different data source as explaind previously in Section~\ref{SourceData}. IBM Watson (as presented in Section~\ref{IBMWatson}) used as an API to extract Big Personality Traits and Basic Emotions, based on the lexicon approach.

\subsubsection{Pearson Correlation}

\begin{figure}[!ht]
\centering
\includegraphics[width=400px,keepaspectratio]{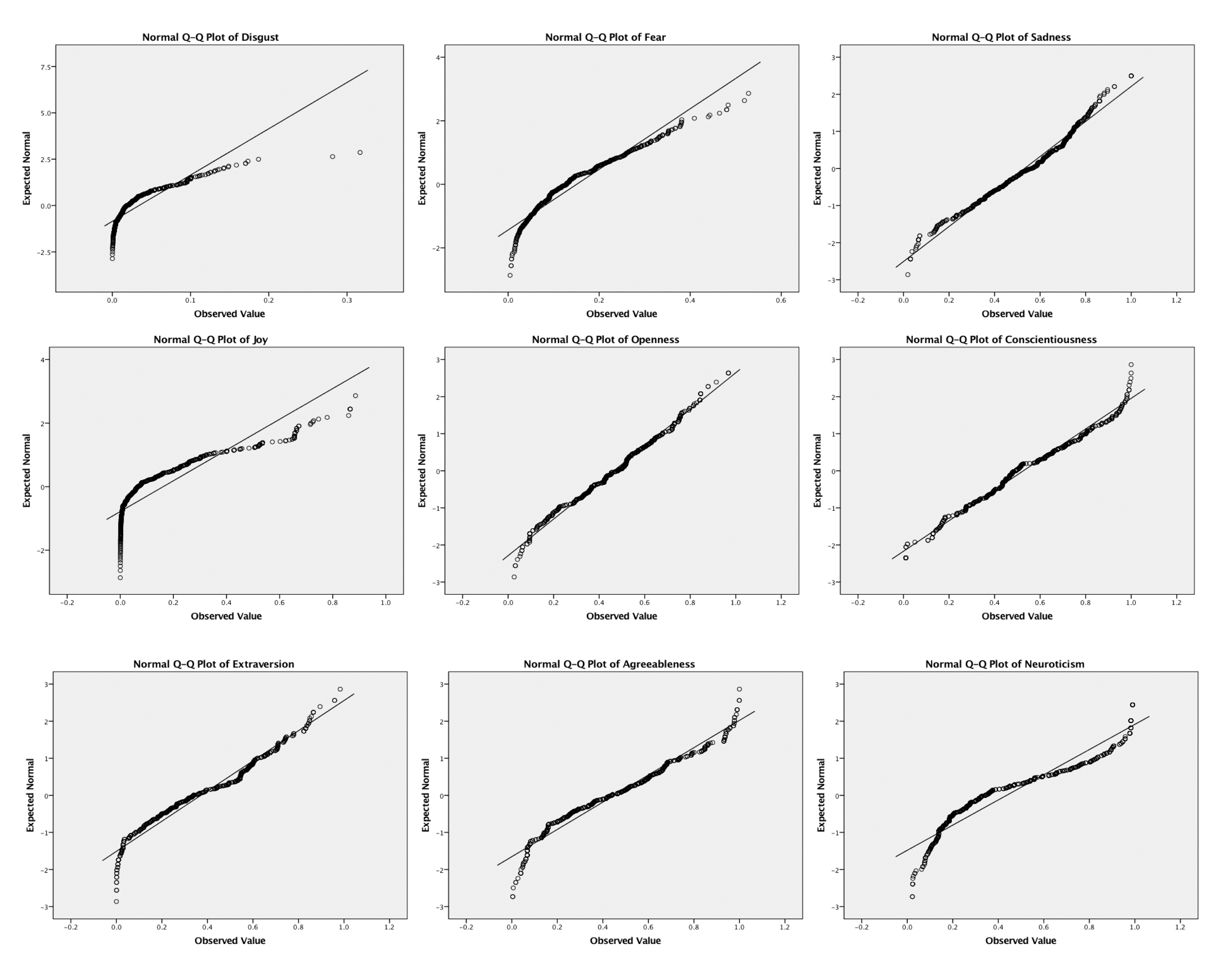}
\caption{Normal Q-Q plot for Big Five traits and basic emotion}
\label{fig:NormalQQBigFiveBasic} 
\end{figure}

Figure~\ref{fig:NormalQQBigFiveBasic}, shows the assumption of normality for Big Five Traits and Basic Emotions was satisfied for most of the variables as assessed by visual inspection of Normal Q-Q Plots.

\begin{table}[!ht]
\centering
\resizebox{\textwidth}{!}{%
\begin{tabular}{@{}lllllll@{}}
\toprule
\multicolumn{7}{c}{\textbf{Pearson's Correlations}}                                                                                                             \\ \midrule
                 &                     & \textbf{Openness} & \textbf{Conscientiousness} & \textbf{Extraversion} & \textbf{Agreeableness} & \textbf{Neuroticism} \\
\textbf{Anger}   & Pearson Correlation & .045              & .014                       & -.002                 & -.068                  & .012                 \\
\textbf{}        & Sig. (2-tailed)     & .330              & .759                       & .963                  & .140                   & .799                 \\
\textbf{Disgust} & Pearson Correlation & .071              & .041                       & .006                  & -.086                  & -.074                \\
\textbf{}        & Sig. (2-tailed)     & .122              & .369                       & .895                  & .062                   & .107                 \\
\textbf{Fear}    & Pearson Correlation & .015              & -.092*                     & .003                  & -.021                  & -.248**              \\
\textbf{}        & Sig. (2-tailed)     & .752              & .044                       & .944                  & .650                   & .000                 \\
\textbf{Joy}     & Pearson Correlation & .083              & .126**                     & -.016                 & -.046                  & .106*                \\
\textbf{}        & Sig. (2-tailed)     & .069              & .006                       & .731                  & .311                   & .021                 \\
\textbf{Sadness} & Pearson Correlation & -.123**           & .025                       & -.053                 & .069                   & .068                 \\
\textbf{}        & Sig. (2-tailed)     & .007              & .590                       & .244                  & .134                   & .136                
\end{tabular}
}
\caption{Pearson correlation coefficient, Big Five and Basic emotions}
\label{PearsonBigFiveBasicEmotions}
\end{table}

Table~\ref{PearsonBigFiveBasicEmotions} shows a Pearson's correlation coefficient was run to assess the relationship between Big Five traits and basic emotions. Preliminary analyses showed the relationship to be linear with both variables normally distributed, as assessed by visual inspection of Normal Q-Q Plots, and there were no outliers. There was a moderate negative correlation between \emph{fear} and \emph{Conscientiousness},\emph{Neuroticism}, ${r(447) = -.092}$, ${p < .05}$, ${r(447) = -.248}$, ${p < .05}$ and moderate positive correlation between \emph{joy} and \emph{Conscientiousness}, \emph{Neuroticism}, ${r(447) = -.126}$, ${p < .05}$, ${r(447) = -.106}$, ${p < .05}$. 

\subsection{Discussion}

The findings from experiment (see Section~\ref{bigfiveEI}) is relevant to, and can be interpreted from the perspective of, the emerging literature on the general factor of personality~\cite{Figueredo2009,Hofstee2001,Musek2007}. The output agrees with the the reported findings in study conducted by McCrae~\cite{McCrae2002}, there are a moderate correlation between \emph{Neuroticism}, \emph{self control} and \emph{well being}.  \emph{Well being} has a moderate correlation to \emph{agreeableness} and \emph{conscientiousness}. Beyond the theoretical value of these data, the results demonstrate the practical equivalence of TEIQue and in relation to their associations with Big Five personality traits. However, according to the literature as previously discussed in Section~\ref{IBMWatson}, emotions can be captured using linguistic analysis, therefore, further experiments presented in Section~\ref{bigfiveBasicEmotions} were conducted to explore the association between Big Five traits and basic emotions. The findings suggested a moderate correlations between the traits as explained previously. Furthermore, as both theories interested in capturing the emotions it was decided to move forward with the \emph{basic emotions (lexion approach)} as it can capture the emotions from the text which fits the direction of this study.

\section{Investigating Behavioural and Emotional Change}\label{emotionSystemStages}

\subsection{Introduction}

The experiments presented in Section~\ref{subs:sentimentStages} suggest an 80.2\% accuracy of the proposed model, which proves the assumption of strong correlation between personality traits, sentiment and the stages of the application form. Therefore, this experiment has been introduced as an extension to the previous model to extend the sentiment instead of being \emph{negative} and \emph{positive} to include the emotion extracted using IBM Watson Tone Analyzer stated on Section~\ref{section:extractEmotions}. This experiment will focus on incorporating the emotion tone \emph{Anger, Fear, Disgust, Joy and Sadness}. Although, according to the literature discussed in Section~\ref{IBMWatson}, it is essential to investigate the fundamental emotions and it is an association with Big Five Personality trait.

The dataset is prepared from the previous experiment in Section~\ref{subs:sentimentStages}, however, an extension preparation is required to cover the new requirements for this experiment. In this respect, the data in Table~\ref{tbl:StagesText} is used again to extract the emotion based on the stages reported in the Table~\ref{stagestablel}. Table~\ref{tbl:StagesEmotions} shows sample of the output from the IBM Watson Tone Analyzer. The column \emph{StagesTextID} associated with the text from Table~\ref{tbl:StagesText}, and the attributes \emph{Anger, Disgust , Fear, Joy and Sadness} all adding to 1.

\begin{table}[!ht]
\centering

\begin{tabular}{@{}llllll@{}}
\toprule
StagesTextID & Anger    & Disgust  & Fear     & Joy      & Sadness  \\ \midrule
1            & 0.236374 & 0.148785 & 0.167954 & 0.116915 & 0.517816 \\
2            & 0.164751 & 0.015199 & 0.263669 & 0.01941  & 0.615725 \\
3            & 0.099684 & 0.005159 & 0.224768 & 0.003826 & 0.740602 \\
4            & 0.322635 & 0.007574 & 0.032424 & 0.000671 & 0.742377 \\
5            & 0.159617 & 0.029623 & 0.130249 & 0.004624 & 0.765329 \\
6            & 0.176493 & 0.02529  & 0.196468 & 0.101357 & 0.493553 \\
7            & 0.237498 & 0.066133 & 0.136312 & 0.185933 & 0.331277 \\
8            & 0.36853  & 0.119263 & 0.175762 & 0.000132 & 0.508752 \\
9            & 0.103784 & 0.014536 & 0.205593 & 0.004806 & 0.748736 \\
10           & 0.109112 & 0.002726 & 0.338946 & 0.002019 & 0.638848 \\
11           & 0.249994 & 0.047278 & 0.137731 & 0.002621 & 0.684495 \\
12           & 0.654401 & 0.012053 & 0.143045 & 0.00026  & 0.333567 \\
13           & 0.127089 & 0.047328 & 0.196604 & 0.003453 & 0.723312 \\
14           & 0.215104 & 0.011095 & 0.106291 & 0.015954 & 0.729163 \\
15           & 0.23544  & 0.004584 & 0.092963 & 0.06229  & 0.624768 \\
16           & 0.23544  & 0.004584 & 0.092963 & 0.06229  & 0.624768 \\ \bottomrule
\end{tabular}
\caption{Sample of the emotion extraction output}
\label{tbl:StagesEmotions}
\end{table}

\subsection{Ordinal Regression Analysis}\label{emotionOrdinal}

The Ordinal regression analysis suggested to be used on this dataset, as noted earlier in Section~\ref{subs:sentimentStages}, the dataset consist of different types of variables (Big Five and Emotions) considered as continues variables however, the stages is a grouped variable (Ordinal Variable). Therefore, the ordinal regression analysis suggested for this experiment.

A critical part of the process involves verification phase to make sure that the data being analyse can actually be analysed using this test. The ordinal logistic regression has four assumptions that you have to consider.  \emph{(a) you have an ordinal dependent variable, (b) dataset consist of one or more independent variables that are continuous. (c) There should be no multicollinearity (d) There should be a proportional odds}. The assumption \emph{(a)} and \emph{(b)} is valid in the dataset, as shown previously the dependent variable \emph{StageID} -- Table~\ref{stagestablel} is type of ordinal values scale from \emph{ 1 to 5}. The \emph{(b)} assumption is valid as The Big Five and emotions values is type of continuous as shown in Tables~\ref{tbl:StagesEmotions} and ~\ref{PearsonBigFiveBasicEmotions}.

\paragraph{Testing the assumption of multicollinearity}
Discovering whether there is multicollinearity is an essential step in ordinal logistic regression. To test the  Multicollinearity, we need to check if two or multiple variables (independent) are correlacted with each other. If this occurs will lead to a lack of understanding which independent variable explains the dependent, and this will give an inaccurate association for the model.

\begin{table}[!ht]
\centering
\begin{tabular}{@{}lll@{}}
\toprule
Model             & Tolerance & VIF   \\ \midrule
Anger             & .674      & 1.484 \\
Disgust           & .927      & 1.079 \\
Fear              & .662      & 1.510 \\
Joy               & .273      & 3.661 \\
Sadness           & .385      & 2.599 \\
Openness          & .782      & 1.279 \\
Conscientiousness & .836      & 1.197 \\
Extraversion      & .715      & 1.398 \\
Agreeableness     & .618      & 1.619 \\
Neuroticism       & .846      & 1.182 \\ \bottomrule
\end{tabular}
\caption{Multicollinearity output - Coefficients (Dependent Variable: stageid)}
\label{tbl:multiEmotionsSta}
\end{table}

In examining the Tolerance values and variance inflation factor (VIF) values, according to Table~\ref{tbl:multiEmotionsSta} all the ${\emph{Tolerance} > 0.1}$ (the lowest is 0.273), and VIF values are much less than 10. Therefore, it is fairly certain that there is no issue with collinearity in this dataset. The findings confirm the validity of the assumption \emph{(c)}.

\paragraph{The assumption of proportional odds} 
Progressing to assumption \emph{(d)}, the assumption of proportional odds is essential to the kind of ordinal logistic regression. Each \emph{independent variable} has the same influence at each developing separation of the ordinal \emph{dependent variable}. This assumption can be examined using two approaches: \emph{(a) with a full likelihood ratio test} associating the fit of the proportional odds model to a model with different location parameters, and \emph{(b) by running separate binomial logistic regressions} on cumulative dichotomous dependent variables. 
\begin{itemize}
\item{\textbf{Full likelihood ratio test}}
This test works by measuring the model fit within two separate models. That two models that we are interested in for this test are \emph{the proportional odds model - null hypothesis row-} and \emph{the proportional odds constraint/assumption (the ``General" row)}. That where the slope coefficients are provided to be different in each logit value.

The assumption of proportional odds will be valid only if the difference in model fit is small and not statistically significant ${p > 0.05}$. On the other hand, the assumption of is invalid ${p < 0.05}$, which means the model fit is substantial and statistically significant.

According to Table~\ref{tbl:fullratio}, \emph{${p = 0.001}$}, which is less than $0.05$. Therefore, the assumption of proportional odds is violated. By violating this assumption, therefore each independent variable  cannot be treat as having the same impact for each cumulative logit.

\begin{table}[!ht]
\centering
\begin{tabular}{@{}ccccc@{}}
\toprule
\textbf{Model}  & \textbf{-2 Log Likelihood} & \textbf{Chi-Square} & \textbf{df} & \textbf{Sig.} \\ \midrule
Null Hypothesis & 1226.624                   &                     &             &               \\
General         & 1167.651                   & 58.974              & 30          & .001          \\ \bottomrule
\end{tabular}
\caption{Full likelihood ratio test - Test of Parallel Lines}
\label{tbl:fullratio}
\end{table}

\item{\textbf{Separate binomial logistic regressions}}

The full likelihood ratio test flagged violations of the assumption of proportional odds that do not exist. Therefore, a wider examination of the assumption of proportional odds needed by running separate binomial logistic regressions on the dichotomous dependent variables. According to Hardy (1993)~\cite{Hardy1993}, many regression procedures, such as linear regression or logistic regression, do not accept categorical variables directly into the model: they have to recoded first. This recording can take many different forms, with the most popular called \emph{indicator coding}~\cite{pasta2005parameterizing}. This process will perform a separation method which will divide the dependent dichotomous variable into separate parameters that number one less than the number of classes of the dichotomous variable. Creating a new variables \emph{Stage1, Stage 2, Stage 3 and Stage 4}.

\begin{table}[]
\centering
\resizebox{\textwidth}{!}{
\begin{tabular}{@{}lllllllll@{}}
\toprule
\textbf{}                     & \multicolumn{4}{l}{\textbf{B (Parameter estimates)}}                  & \multicolumn{4}{l}{\textbf{Exp(B) (Odds Ratio, OR)}}                      \\ 
\textbf{Independent variable} & \textbf{Stage1} & \textbf{Stage2} & \textbf{Stage3} & \textbf{Stage4} & \textbf{Stage 1} & \textbf{Stage 2} & \textbf{Stage 3} & \textbf{Stage 4} \\ \midrule
Anger                         & 2.246           & .775            & .620            & -.316           & 9.455            & 2.171            & 1.859            & .729             \\
Disgust                       & 7.918           & 6.394           & 8.632           & 8.630           & 2745.452         & 598.318          & 5606.564         & 5596.071         \\
Fear                          & -1.559          & -.517           & -.241           & .317            & .210             & .596             & .786             & 1.373            \\
Joy                           & -.770           & -.130           & .458            & 1.039           & .463             & .878             & 1.581            & 2.825            \\
Sadness                       & -2.869          & -1.689          & -1.147          & -1.383          & .057             & .185             & .318             & .251             \\
Openness                      & -.257           & .121            & .204            & .312            & .773             & 1.129            & 1.227            & 1.366            \\
Conscientiousness             & -.373           & -.005           & -.595           & -.810           & .689             & .995             & .551             & .445             \\
Extraversion                  & -.531           & -.179           & -.360           & -.321           & .588             & .836             & .698             & .726             \\
Agreeableness                 & .070            & .035            & .166            & .320            & 1.072            & 1.036            & 1.181            & 1.377            \\
Neuroticism                   & -.391           & -.220           & -.183           & -.687           & .676             & .802             & .833             & .503             \\
Constant                      & .491            & .601            & .702            & 1.544           & 1.634            & 1.824            & 2.018            & 4.684            \\ \bottomrule
\end{tabular}
}
\caption{Parameter estimates and odd ratios for the dichotomised cumulative categories of the dependent variable}
\label{tbl:binomialeq}
\end{table}

Table~\ref{tbl:binomialeq} presents the information extracted from the Variables in the Equation tables to allow more straightforward comparison. Essentially, the assumption of proportional odds states that the estimated parameters, except the intercept (i.e., just the slope coefficients), are equal for each binomial logistic regression run on each dichotomised cumulative category; only the intercept –- called the threshold in ordinal regression –- is free to vary. If this assumption is tenable, the coefficients above should be similar for \emph{Stage1, Stage 2, Stage 3 and Stage 4}. However, it usually makes more sense to look at the differences or similarities between the odds ratios.

\emph{Openness,Extraversion and Agreeableness} in the Table~\ref{tbl:binomialeq}, odd ratios for four different binomial logistic regression are similar (i.e, 1.072, 1.036, 1.181 and 1.377). It would appear that, for this variable, the assumption of proportional odds appears tenable. However, consider \emph{anger, disgust, fear, joy, sadness, conscientiousness, extraversion, agreeableness and neuroticism} (i.e, 2745.452, 598.318	5606.564 and 5596.071). The assumption of similar odds for this variables might not be tenable. Therefore, treating those variable in the final ordinal regression with more caution.
\end{itemize}

\begin{table}[!ht]
\centering
\begin{tabular}{@{}llll@{}}
\toprule
         & \textbf{Chi-Square} & \textbf{df} & \textbf{Sig.} \\ \midrule
Pearson  & 1299.332            & 1326        & .694          \\
Deviance & 1098.406            & 1326        & 1.000         \\ \bottomrule
\end{tabular}
\caption{Goodness-of-Fit}
\label{tbl:ordinalgoodfit}
\end{table}

\begin{table}[!ht]
\centering
\begin{tabular}{@{}llll@{}}
\toprule
\textbf{Source}   & \textbf{Wald Chi-Square} & \textbf{df} & \textbf{Sig.} \\ \midrule
Anger             & 1.341                    & 1           & .247          \\
Disgust           & 12.046                   & 1           & .001          \\
Fear              & .102                     & 1           & .749          \\
Joy               & .127                     & 1           & .721          \\
Sadness           & 6.986                    & 1           & .008          \\
Openness          & .054                     & 1           & .816          \\
Conscientiousness & 1.556                    & 1           & .212          \\
Extraversion      & .612                     & 1           & .434          \\
Agreeableness     & .203                     & 1           & .652          \\
Neuroticism       & 1.214                    & 1           & .271          \\ \bottomrule
\end{tabular}
\caption{Tests of Model Effects}
\label{tbl:ordinaltest}	
\end{table}

\paragraph{Ordinal Regression Key Findings}. 
A cumulative odds ordinal logistic regression with partial proportional odds was run to understand the impact of Big Five traits and emotions on different stages of the system reported in Table~\ref{stagestablel}. The proportional odds were violated, as assessed by a full likelihood ratio test comparing the fitted model to a model with varying location parameters, ${x^2}{(40) = 58.974}$, ${p = .001}$. The deviance goodness-of-fit test indicated that the model was a good fit to the observed data, ${x^2}{(1326) = 1098.406136}$, ${p = 1.0}$ as shown table ~\ref{tbl:ordinalgoodfit}, but most cells were sparse with zero frequencies in 74.8\% of cells. However, the final model statistically significantly predicted the dependent variable over and above the intercept-only model, ${x^2}{(10) = 42.840503}$, ${p < .001}$. The \emph{disgust and sadness} parameters had a statistically significant effect on the prediction of stageID, ${x^2}{(1) = 12.046}$, ${p = .001}$ and ${x^2}{(1) = 6.086}$, ${p = .008}$ as shown in Table~\ref{tbl:ordinaltest}. Table~\ref{tbl:ordinalfindings} shows an increase in \emph{disgust} is associated with an increase in the odds of being in later stages (4 or 5), with an odds ratio of .041 (95\% CI, 5.04 to .034), Wald ${x^2}{(1) = 12.046}$, ${p <.005}$. Furthermore, an increase in \emph{Sadness} was associated with an increase in the odds of being in later stages (4 or 5) with an odds ratio of 5.356 (95\% CI, 1.543 to 18.589), Wald ${x^2}{(1) = 6.986}$, ${p < .008}$.

\begin{table}[!ht]
\centering
\resizebox{\textwidth}{!}{\begin{tabular}{@{}lllllllllll@{}}
\toprule
\textbf{Parameter} & \textbf{B} & \textbf{Std. Error} & \multicolumn{2}{l}{\textbf{95\% Wald Confidence Interval}} & \textbf{Hypothesis Test} & \textbf{}   & \textbf{}     & \textbf{Exp(B)} & \multicolumn{2}{l}{\textbf{95\% Wald Confidence Interval for Exp(B)}} \\ 
\textbf{Threshold} & \textbf{}  & \textbf{}           & \textbf{Lower}               & \textbf{Upper}              & \textbf{Wald Chi-Square} & \textbf{df} & \textbf{Sig.} & \textbf{}       & \textbf{Lower}                    & \textbf{Upper}                    \\ \midrule
Stage=1            & -.489      & .6809               & -1.823                                 & .846           & .515                     & 1           & .473          & .614            & .162                                              & 2.330          \\
Stage=2            & .707       & .6810               & -.627                                  & 2.042          & 1.079                    & 1           & .299          & 2.028           & .534                                              & 7.706          \\
Stage=3            & 1.016      & .6818               & -.321                                  & 2.352          & 2.220                    & 1           & .136          & 2.761           & .726                                              & 10.507         \\
Stage=4            & 1.536      & .6837               & .196                                   & 2.876          & 5.048                    & 1           & .025          & 4.647           & 1.217                                             & 17.747         \\
Anger              & -1.018     & .8790               & -2.741                                 & .705           & 1.341                    & 1           & .247          & .361            & .065                                              & 2.024          \\
Disgust            & -7.796     & 2.2462              & -12.198                                & -3.393         & 12.046                   & 1           & .001          & .041            & 5.040E-06                                         & .034           \\
Fear               & .312       & .9753               & -1.600                                 & 2.223          & .102                     & 1           & .749          & 1.366           & .202                                              & 9.238          \\
Joy                & -.274      & .7679               & -1.779                                 & 1.231          & .127                     & 1           & .721          & .760            & .169                                              & 3.425          \\
Sadness            & 1.678      & .6349               & .434                                   & 2.923          & 6.986                    & 1           & .008          & 5.356           & 1.543                                             & 18.589         \\
Openness           & -.108      & .4639               & -1.017                                 & .802           & .054                     & 1           & .816          & .898            & .362                                              & 2.229          \\
Conscientiousness  & .470       & .3769               & -.269                                  & 1.209          & 1.556                    & 1           & .212          & 1.600           & .765                                              & 3.349          \\
Extraversion       & .314       & .4013               & -.473                                  & 1.101          & .612                     & 1           & .434          & 1.369           & .623                                              & 3.006          \\
Agreeableness      & -.175      & .3895               & -.939                                  & .588           & .203                     & 1           & .652          & .839            & .391                                              & 1.800          \\
Neuroticism        & .339       & .3076               & -.264                                  & .942           & 1.214                    & 1           & .271          & 1.403           & .768                                              & 2.565          \\ \bottomrule
\end{tabular}
}
\caption{Parameter estimates using the GENLIN procedure}
\label{tbl:ordinalfindings}
\end{table}

\subsection{Multinomial Logistics Regression}
The ordinal regression analysis from the experiment in Section~\ref{emotionOrdinal} reported \emph{disgust, sadness and conscientiousness} as statistically signification with StageID. However, the proportional odd assumption treated with very caution and it suggested partial proportional odd relationship with some parameters. Therefore, it is suggested to use multinomial logistics regression analysis to predict a nominal dependent variable with more than one value. Multinomial Regression required the data to pass six assumption \emph{1-dependent variable should be measured at the nominal level, 2- One or more independent variables that are continuous, 3-Independence of observations and the dependent variable should have mutually exclusive and exhaustive categories -- according to Table~\ref{stagestablel} the dependent variable \emph{stageID} are mutually exclusive and exhaustive (i.e. each record fall into one and only one category) , 4-There should be no multicollinearity, 5-linear relationship between any continuous independent variables and the logit transformation of the dependent variable, 6- No outliers, high leverage values or highly influential points}. From the previous experiment in Section~\ref{emotionOrdinal} the data passes assumption \emph{1,2,3,4 and 6}.

\paragraph{Assumptions}
Assumption \emph{number 5}, linear relationship between any continuous independent variables and the logit transformation of the dependent variable (see Section~\ref{logittranformation}). 
\begin{equation}
    logit(StageID)= ln\frac{StageID}{1-StageID}\label{logittranformation}
\end{equation}

Using linear regressing to investigate the relationship between the Big Five, Emotions and out dependent value (StageID). Table~\ref{tbl:lineoutput} shows the output of the linear regression. According to the table, there is a significant correlation with \emph{Disgust} (B=-.158, p-value=0.001) and \emph{Sadness} (B=.183, p-value=.011). The assumption number 5 is not violated and passed since the two independent variable have significant statistics correlation with the logit of StageID.

\begin{table}[!ht]
\centering
\resizebox{\textwidth}{!}{\begin{tabular}{@{}lllllllll@{}}
\toprule
\textbf{}         & \multicolumn{2}{l}{\textbf{Unstandardised Coefficients}} & \textbf{Standardised Coefficients} & \textbf{t} & \textbf{Sig.} & \multicolumn{3}{l}{\textbf{Correlations}}              \\ 
\textbf{}         & \textbf{B}             & \textbf{Std. Error}             & \textbf{Beta}                      & \textbf{}  & \textbf{}     & \textbf{Zero-order} & \textbf{Partial} & \textbf{Part} \\ \midrule
(Constant)        & 2.373                  & .588                            &                                    & 4.033      & .000          &                     &                  &               \\
Anger             & -.698                  & .760                            & -.050                              & -.918      & .359          & -.067               & -.043            & -.041         \\
Disgust           & -6.392                 & 1.863                           & -.158                              & -3.432     & .001          & -.173               & -.157            & -.153         \\
Fear              & .294                   & .842                            & .019                               & .349       & .727          & .005                & .016             & .016          \\
Joy               & -.207                  & .665                            & -.026                              & -.311      & .756          & -.112               & -.014            & -.014         \\
Sadness           & 1.394                  & .546                            & .183                               & 2.551      & .011          & .213                & .117             & .113          \\
Openness          & -.095                  & .401                            & -.012                              & -.238      & .812          & -.034               & -.011            & -.011         \\
Conscientiousness & .370                   & .324                            & .056                               & 1.144      & .253          & .051                & .053             & .051          \\
Extraversion      & .272                   & .345                            & .041                               & .788       & .431          & .009                & .036             & .035          \\
Agreeableness     & -.114                  & .336                            & -.019                              & -.340      & .734          & .043                & -.016            & -.015         \\
Neuroticism       & .292                   & .265                            & .053                               & 1.100      & .272          & .070                & .051             & .049          \\
                  &                        &                                 &                                    &            &               &                     &                  &               \\ \bottomrule
\end{tabular}
}
\caption{Linear Regression Coefficients output (dependent variable: Stageid)}
\label{tbl:lineoutput}
\end{table}

All assumption for multinomial regression analysis is validated for the dataset, the next section will show the output of the multinomial logistics analysis.

\paragraph{Analysis Findings}

\begin{table}[!ht]
\centering
\begin{tabular}{@{}lllll@{}}
\toprule
\textbf{Model} & \textbf{Model Fitting Criteria} & \multicolumn{3}{l}{\textbf{Likelihood Ratio Tests}} \\ \midrule
               & -2 Log Likelihood               & Chi-Square           & df           & Sig.          \\
Intercept Only & 1269.465                        &                      &              &               \\
Final          & 1186.030                        & 83.435               & 40           & .000          \\ \bottomrule
\end{tabular}
\caption{Multinomial regression output - Model Fitting}
\label{tbl:multifitting}
\end{table}

According to Table~\ref{tbl:multifitting}, \emph{p-value=.000}, the model fits the data significantly better  than the null model. Table~\ref{tbl:multilike}, variables \emph{Disgust, Sadness and Conscientiousness} with  \emph{p-values .006,0.017 and 0.051} has a significant overall effect on the dependent \emph{StageID}.

\begin{table}[!ht]
\centering
\begin{tabular}{@{}lllll@{}}
\toprule
\textbf{Effect}   & \textbf{Model Fitting Criteria}             & \multicolumn{3}{l}{\textbf{Likelihood Ratio Tests}} \\ 
\textbf{}         & \textbf{-2 Log Likelihood of Reduced Model} & \textbf{Chi-Square}  & \textbf{df}  & \textbf{Sig.} \\ \midrule
Intercept         & 1190.707                                    & 4.678                & 4            & .322          \\
Anger             & 1191.875                                    & 5.846                & 4            & .211          \\
Disgust           & 1200.616                                    & 14.587               & 4            & .006          \\
Fear              & 1188.317                                    & 2.287                & 4            & .683          \\
Joy               & 1189.467                                    & 3.438                & 4            & .487          \\
Sadness           & 1198.063                                    & 12.033               & 4            & .017          \\
Openness          & 1187.017                                    & .987                 & 4            & .912          \\
Conscientiousness & 1195.475                                    & 9.445                & 4            & .051          \\
Extraversion      & 1187.510                                    & 1.481                & 4            & .830          \\
Agreeableness     & 1186.722                                    & .692                 & 4            & .952          \\
Neuroticism       & 1192.894                                    & 6.864                & 4            & .143          \\ \bottomrule
\end{tabular}
\caption{Multinomial Regression - Likelihood Ratio Tests}
\label{tbl:multilike}
\end{table}

According to Table~\ref{tbl:multioverall} \emph{Disgust} likely to increase ratio odds  while in \emph{Stage 1} over being in \emph{Stage 5}\footnote{Reference outcome} with in odds ratio of \textbf{194325.6} (95\% CI, 156.670380 to 241031239.97), Wald ${x^2}{(1) = 11.227}$, ${p < .001}$. Output suggested \emph{Sadness} more likely to increase with ratio odds of \textbf{.040} (95\% CI, .005 to .355), Wald ${x^2}{(1) = 8.358}$, ${p < .004}$, while in \emph{Stage 1} decrease over being in \emph{Stage 5}. Findings reported, odds increase \emph{Disgust} with odd ratio 179275.939 to be in \emph{Stage 3} (95\% CI, 9.041 to 3554947303.3), Wald ${x^2}{(1) = 5.741}$, ${p < .017}$. For every one-unit increase in the \emph{Conscientiousness} the ratio odds of being in \emph{Stage 3} decrease by \textbf{.081} than on \emph{Stage 5} (95\% CI,0.14 to .460), Wald ${x^2}{(1) = 8.028}$, ${p < .005}$. For every one-unit increase in the \emph{Neuroticism} the ratio odds of being in \emph{Stage 4} increase by \textbf{.225} than on \emph{Stage 5} (95\% CI,0.066 to .769), Wald ${x^2}{(1) = 5.656}$, ${p < .017}$. 

\begin{table}[!ht]
\centering
\resizebox{\textwidth}{!}{
\begin{tabular}{@{}llllllllll@{}}
\toprule
\multirow{2}{*}{\textbf{stageid (Stage 5 is reference)}} & \multirow{2}{*}{\textbf{}} & \multirow{2}{*}{\textbf{B}} & \multirow{2}{*}{\textbf{Std. Error}} & \multirow{2}{*}{\textbf{Wald}} & \multirow{2}{*}{\textbf{df}} & \multirow{2}{*}{\textbf{Sig.}} & \multirow{2}{*}{\textbf{Exp(B)}} & \multicolumn{2}{l}{\textbf{95\% Confidence Interval for Exp(B)}} \\ \cmidrule(l){9-10} 
                                                         &                            &                             &                                      &                                &                              &                                &                                  & \textbf{Lower Bound}            & \textbf{Upper Bound}           \\ \cmidrule(r){1-8}
\multirow{11}{*}{Stage 1}                                & Intercept                  & 1.603                       & 1.165                                & 1.892                          & 1                            & .169                           &                                  &                                 &                                \\
                                                         & Anger                      & 1.436                       & 1.346                                & 1.139                          & 1                            & .286                           & 4.205                            & .301                            & 58.815                         \\
                                                         & Disgust                    & 12.177                      & 3.634                                & 11.227                         & 1                            & .001                           & 194325.644                       & 156.670                         & 241031239.977                  \\
                                                         & Fear                       & -1.157                      & 1.623                                & .509                           & 1                            & .476                           & .314                             & .013                            & 7.562                          \\
                                                         & Joy                        & .090                        & 1.377                                & .004                           & 1                            & .948                           & 1.094                            & .074                            & 16.268                         \\
                                                         & Sadness                    & -3.217                      & 1.113                                & 8.358                          & 1                            & .004                           & .040                             & .005                            & .355                           \\
                                                         & Openness                   & .001                        & .739                                 & .000                           & 1                            & .999                           & 1.001                            & .235                            & 4.265                          \\
                                                         & Conscientiousness          & -.851                       & .603                                 & 1.992                          & 1                            & .158                           & .427                             & .131                            & 1.392                          \\
                                                         & Extraversion               & -.645                       & .653                                 & .977                           & 1                            & .323                           & .524                             & .146                            & 1.886                          \\
                                                         & Agreeableness              & .288                        & .624                                 & .213                           & 1                            & .644                           & 1.334                            & .393                            & 4.529                          \\
                                                         & Neuroticism                & -.789                       & .495                                 & 2.544                          & 1                            & .111                           & .454                             & .172                            & 1.198                          \\
\multirow{11}{*}{Stage 2}                                & Intercept                  & -.035                       & 1.009                                & .001                           & 1                            & .972                           &                                  &                                 &                                \\
                                                         & Anger                      & -1.245                      & 1.354                                & .845                           & 1                            & .358                           & .288                             & .020                            & 4.093                          \\
                                                         & Disgust                    & 6.481                       & 3.806                                & 2.899                          & 1                            & .089                           & 652.527                          & .376                            & 1133879.043                    \\
                                                         & Fear                       & .584                        & 1.438                                & .165                           & 1                            & .684                           & 1.794                            & .107                            & 30.034                         \\
                                                         & Joy                        & .967                        & 1.127                                & .736                           & 1                            & .391                           & 2.631                            & .289                            & 23.971                         \\
                                                         & Sadness                    & -.812                       & .925                                 & .771                           & 1                            & .380                           & .444                             & .072                            & 2.720                          \\
                                                         & Openness                   & .522                        & .696                                 & .563                           & 1                            & .453                           & 1.685                            & .431                            & 6.590                          \\
                                                         & Conscientiousness          & -.308                       & .552                                 & .312                           & 1                            & .576                           & .735                             & .249                            & 2.167                          \\
                                                         & Extraversion               & -.090                       & .581                                 & .024                           & 1                            & .877                           & .914                             & .293                            & 2.852                          \\
                                                         & Agreeableness              & .219                        & .574                                 & .146                           & 1                            & .703                           & 1.245                            & .404                            & 3.836                          \\
                                                         & Neuroticism                & -.468                       & .455                                 & 1.058                          & 1                            & .304                           & .626                             & .257                            & 1.528                          \\
\multirow{11}{*}{Stage 3}                                & Intercept                  & -1.535                      & 1.418                                & 1.173                          & 1                            & .279                           &                                  &                                 &                                \\
                                                         & Anger                      & -.944                       & 2.172                                & .189                           & 1                            & .664                           & .389                             & .006                            & 27.492                         \\
                                                         & Disgust                    & 12.097                      & 5.049                                & 5.741                          & 1                            & .017                           & 179275.939                       & 9.041                           & 3554947303.303                 \\
                                                         & Fear                       & .945                        & 2.161                                & .191                           & 1                            & .662                           & 2.572                            & .037                            & 177.816                        \\
                                                         & Joy                        & 2.562                       & 1.534                                & 2.790                          & 1                            & .095                           & 12.964                           & .641                            & 262.113                        \\
                                                         & Sadness                    & .834                        & 1.315                                & .402                           & 1                            & .526                           & 2.302                            & .175                            & 30.320                         \\
                                                         & Openness                   & .565                        & 1.035                                & .298                           & 1                            & .585                           & 1.759                            & .231                            & 13.379                         \\
                                                         & Conscientiousness          & -2.517                      & .888                                 & 8.028                          & 1                            & .005                           & .081                             & .014                            & .460                           \\
                                                         & Extraversion               & -.751                       & .958                                 & .616                           & 1                            & .433                           & .472                             & .072                            & 3.082                          \\
                                                         & Agreeableness              & .527                        & .902                                 & .341                           & 1                            & .559                           & 1.693                            & .289                            & 9.921                          \\
                                                         & Neuroticism                & -.507                       & .654                                 & .601                           & 1                            & .438                           & .602                             & .167                            & 2.171                          \\
\multirow{11}{*}{Stage 4}                                & Intercept                  & -.150                       & 1.288                                & .013                           & 1                            & .908                           &                                  &                                 &                                \\
                                                         & Anger                      & -2.805                      & 1.994                                & 1.979                          & 1                            & .159                           & .060                             & .001                            & 3.014                          \\
                                                         & Disgust                    & 3.907                       & 5.125                                & .581                           & 1                            & .446                           & 49.763                           & .002                            & 1147335.006                    \\
                                                         & Fear                       & 1.603                       & 1.820                                & .776                           & 1                            & .378                           & 4.970                            & .140                            & 175.957                        \\
                                                         & Joy                        & 1.551                       & 1.467                                & 1.117                          & 1                            & .291                           & 4.715                            & .266                            & 83.649                         \\
                                                         & Sadness                    & -1.002                      & 1.210                                & .686                           & 1                            & .408                           & .367                             & .034                            & 3.934                          \\
                                                         & Openness                   & .497                        & .887                                 & .314                           & 1                            & .575                           & 1.644                            & .289                            & 9.346                          \\
                                                         & Conscientiousness          & -.731                       & .765                                 & .915                           & 1                            & .339                           & .481                             & .108                            & 2.154                          \\
                                                         & Extraversion               & -.074                       & .753                                 & .010                           & 1                            & .922                           & .929                             & .212                            & 4.063                          \\
                                                         & Agreeableness              & .510                        & .744                                 & .470                           & 1                            & .493                           & 1.665                            & .388                            & 7.149                          \\
                                                         & Neuroticism                & -1.493                      & .628                                 & 5.656                          & 1                            & .017                           & .225                             & .066                            & .769                           \\ \cmidrule(l){2-10} 
\end{tabular}
}
\caption{Multinomial Regression - Parameter Estimates}
\label{tbl:multioverall}
\end{table}

\subsection{Key Findings and Discussion}

The aim of the experiment to extend the previous model as presented in Section~\ref{subs:sentimentStages} to include the emotion parameters to investigate how a model would fit and if there is any significant statistical correlation between the independent variables (BigFive+emotions) and the dependent variable (StageID) as shown in table\ref{stagestablel}. The Ordinal Regression suggested a final model statistically significantly predicted the dependent variable over and above the intercept-only model, χ2(10) = 42.840503, ${p < .001}$. The \emph{disgust and sadness} parameters had a statistically significant effect on the prediction of stageID, ${χ^2}{(1) = 12.046}$, ${p = .001}$ and ${χ^2}{(1) = 6.086}$, ${p = .008}$ as shown in table ~\ref{tbl:ordinaltest}. However, the ordinal assumption Partial Proportion Odds have been violated in some parameters in the model which suggested an inaccurate outcome. Therefore, Multinomial Regression has been used to confirm the output since all six assumptions of the multinomial regression is passed. The output reported is from Table~\ref{tbl:multilike}, variables \emph{Disgust, Sadness and Conscientiousness} with  \emph{p-values .006,0.017 and 0.051} has a significant overall effect on the dependent \emph{StageID}. The output from Ordinal and Multinomial Regression analysis agreed on that the variables \emph{Disgust, Sadness, Conscientiousness, and Neuroticism} has a statistically significant impact on the dependent value \emph{StageID}.


\section{Incorporating Emotion and Personality-Based Analysis in User-centered Modelling}\label{Emotionscentre}

\subsection{Introduction}

As computer systems and applications have become more widespread and complex, with increasing demands and expectations of ever-more intuitive human-computer interactions, research in modelling, understanding and predicting user behaviour demands has become a priority across a number of domains~\cite{mostafa-et-al-ai2016}. In these application domains, it is useful to obtain knowledge about user profiles or models of software applications, including intelligent agents, adaptive systems, intelligent tutoring systems, recommender systems, e-commerce applications and knowledge management systems~\cite{devos-et-al:2005,devos-et-al:2006,schiaffino+amandi:2009}. Furthermore, understanding user behaviour during system events leads to a better informed predictive model capability, allowing the construction of more intuitive interfaces and an improved user experience. This work further builds upon our research published in 2016~\cite{mostafa-et-al-ai2016}.

We are particularly interested in the relationship between digital footprint and behaviour and personality~\cite{oatley+crick:2014,oatley-et-al_dasc2015,blamey-et-al-2012,blamey-et-al-2013}. A wide range of pervasive and often publicly available datasets encompassing digital footprints, such as social media activity, can be used to infer personality~\cite{lambiotte+kosinski:2014,oatley-et-al-soccogcomp2015}. Big social data offers the potential for new insights into human behaviour and development of robust models capable of describing individuals and societies~\cite{lazer-et-al:2009}. Social media has been used in varying computer system approaches; in the past this has mainly been the textual information contained in blogs, status posts and photo comments~\cite{blamey-et-al-2012,blamey-et-al-2013}, but there is also a wealth of information in the other ways of interacting with online artefacts. Research in an image or video analysis includes promising studies on YouTube videos for classification of specific behaviours and indicators of personality
traits~\cite{biel+gatica-perez:2012}. This work uses crowdsourced impressions, social attention, and audio-visual behavioural analysis on slices of conversational video blogs extracted from YouTube. From sharing and gathering of information and data to catering for marketing and business needs; it is now widely used as technical support for computer system platforms.

The work presented in this experiment is based on previous work in psycholinguistic science and aims to provide further insight into how the words and constructs we use in our daily life and online interactions reflect our personalities and our underlying emotions. As part of this active research field, it is widely accepted that written text reflects more than the words and syntactic constructs, but also conveys emotion and personality traits~\cite{pennebaker+king:1999}. As part of our work, the IBM Watson Tone Analyzer (part of the IBM Watson Developer Cloud toolchain) has been used to identify emotion tones in the textual interactions in an online system, building on previous work in this area that shows a strong correlation between the word choice and personality, emotions, attitude and cognitive processes, providing further evidence that it is possible to profile and potentially predict users’ identity~\cite{fast+funder:2008}. The {\emph{Linguistic Inquiry and Word Count}} (LIWC) psycholinguistics dictionary~\cite{Pennebaker2001,tausczik+pennebaker:2010} is used to find psychologically meaningful word categories from word usage in writing; the work presented here provides a modelling and analysis framework, as well as associated toolchain, for further application to larger datasets to support the research goal of improving user-centered modelling~\cite{mostafa-et-al-ai2016}.

\begin{figure}[!ht]
\centering
\includegraphics[height=60px,keepaspectratio]{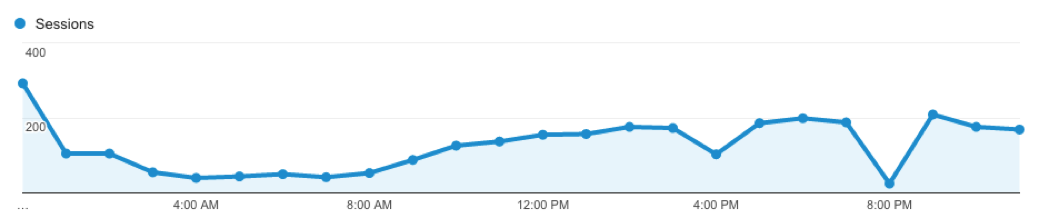}
\caption[Google Analytics profile showing behaviour of the system over a
  24 hour period]{Google Analytics profile showing behaviour of the system over a
  24 hour period (timeline during the day vs. number of active sessions)}
\label{fig:googleanalytics} 
\end{figure}

The dataset used in this experiment (see Section~\ref{SourceData}) consists of users (N=391), interactions and comments (N=1390) as responses to system status and reporting their experience with using the system. Google Analytics has been used to track user behaviour and web statistics (such as impressions); this data from has been used to identify the server's status and categorised the status as two stages: \emph{Idle}, where the system had a higher number of active sessions; and marked as \emph{Failure}, where the system had a lower number of sessions engaged. Figure~\ref{fig:googleanalytics} provides a plot of web traffic from Google Analytics over a specific day, clearly showing the drop at 20:00 where the system had been identified as in the \emph{Failure} state.

\subsection{Analysis}

\begin{figure}[!ht]
\centering
\includegraphics[height=150px,keepaspectratio]{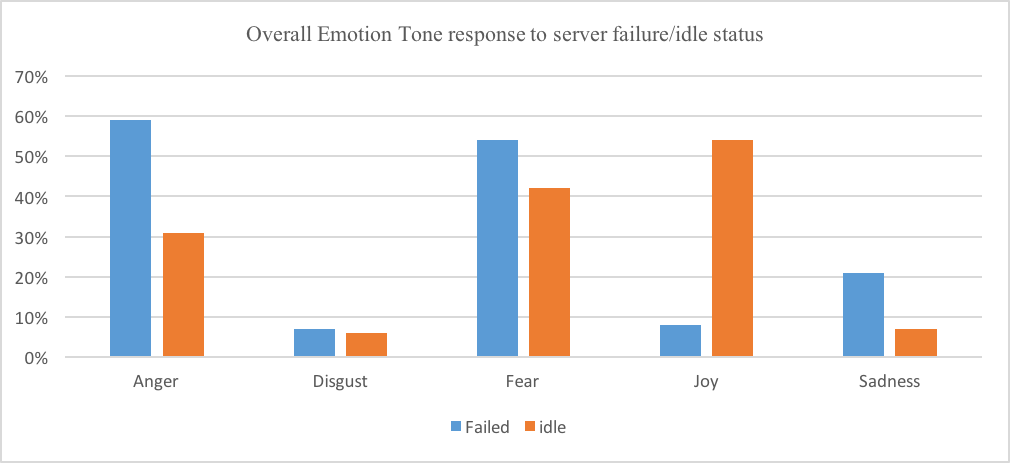}
\caption{Overall emotion tone response to server failure/idle status}
\label{fig:emotiontone} 
\end{figure}

All communications had been collected and grouped by server status, then sent to the IBM Watson Tone Analyzer to produce the emotion social tone scores, to present an overview of the system behaviour and user’s interaction with Facebook at the same time. Figure~\ref{fig:emotiontone} shows the association between the server behaviour and emotions of the users; in the system, {\emph{Failure}} status gives a significant difference in overall {\emph{Anger}} in different status; furthermore, the {\emph{Joy}} parameter shows a significant difference with the system in {\emph{Idle}} and {\emph{Failure}} status. However {\emph{Fear}} and {\emph{Sadness}} parameters is about the same, even with the system in {\emph{Idle}} status.

We recognised the user's personality based on analysis of their Facebook interactions, namely by collecting all comments from the users, again using the IBM Watson Personality Insights tool. However, some users in the dataset had completed the \emph{Big Five Questionnaire} ($N=44$); for these users, their ``Big Five" scores have been used instead. The second stage involved grouping the comments based on server status and segmenting these communications by the user; this allowed us to investigate the impact of server status in the emotion of the user and investigate the Big Five dimension as a constant parameter. By investigating the association between the personality trait dimensions and the social emotion tones, we can find the highest correlation to identify the critical elements of the potential model by applying linear regression and Pearson correlation. The output will allow the building of a neural network multilayer perception using the possible significant aspects with higher associations~\cite{mostafa-et-al-ai2016}.

The previous overview encourages further study to understand the correlation between user's behaviour and complex computer system behaviours. The data collected from the social media communications have been grouped by users and using the IBM Watson Personality Insights, we were able to identify the \emph{Big Five personality traits} for each user. Using the IBM Watson Tone Analyzer, the data has been grouped by user's comments and server status ({\emph{Failure}}, {\emph{Idle}}) to identify social emotion tone for each user. Table~\ref{tab:sample} shows an example of data used in this investigation, with each row representing a unique user, and each column represents the \emph{Big Five traits}, social emotion tones, and server status.

\subsection{Key Findings}

\begin{table}[]
\centering
\resizebox{\textwidth}{!}{\begin{tabular}{lrrrrrrrrrr}
\hline
\multicolumn{1}{c}{Openness} & \multicolumn{1}{c}{Conscientiousness} & \multicolumn{1}{c}{Extraversion} & \multicolumn{1}{c}{Agreeableness} & \multicolumn{1}{c}{Neuroticism} & \multicolumn{1}{c}{anger} & \multicolumn{1}{c}{disgust} & \multicolumn{1}{c}{fear} & \multicolumn{1}{c}{joy} & \multicolumn{1}{c}{sadness} & \multicolumn{1}{c}{Server  Status} \\ \hline
0.528                        & 0.523                                 & 0.537                            & 0.653                             & 0.511                           & 0.217821                  & 0.793375                    & 0.501131                 & 0.031477                & 0.284936                    & Failure                            \\
0.252                        & 0.063                                 & 0.037                            & 0.266                             & 0.989                           & 0.542857                  & 0.084615                    & 0.178302                 & 0.224453                & 0.264283                    & Failure                            \\
0.817                        & 0.571                                 & 0.157                            & 0.012                             & 0.401                           & 0.162798                  & 0.166694                    & 0.213870                 & 0.410916                & 0.220049                    & Failure                            \\
0.197                        & 0.130                                 & 0.180                            & 0.419                             & 0.990                           & 0.468938                  & 0.259794                    & 0.350803                 & 0.037265                & 0.636412                    & Failure                            \\
0.155                        & 0.079                                 & 0.081                            & 0.226                             & 0.975                           & 0.539162                  & 0.219993                    & 0.431932                 & 0.011625                & 0.642158                    & Failure                            \\
0.158                        & 0.281                                 & 0.332                            & 0.510                             & 0.869                           & 0.419015                  & 0.162022                    & 0.213941                 & 0.066892                & 0.686369                    & Failure                            \\
0.817                        & 0.571                                 & 0.157                            & 0.012                             & 0.401                           & 0.041602                  & 0.026298                    & 0.141606                 & 0.651962                & 0.106500                    & Failure                            \\
0.058                        & 0.038                                 & 0.147                            & 0.375                             & 0.989                           & 0.449222                  & 0.057946                    & 0.181654                 & 0.158412                & 0.547968                    & Idle                               \\
0.178                        & 0.138                                 & 0.800                            & 0.564                             & 0.828                           & 0.207497                  & 0.096643                    & 0.093218                 & 0.769316                & 0.162241                    & Idle                               \\
0.105                        & 0.463                                 & 0.792                            & 0.704                             & 0.041                           & 0.134487                  & 0.257145                    & 0.195858                 & 0.181699                & 0.509379                    & Idle                               \\
0.589                        & 0.479                                 & 0.147                            & 0.339                             & 0.828                           & 0.360527                  & 0.240875                    & 0.321188                 & 0.117492                & 0.212762                    & Idle                               \\
0.338                        & 0.235                                 & 0.104                            & 0.304                             & 0.869                           & 0.164107                  & 0.015058                    & 0.230148                 & 0.629562                & 0.356028                    & Idle                               \\
0.204                        & 0.203                                 & 0.480                            & 0.329                             & 0.892                           & 0.625891                  & 0.193692                    & 0.242459                 & 0.153679                & 0.166561                    & Idle                               \\
0.689                        & 0.968                                 & 0.805                            & 0.465                             & 0.029                           & 0.246246                  & 0.080353                    & 0.123761                 & 0.807537                & 0.135646                    & Idle                               \\
0.093                        & 0.175                                 & 0.642                            & 0.563                             & 0.875                           & 0.279503                  & 0.045658                    & 0.207278                 & 0.088724                & 0.505607                    & Idle                               \\
0.277                        & 0.296                                 & 0.276                            & 0.332                             & 0.892                           & 0.499199                  & 0.143897                    & 0.269725                 & 0.188664                & 0.285462                    & Idle                               \\
0.055                        & 0.095                                 & 0.783                            & 0.699                             & 0.935                           & 0.450997                  & 0.153940                    & 0.263070                 & 0.350778                & 0.116282                    & Idle                               \\ \hline
\end{tabular}
}
\caption{Snapshot of the data used in the analysis}
\label{tab:sample}
\end{table}

As part of modelling the users' responses and behaviour, one of the approaches to building the conceptual framework model is to apply linear regression to investigate the relationship between the Big Five personality dimensions and the emotion tones features.

\begin{figure}[!ht]
\centering
\includegraphics[height=200px,keepaspectratio]{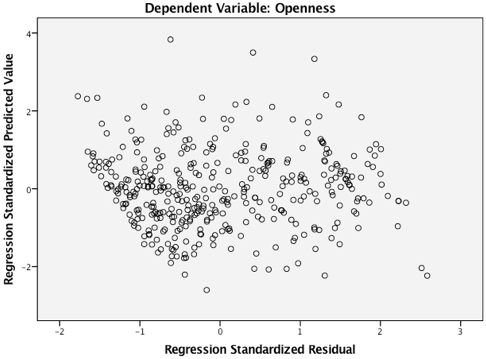}
\caption[Scatter plot of Big Five dimension ``Openness" and
  social emotion tones]{Scatter plot of Big Five dimension ``Openness" (dependent variable) and social emotion tones (independent variables)}
\label{fig:opennessplot} 
\end{figure}

\begin{figure}[!ht]
\centering
\includegraphics[height=200px,keepaspectratio]{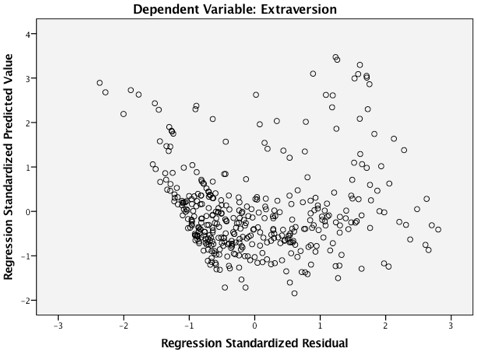}
\caption[Scatter plot of Big Five dimension ``Extraversion" and social emotion tones]{Scatter plot of Big Five dimension ``Extraversion" (dependent variable) and social emotion tones (independent variables)}
\label{fig:extraversioNLPot}
\end{figure}

\begin{figure}[!ht]
\centering
\includegraphics[height=200px,keepaspectratio]{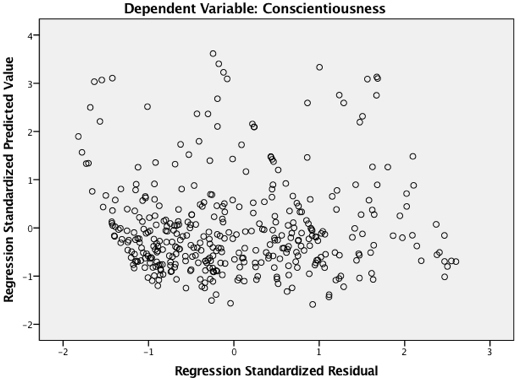}
\caption[Scatter plot of Big Five dimension ``Conscientiousness" and social emotion tones]{Scatter plot of Big Five dimension ``Conscientiousness" (dependent variable) and social emotion tones (independent variables)}
\label{fig:conscientiousnessplot} 
\end{figure}

\begin{figure}[!ht]
\centering
\includegraphics[height=200px,keepaspectratio]{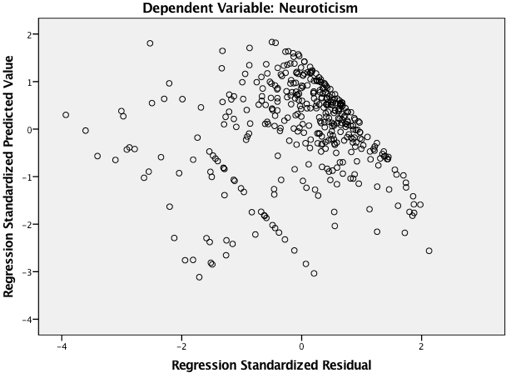}
\caption[Scatter plot of Big Five dimension ``Neuroticism" and social emotion tones]{Scatter plot of Big Five dimension ``Neuroticism" (dependent variable) and social emotion tones (independent variables)}
\label{fig:neuroticismplot} 
\end{figure}

\begin{figure}[!ht]
\centering
\includegraphics[height=200px,keepaspectratio]{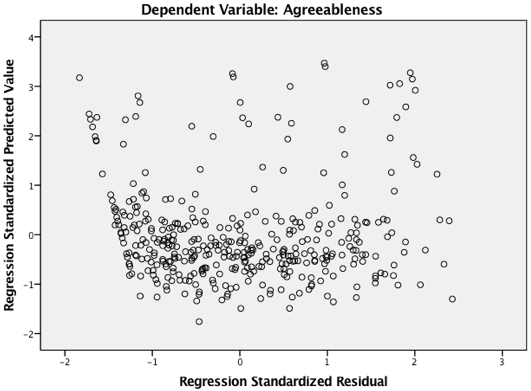}
\caption[Scatter plot of Big Five dimension ``Agreeableness" and social emotion tones]{Scatter plot of Big Five dimension ``Agreeableness" (dependent variable) and social emotion tones (independent variables)}
\label{fig:agreeablenessplot} 
\end{figure}

\begin{table}[!ht]
\centering
\resizebox{\textwidth}{!}{\begin{tabular}{llllllllllllllll}
\hline
           & \multicolumn{3}{c}{Openness}                                            & \multicolumn{3}{c}{Extraversion}                                              & \multicolumn{3}{c}{Conscientiousness}                                        & \multicolumn{3}{c}{Agreeableness}                                             & \multicolumn{3}{c}{Neuroticism}                                               \\ \hline
           & \multicolumn{1}{c}{B} & \multicolumn{1}{c}{t} & \multicolumn{1}{c}{Sig} & \multicolumn{1}{c|}{B}      & \multicolumn{1}{c}{t} & \multicolumn{1}{c}{Sig} & \multicolumn{1}{c|}{B}     & \multicolumn{1}{c}{t} & \multicolumn{1}{c}{Sig} & \multicolumn{1}{c|}{B}      & \multicolumn{1}{c}{t} & \multicolumn{1}{c}{Sig} & \multicolumn{1}{c|}{B}      & \multicolumn{1}{c}{t} & \multicolumn{1}{c}{Sig} \\
(constant) & 0.356                 & 3.282                 & 0.001                   & \multicolumn{1}{l|}{0.162}  & 1.642                 & 0.101                   & \multicolumn{1}{l|}{0.16}  & 1.623                 & 0.105                   & \multicolumn{1}{l|}{0.297}  & 2.831                 & 0.005                   & \multicolumn{1}{l|}{0.828}  & 9.934                 & 0                       \\
anger      & -0.063                & -0.735                & 0.463                   & \multicolumn{1}{l|}{0.064}  & 0.831                 & 0.406                   & \multicolumn{1}{l|}{0.124} & 1.592                 & 0.112                   & \multicolumn{1}{l|}{0.024}  & 0.293                 & 0.769                   & \multicolumn{1}{l|}{0.116}  & 1.767                 & 0.078                   \\
disgust    & 0.478                 & 4.354                 & 0                       & \multicolumn{1}{l|}{0.114}  & 1.142                 & 0.253                   & \multicolumn{1}{l|}{0.255} & 2.551                 & 0.011                   & \multicolumn{1}{l|}{-0.061} & -0.574                & 0.566                   & \multicolumn{1}{l|}{-0.363} & -4.303                & 0                       \\
fear       & 0.065                 & 0.534                 & 0.594                   & \multicolumn{1}{l|}{0.172}  & 1.549                 & 0.122                   & \multicolumn{1}{l|}{0.04}  & 0.356                 & 0.722                   & \multicolumn{1}{l|}{0.093}  & 0.783                 & 0.434                   & \multicolumn{1}{l|}{-0.023} & -0.241                & 0.81                    \\
joy        & 0.066                 & 0.561                 & 0.575                   & \multicolumn{1}{l|}{0.446}  & 4.179                 & 0                       & \multicolumn{1}{l|}{0.436} & 4.058                 & 0                       & \multicolumn{1}{l|}{0.188}  & 1.652                 & 0.099                   & \multicolumn{1}{l|}{-0.487} & -5.39                 & 0                       \\
sadness    & -0.226                & -2.118                & 0.035                   & \multicolumn{1}{l|}{-0.185} & -1.906                & 0.057                   & \multicolumn{1}{l|}{-0.03} & -0.313                & 0.754                   & \multicolumn{1}{l|}{0.014}  & 0.132                 & 0.895                   & \multicolumn{1}{l|}{0.233}  & 2.841                 & 0.005                   \\ \hline
\end{tabular}
}
\caption{Linear regression coefficients}
\label{tbl:linreg}
\end{table}

During the analysis, the linear regressions (presented in
Table~\ref{tbl:linreg} and Figures~\ref{fig:opennessplot},
\ref{fig:extraversioNLPot}, \ref{fig:conscientiousnessplot},
\ref{fig:neuroticismplot} and \ref{fig:agreeablenessplot}) does  show
significant correlations between the Big Five dimensions and the
social emotion tones; however, certain correlations can be highlighted
and used as key elements for the model. The correlation
of \emph{Openness} and \emph{Disgust}, is 0.479; the correlation of
\emph{Extraversion} and {\emph{Joy}} is 0.446 with p-value of
zero. \emph{Conscientiousness} and {\emph{Joy}} with 0.436
correlation and \emph{Disgust} with 0.255. \emph{Agreeableness},
does not appear to have a high impact in the social emotion
parameters, with the highest correlation being 0.188 with
\emph{Joy}, which can be overlooked as a useful factor in the
model. \emph{Neuroticism} and \emph{Disgust} is -0.363, \emph{Joy}
is -0.487 and p-value is zero is both cases; and {\emph{Sadness}} with
0.233. All correlation values are $<0.5$; however, it is noticed that
{\emph{Agreeableness}} does not have a linear relationship with any of
the social emotion tones. Furthermore, the social emotion tones that
have a potential linear relationship are {\emph{Disgust}},
{\emph{Joy}} and {\emph{Sadness}}, since the three tones have a
correlation between $>0.3$ and $<0.5$.

Previous linear regression analysis suggested that the following Big
Five dimensions ({\emph{Openness}}, {\emph{Extraversion}},
{\emph{Conscientiousness}} and {\emph{Neuroticism}}) have the highest
correlation with the social emotion tones ({\emph{Joy}},
{\emph{Sadness}} and {\emph{Disgust}}). For further analysis, the
Pearson correlation for the same dataset has been performed to compare
the output with the linear regression correlations. As you can see in
Table~\ref{tab:pearson}, there is no significant correlation in both;
however, in the Pearson correlation, {\emph{Neuroticism}} has the
highest correlation values across emotion tones, especially
{\emph{Anger}}, {\emph{Joy}} and {\emph{Sadness}}. {\emph{Joy}} does
have a correlation with all Big Five dimensions except for
{\emph{Agreeableness}} which agrees with the previous
analysis. However, {\emph{Disgust}} does not have a strong correlation
with any of the Big Five dimensions, which deviates from the previous
analysis.

\begin{table}[!ht]
\centering
\begin{tabular}{lrrrrr}
\hline
                  & \multicolumn{1}{c}{Anger} & \multicolumn{1}{c}{Disgust} & \multicolumn{1}{c}{Fear} & \multicolumn{1}{c}{Joy} & \multicolumn{1}{c}{Sadness} \\ \hline
Openness          & -0.098                    & 0.231                       & 0.043                    & 0.035                   & -0.151                      \\
Conscientiousness & -0.111                    & -0.001                      & -0.113                   & 0.267                   & -0.19                       \\
Extraversion      & -0.175                    & -0.077                      & -0.071                   & 0.349                   & -0.291                      \\
Agreeableness     & -0.068                    & -0.089                      & -0.027                   & 0.14                    & -0.069                      \\
Neuroticism       & 0.375                     & -0.037                      & 0.153                    & -0.488                  & 0.379                       \\ \hline
\end{tabular}
\caption{Pearson correlations}
\label{tab:pearson}
\end{table}

\paragraph{Key Elements of the Model}\label{model}

According to the output of the statistical analysis presented in
Table~\ref{tbl:linreg} (linear regression) and Table~\ref{tab:pearson}
(Pearson correlation), the Big Five dimension identified as the key
elements from the personality traits are: {\emph{Openness}},
{\emph{Extraversion}}, {\emph{Conscientiousness}} and
{\emph{Neuroticism}}. The statistical analysis agrees that
{\emph{Agreeableness}} does not have a significant correlation across
any of the social emotion tones. The social emotion tones to be used
as key input elements for the proposed model are {\emph{Joy}},
{\emph{Sadness}}, {\emph{Anger}} and {\emph{Disgust}}; although the
{\emph{Anger}} tone did not show any significant correlation in linear
regression analysis, the value of the Pearson correlation coefficient
is between 0.3 and 0.5 which can be used as input for the model.

\begin{table}[!ht]
\centering
\begin{tabular}{lrr}
Correctly classified instances:   & 43     & ({\emph{75.44\%}}) \\
Incorrectly classified instances: & 14     & ({\emph{24.56\%}}) \\
Kappa statistic:                  & 0.5295 &         \\
Mean absolute error:              & 0.3432 &         \\
Root mean squared error:          & 0.4246 &         \\
Total number of instances:        & 57     &        
\end{tabular}
\caption{Re-evaluation output of proposed model}
\label{tab:reeval}
\end{table}

The dataset used to build this model is based upon a number of users (N=391), eight inputs ({\emph{Openness}}, {\emph{Extraversion}}, {\emph{Conscientiousness}}, {\emph{Neuroticism}}, {\emph{Joy}}, {\emph{Sadness}}, {\emph{Anger}} and {\emph{Disgust}}) and the class/output variable as the server status (where No: System {\emph{Failure}} and Yes: System {\emph{Idle}}). As shown in Table~\ref{tab:reeval}, the total number of the instances for the testing set is 57. The output of the model shows a 75.44\% corrected predicted instances and 24.56\% incorrectly classified instances. As this has been performed on a small subset of the overall larger project dataset, the output data is encouraging and provides the infrastructure for further analysis and research to exploit the full dataset~\cite{mostafa-et-al-ai2016}.

This experiment presents preliminary results from the previous flow of experiments~\cite{oatley+crick:2014,oatley+crick-gisruk2015,oatley-et-al_dasc2015,mostafa-et-al-ai2016}, which could provide the conceptual framework to improve user experience (UX) and computer system architecture design. Social media is now not only being used as a content and sharing platform but also as a platform for technical support for several of online applications and services. We have produced a model that can predict server status based on personality traits and social emotion tones, by investigating the linear regression and Pearson correlation to identify the key components to be used as input for the neural network to build this model ({\emph{Openness}}, {\emph{Extraversion}}, {\emph{Conscientiousness}}, {\emph{Neuroticism}}, {\emph{Joy}}, {\emph{Sadness}}, {\emph{Anger}} and {\emph{Disgust}}). The model developed shows a good potential starting point for further data analysis, with 75\% accuracy in prediction based on 57 test cases.

\subsection{Model Evaluation}
\paragraph{Using python}

\begin{table}[!ht]
\centering
\begin{tabular}{llll}
Model                            & Score learn & Result mean & Result std  \\
Logistic Regression              & 0.703389831 & 0.568717949 & 0.229815122 \\
Linear Discriminant Analysis     & 0.711864407 & 0.581538462 & 0.189468963 \\
KNeighborsClassifier             & 0.559322034 & 0.565769231 & 0.125269007 \\
DecisionTreeClassifier           & 0.669491525 & 0.583782051 & 0.116959357 \\
GaussianNB                       & 0.584745763 & 0.555192308 & 0.121590399 \\
C-Support Vector Classification. & 0.63559322  & 0.48974359  & 0.359056767 \\ 
\end{tabular}
\caption{Evaluate the limit model }
\label{tab:evlimitedmodel}
\end{table}

Table~\ref{tab:evlimitedmodel} shows what would be the best classifier to build the model, the score learn shows Logistic Regression and Linear Discriminant analysis




\section{Summary}\label{featureSelectionProcess}

This section provides a summary of the feature extraction process grounding the selection process on the experiments conducted in Chapter~\ref{pmsys}.

\begin{itemize}

\item The experiment presented in Section~\ref{subs:ProfilingComplex}, \emph{Openness to Experience} reported a strong correlation with the ``Accepted" groups within the dataset sample, however, the ``final selection" result parameter for the scholarship system has a different other evaluation criteria such as Personality of the evaluator, Language level, Academic qualification, number of spaces available, etc which was not included in the analysis, therefore, \emph{Openness to Experience} was not selected as part of the features.

\item In the experiment presented in Section~\ref{subs:sentimentStages}, there were statistically significant: Stages (1,4), \emph{Extraversion}, \emph{Agreeableness} and \emph{Conscientiousness}, however, \emph{Conscientiousness} and \emph{Agreeableness} reported a higher correlation value of \emph{0.01} and \emph{0.04}, therefore,  \emph{Conscientiousness} and \emph{Agreeableness} selected from this experiment.

\item In the experiment presented in Section~\ref{emotionSystemStages}, two analysis methods were used to confirm the result, Ordinal and Multinomial Regression analysis, the output reported from ordinal that variables \emph{Disgust}, \emph{Sadness} and \emph{Conscientiousness}with p-values
\emph{.006},\emph{0.017} and \emph{0.051} has a significant overall effect on the dependent \emph{StageID}. While, the two analysis methods agreed on  \emph{Disgust}, \emph{Sadness} and \emph{Conscientiousness}, the parameter \emph{Neuroticism} was excluded from the selection as it was has not significant contribution to the model in Ordinal Regression and low p-value on Multinomial Regression Analysis of \emph{0.143}. Therefore, only  \emph{Disgust}, \emph{Sadness} and \emph{Conscientiousness} were selected from this experiment.

\item The experiment presented in Section~\ref{personalityTraitsTemporal} suggested a weak correlation between personality traits and temporal behaviour using the dataset extracted from the system and the questionnaire filled with the same users. Therefore, no features selected from this experiment.

\item The experiment presented in Section~\ref{Emotionscentre} suggested that \emph{Neuroticism} a very strong Pearson's correlation values with \emph{Anger}, \emph{Joy} and \emph{Sadness}. While \emph{Joy} does have a correlation with all Big Five Traits except for \emph{Agreeableness}. Therefore, \emph{Neuroticism}, \emph{Anger}, \emph{Joy} and \emph{Sadness} were selected from this experiment.

\end{itemize}


\newpage
\chapter{Developing the Conceptual Framework for the {\emph{PMSys}} Engine}\label{modellingphase}

\section{Introduction}

\begin{figure}[!ht]
\centering
\includegraphics[width=\textwidth,height=750px,keepaspectratio]{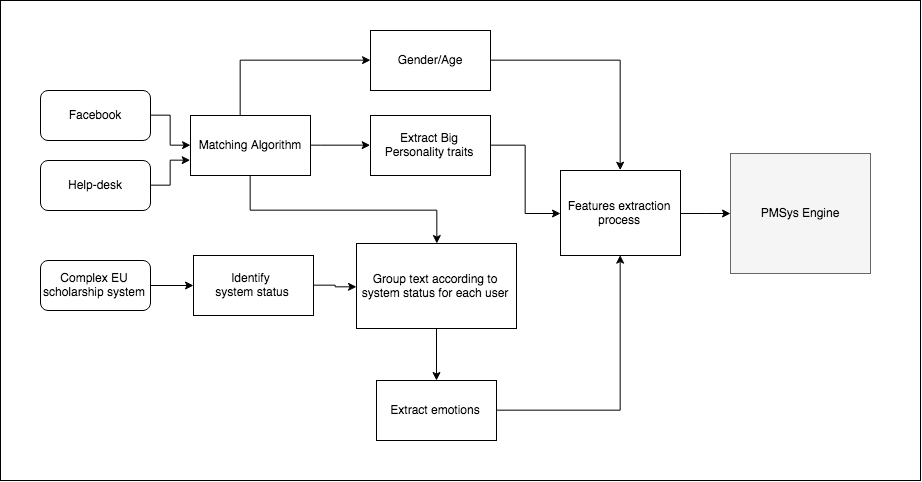}
\caption{Developing PMsys engine stages}
\label{fig:flowchartPMSysProcess}
\end{figure}

During this research project, several experiments and studies were conducted to explore and understand the correlation between personality traits, emotions, and server status. To enhance user's experience and to understand more how user's behaviour changes according to the system's response. The results of these experiments and studies provided us with a good understanding of how user's traits and emotions changes in different server status from the perspective of the categories presented in our classification (Section~\ref{serverStatus}). The flow of experiments leads us to build our conceptual model, using the features extracted from the analyses as presented in Chapter~\ref{methodology}. In this chapter, we offer a novel conceptual framework to model user's behaviour in different computer status (see Figure~\ref{fig:flowchartPMSysProcess}.

\section{Personality Traits vs. Emotions, Gender and Age}\label{bigFiveEmotionsGenderAge}

\subsection{Introduction}

The previous analyses suggested a strong association between personality traits and emotions. Furthermore, the attempt of modelling server status, suggested a strong and possible method to model the user behaviour in different complex system behaviour. This analysis to explore the big personality traits and emotions association and correlation, a further correlation between \emph{Gender/Age} and \emph{Personality traits} -- \emph{Emotions}. According to research by Schwartz (2013)\cite{10.1371/journal.pone.0073791}, Gender and Age correlate with the personality traits. However, the same study did not mention the emotions. Therefore, this experiment is essential a cross-validating the methodology used in Schwartz study~\cite{10.1371/journal.pone.0073791} and with extended features to the model to verify the connection between the emotions and gender/age.

The same dataset (See section: ~\ref{SourceData}) will be used in the analysis as Personality Traits retrieved from the Motivation letter and emotions extracted from a different platform that was used in communication (i.e., Facebook and HelpDesk). The necessary information will be retrieved to include the gender and age as extra parameters, and the analysis will be running separately.

\subsection{Binomial Logistic Regression}\label{BioGenderAge}
The dataset combination suggested using Binomial Logistic Regression, as the dataset similar to what is presented in Section~\ref{subs:sentimentStages}, with \emph{Gender} instead of \emph{StageID} as a dependent variable. The experiment is to examine the probability of being able to predicate the gender based on Big Five traits and Emotions, and according to the result, it will be decided whether to include the Gender as controlling variable in the conceptual model. Adding the Age variable alongside with Big Five Traits and Emotions to examine if it would improve the model or not.

In order to apply, Binomial Logistic Regression, the data needs to pass the following assumptions:

\begin{itemize}
\item
Linear relationship between the Big Five Traits, Emotions and logit transformation of the gender variable. 
\item
Data must not show multicollinearity
\item
There should be no significant outliers, high leverage points or highly influential points.
\end{itemize}

\paragraph{Linear relationship between the Big Five Traits, Emotions and logit transformation of the \textbf{gender} variable}

The first part of the Box-Tidwell (1962)~\cite{Box1962} method expects that all continuous independent variables to transformed into their natural logs, this means that we need to perform natural log transformations on our continuous independent variables: Big Five Traits and Emotions. The second part of the Box-Tidwell (1962) procedure requires creating interaction terms for each of your continuous independent variables and their respective natural log-transformed variables. Since we have 12 continuous independent variables in this study, this means that we have to create Big Five Trait and Emotions -- interaction terms: \verb|ln_sadness| \mbox{*} sadness (i.e., the product of \verb|ln_sadness| by sadness –). Moreover, need to be entered into the Binomial Logistic Regression procedure, together with the \emph{Gender} and \emph{Age}.

According to Tabachnick (2001) ~\cite{Tabachnick2001}, to calculate the new alpha (α) level (i.e., p-value) for current dataset, it is by dividing the alpha level (${p < .05}$) by the number of terms in the model. Formulaically, this is:

\begin{equation}
adjusted alpha level = \frac{OriginalAlphaLevel}{numberofComparisons}
\end{equation}

The new adjusted alpha level in this case is \textit{0.002}, (i.e., 0.05 / 23= 0.002). The linearity of the Big Five Traits and Emotions concerning the logit of the \textit{Gender} variable was assessed via the Box-Tidwell (1962)~\cite{Box1962} procedure. A Bonferroni correction was applied using all twenty-one terms in the model resulting in statistical significance being accepted when p \mbox{<} 0.002~\cite{Tabachnick2001}. According to Table~\ref{tbl:LineraityCheckGender}, all continuous independent variables were found to be linearly related to the logit of the dependent variable.

\begin{table}[!ht]
\centering
\resizebox{\textwidth}{!}{
\begin{tabular}{@{}lllllll@{}}
\toprule
\multicolumn{7}{c}{\textbf{Variables in the Equation}}                                          \\ \midrule
                                           & B      & S.E.   & Wald  & df    & Sig.  & Exp(B)   \\
Step 1a                                    & Anger  & 1.248  & 2.244 & 0.309 & 1     & 0.578    \\
Disgust                                    & -0.332 & 23.43  & 0     & 1     & 0.989 & 0.717    \\
Fear                                       & 1.667  & 2.822  & 0.349 & 1     & 0.555 & 5.294    \\
Joy                                        & 1.961  & 1.496  & 1.719 & 1     & 0.19  & 7.107    \\
Sadness                                    & -1.251 & 0.854  & 2.145 & 1     & 0.143 & 0.286    \\
Openness                                   & -2.494 & 1.152  & 4.691 & 1     & 0.03  & 0.083    \\
Conscientiousness                          & 2.303  & 1.266  & 3.311 & 1     & 0.069 & 10.008   \\
Extraversion                               & 0.484  & 0.911  & 0.282 & 1     & 0.595 & 1.622    \\
Agreeableness                              & 0.155  & 0.954  & 0.026 & 1     & 0.871 & 1.167    \\
Neuroticism                                & 0.166  & 0.832  & 0.04  & 1     & 0.842 & 1.181    \\
Age                                        & 0.289  & 0.584  & 0.245 & 1     & 0.621 & 1.335    \\
Anger by ln\_anger                         & 1.945  & 3.032  & 0.412 & 1     & 0.521 & 6.992    \\
Disgust by ln\_disgust                     & 2.757  & 11.198 & 0.061 & 1     & 0.806 & 15.75    \\
Fear by ln\_fear                           & -0.623 & 3.262  & 0.036 & 1     & 0.849 & 0.536    \\
Joy by ln\_joy                             & 4.399  & 2.49   & 3.121 & 1     & 0.077 & 81.354   \\
Sadness by ln\_sadness                     & -1.914 & 2.128  & 0.81  & 1     & 0.368 & 0.147    \\
Openness by ln\_openness                   & 7.291  & 3.366  & 4.693 & 1     & 0.03  & 1466.912 \\
Conscientiousness by ln\_conscientiousness & 1.096  & 2.669  & 0.169 & 1     & 0.681 & 2.992    \\
Extraversion by ln\_extraversion           & -3.599 & 2.282  & 2.487 & 1     & 0.115 & 0.027    \\
Agreeableness by ln\_agreeableness         & -0.662 & 2.413  & 0.075 & 1     & 0.784 & 0.516    \\
Neuroticism by ln\_neuroticism             & 1.957  & 2.622  & 0.557 & 1     & 0.456 & 7.076    \\
Age by ln\_age                             & -0.053 & 0.129  & 0.17  & 1     & 0.68  & 0.948    \\
Constant                                   & -0.39  & 4.891  & 0.006 & 1     & 0.936 & 0.677    \\ \bottomrule
\end{tabular}
}
\caption{Variables in the Equation -  Gender}
\label{tbl:LineraityCheckGender}
\end{table}

\paragraph{Data must not show multicollinearity}, next step to investigate if the data shows or does not show multicollinearity to validate the possibility of applying binomial logistic regression. According to Table~\ref{tbl:GenderCasewiseDiagnostics} there was one studentized residual with a value of -2.376743 standard deviations, which was kept in the analysis.

\begin{table}[!ht]
\centering
\resizebox{\textwidth}{!}{

\begin{tabular}{@{}lllllll@{}}
\toprule
\multicolumn{7}{c}{\textbf{Casewise Listb}}                                                      \\ \midrule
Case   & Selected Statusa & Observed & Predicted & Predicted Group & Temporary Variable &        \\
gender &                  &          &           &                 & Resid              & ZResid \\
7      & S                & F**      & 0.85      & M               & -0.85              & -2.377 \\ \bottomrule
\end{tabular}
}
\caption{Casewise Diagnostics}
\label{tbl:GenderCasewiseDiagnostics}
\end{table}

\subsubsection{Bionomial Findings}
This experiment aims to investigate which variable is statistically significant of the Big Five Traits, Emotions and Age concerning the Gender; only three were statistically significant: Openness (${p<0.072}$), Conscientiousness and Age (as shown in Table~\ref{tbl:VariablesintheEquation}). The result reported does not give enough accuracy regarding the correlation between Big Five Traits, Emotions and Age to predict the Gender. Therefore, another form of analysis is applied next to explore and investigate a potential association between the above variables.

\begin{table}[!ht]
\centering
\resizebox{\textwidth}{!}{

\begin{tabular}{@{}lllllllll@{}}
\toprule
\multicolumn{9}{c}{\textbf{Variables in the Equation}}                                                                                                                                          \\ \midrule
                  & \multirow{2}{*}{B} & \multirow{2}{*}{S.E.} & \multirow{2}{*}{Wald} & \multirow{2}{*}{df} & \multirow{2}{*}{Sig.} & \multirow{2}{*}{Exp(B)} & 95\% C.I.for EXP(B) &          \\
                  &                    &                       &                       &                     &                       &                         & Lower               & Upper    \\
Anger             & 0.329              & 1.518                 & 0.047                 & 1                   & 0.828                 & 1.39                    & 0.071               & 27.209   \\
Disgust           & -5.201             & 7.187                 & 0.524                 & 1                   & 0.469                 & 0.006                   & 0                   & 7219.283 \\
Fear              & 1.405              & 1.635                 & 0.739                 & 1                   & 0.39                  & 4.075                   & 0.165               & 100.404  \\
Joy               & 0.169              & 0.951                 & 0.031                 & 1                   & 0.859                 & 1.184                   & 0.184               & 7.631    \\
Sadness           & -0.567             & 0.731                 & 0.6                   & 1                   & 0.438                 & 0.567                   & 0.135               & 2.379    \\
Openness          & -1.556             & 0.866                 & 3.229                 & 1                   & 0.072                 & 0.211                   & 0.039               & 1.152    \\
Conscientiousness & 1.261              & 0.764                 & 2.722                 & 1                   & 0.099                 & 3.529                   & 0.789               & 15.786   \\
Extraversion      & 0.389              & 0.78                  & 0.249                 & 1                   & 0.618                 & 1.476                   & 0.32                & 6.801    \\
Agreeableness     & 0.262              & 0.778                 & 0.113                 & 1                   & 0.736                 & 1.3                     & 0.283               & 5.974    \\
Neuroticism       & 0.234              & 0.644                 & 0.132                 & 1                   & 0.717                 & 1.263                   & 0.358               & 4.46     \\
Age               & 0.059              & 0.031                 & 3.664                 & 1                   & 0.056                 & 1.061                   & 0.999               & 1.127    \\
Constant          & -1.395             & 1.175                 & 1.41                  & 1                   & 0.235                 & 0.248                   &                     &          \\ \bottomrule
\end{tabular}
}
\caption{Binomial Log -- variables in the equation}
\label{tbl:VariablesintheEquation}
\end{table}

\subsection{Pearson's Partial Correlation}

As the Binomial Logistic Regression, suggested a correlation between \emph{Openness}, \emph{Conscientiousness} and \emph{Age} to predict the \emph{Gender}, the Pearson's partial correlation was run to assess the relationship between Big Five Traits, Emotions, Age and Gender and to confirm the output of the Binomial or include more variable as a strong association.

According to the analysis performed in Section~\ref{subs:ProfilingComplex}, there were linear relationships between Big Five Traits and Emotions, as assessed by scatterplots and partial regression plots. There was univariate normality, as evaluated by Shapiro-Wilk's test (${p > .05}$), and there were no univariate or multivariate outliers, as assessed by Mahalanobis Distance respectively -- see Figure~\ref{fig:normalpp} and ~\ref{fig:scatterplot}. 

\begin{table}[!htbp]
\centering
\resizebox{!}{!}{
\begin{tabular}{@{}llrrrrr@{}}
\toprule
\multicolumn{7}{c}{Pearson's partial correlation}                                       \\ \midrule
                  &                         & Anger & Disgust & Fear  & Joy   & Sadness \\
\multicolumn{7}{c}{\textbf{Controlling Variable: None}}                                 \\
Openness          & Correlation             & .044  & -.008   & -.038 & .024  & -.015   \\
                  & Signif. (2-tailed) & .529  & .913    & .586  & .729  & .833    \\
                  & df                      & 204   & 204     & 204   & 204   & 204     \\
Conscientiousness & Correlation             & -.141 & -.093   & -.099 & .057  & -.040   \\
                  & Signif. (2-tailed) & .043  & .183    & .155  & .418  & .566    \\
                  & df                      & 204   & 204     & 204   & 204   & 204     \\
Extraversion      & Correlation             & .040  & .056    & -.035 & -.079 & .003    \\
                  & Signif. (2-tailed) & .567  & .421    & .616  & .262  & .960    \\
                  & df                      & 204   & 204     & 204   & 204   & 204     \\
Agreeableness     & Correlation             & .010  & .041    & .026  & -.094 & .053    \\
                  & Signif. (2-tailed) & .881  & .556    & .715  & .178  & .447    \\
                  & df                      & 204   & 204     & 204   & 204   & 204     \\
Neuroticism       & Correlation             & -.038 & -.058   & -.166 & -.006 & .030    \\
                  & Signif. (2-tailed) & .585  & .407    & .017  & .937  & .664    \\
                  & df                      & 204   & 204     & 204   & 204   & 204     \\
Age               & Correlation             & -.042 & -.082   & -.187 & .170  & -.085   \\
                  & Signif. (2-tailed) & .551  & .241    & .007  & .015  & .222    \\
                  & df                      & 204   & 204     & 204   & 204   & 204     \\
\multicolumn{7}{c}{\textbf{Controlling Variable: Gender}}                               \\
Openness          & Correlation             & .041  & -.014   & -.039 & .029  & -.020   \\
                  & Signif. (2-tailed) & .558  & .844    & .579  & .681  & .772    \\
                  & df                      & 203   & 203     & 203   & 203   & 203     \\
Conscientiousness & Correlation             & -.139 & -.086   & -.100 & .052  & -.034   \\
                  & Signif. (2-tailed) & .047  & .219    & .155  & .463  & .632    \\
                  & df                      & 203   & 203     & 203   & 203   & 203     \\
Extraversion      & Correlation             & .041  & .058    & -.035 & -.080 & .005    \\
                  & Signif. (2-tailed) & .560  & .410    & .617  & .256  & .946    \\
                  & df                      & 203   & 203     & 203   & 203   & 203     \\
Agreeableness     & Correlation             & .013  & .046    & .026  & -.098 & .057    \\
                  & Signif. (2-tailed) & .855  & .514    & .711  & .163  & .413    \\
                  & df                      & 203   & 203     & 203   & 203   & 203     \\
Neuroticism       & Correlation             & -.036 & -.054   & -.166 & -.008 & .034    \\
                  & Signif. (2-tailed) & .606  & .439    & .017  & .904  & .627    \\
                  & df                      & 203   & 203     & 203   & 203   & 203     \\
Age               & Correlation             & -.038 & -.075   & -.188 & .166  & -.080   \\
                  & Signif. (2-tailed) & .587  & .282    & .007  & .017  & .257    \\
                  & df                      & 203   & 203     & 203   & 203   & 203     \\
                  &                         &       &         &       &       &         \\ \bottomrule
\end{tabular}
}
\caption{Pearson's Partial Correlation (controlling variable: Gender)}
\label{tbl: PearControllingGender}
\end{table}

\begin{table}[!htbp]
\centering
\resizebox{\textwidth}{!}{
\begin{tabular}{@{}llrrrrr@{}}
\hline
\multicolumn{7}{c}{Pearson's partial correlation}                                       \\ \hline
                  &                         & Anger & Disgust & Fear  & Joy   & Sadness \\
\multicolumn{7}{c}{Controlling Variable: None}                                          \\
Openness          & Correlation             & .044  & -.008   & -.038 & .024  & -.015   \\
                  & Signif. (2-tailed) & .529  & .913    & .586  & .729  & .833    \\
                  & df                      & 204   & 204     & 204   & 204   & 204     \\
Conscientiousness & Correlation             & -.141 & -.093   & -.099 & .057  & -.040   \\
                  & Signif. (2-tailed) & .043  & .183    & .155  & .418  & .566    \\
                  & df                      & 204   & 204     & 204   & 204   & 204     \\
Extraversion      & Correlation             & .040  & .056    & -.035 & -.079 & .003    \\
                  & Signif. (2-tailed) & .567  & .421    & .616  & .262  & .960    \\
                  & df                      & 204   & 204     & 204   & 204   & 204     \\
Agreeableness     & Correlation             & .010  & .041    & .026  & -.094 & .053    \\
                  & Signif. (2-tailed) & .881  & .556    & .715  & .178  & .447    \\
                  & df                      & 204   & 204     & 204   & 204   & 204     \\
Neuroticism       & Correlation             & -.038 & -.058   & -.166 & -.006 & .030    \\
                  & Signif. (2-tailed) & .585  & .407    & .017  & .937  & .664    \\
                  & df                      & 204   & 204     & 204   & 204   & 204     \\
Gender            & Correlation             & .030  & .056    & .005  & -.041 & .050    \\
                  & Signif. (2-tailed) & .668  & .423    & .947  & .558  & .473    \\
                  & df                      & 204   & 204     & 204   & 204   & 204     \\
Age               & Correlation             & -.042 & -.082   & -.187 & .170  & -.085   \\
                  & Signif. (2-tailed) & .551  & .241    & .007  & .015  & .222    \\
                  & df                      & 204   & 204     & 204   & 204   & 204     \\
\multicolumn{7}{c}{Controlling Variable Age}                                            \\
Openness          & Correlation             & .050  & .003    & -.014 & .002  & -.004   \\
                  & Signif. (2-tailed) & .475  & .964    & .842  & .978  & .960    \\
                  & df                      & 203   & 203     & 203   & 203   & 203     \\
Conscientiousness & Correlation             & -.138 & -.084   & -.080 & .038  & -.031   \\
                  & Signif. (2-tailed) & .049  & .229    & .257  & .592  & .664    \\
                  & df                      & 203   & 203     & 203   & 203   & 203     \\
Extraversion      & Correlation             & .041  & .058    & -.032 & -.083 & .005    \\
                  & Signif. (2-tailed) & .559  & .407    & .650  & .235  & .940    \\
                  & df                      & 203   & 203     & 203   & 203   & 203     \\
Agreeableness     & Correlation             & .001  & .023    & -.018 & -.057 & .035    \\
                  & Signif. (2-tailed) & .990  & .743    & .796  & .413  & .622    \\
                  & df                      & 203   & 203     & 203   & 203   & 203     \\
Neuroticism       & Correlation             & -.034 & -.051   & -.152 & -.022 & .039    \\
                  & Signif. (2-tailed) & .623  & .471    & .030  & .753  & .580    \\
                  & df                      & 203   & 203     & 203   & 203   & 203     \\
Gender            & Correlation             & .025  & .046    & -.020 & -.019 & .040    \\
                  & Signif. (2-tailed) & .724  & .513    & .772  & .785  & .574    \\
                  & df                      & 203   & 203     & 203   & 203   & 203     \\
\multicolumn{7}{c}{\textbf{Cells contain zero-order (Pearson) correlations.}}           \\ \hline
\end{tabular}
}
\caption{Pearson's Partial Correlation (controlling variable: Age)}
\label{tbl: PearControllingAge}
\end{table}

\begin{table}[!htbp]
\centering
\begin{tabular}{@{}llrrrrr@{}}
\toprule
\multicolumn{7}{c}{\textbf{Pearson's partial correlation}}                              \\ \midrule
Correlation       &                         & Anger & Disgust & Fear  & Joy   & Sadness \\
\multicolumn{7}{c}{\textbf{Controlling Variable: None}}                                 \\
Openness          & Correlation             & .044  & -.008   & -.038 & .024  & -.015   \\
                  & Signif. (2-tailed) & .529  & .913    & .586  & .729  & .833    \\
                  & df                      & 204   & 204     & 204   & 204   & 204     \\
Conscientiousness & Correlation             & -.141 & -.093   & -.099 & .057  & -.040   \\
                  & Signif. (2-tailed) & .043  & .183    & .155  & .418  & .566    \\
                  & df                      & 204   & 204     & 204   & 204   & 204     \\
Extraversion      & Correlation             & .040  & .056    & -.035 & -.079 & .003    \\
                  & Signif. (2-tailed) & .567  & .421    & .616  & .262  & .960    \\
                  & df                      & 204   & 204     & 204   & 204   & 204     \\
Agreeableness     & Correlation             & .010  & .041    & .026  & -.094 & .053    \\
                  & Signif. (2-tailed) & .881  & .556    & .715  & .178  & .447    \\
                  & df                      & 204   & 204     & 204   & 204   & 204     \\
Neuroticism       & Correlation             & -.038 & -.058   & -.166 & -.006 & .030    \\
                  & Signif. (2-tailed) & .585  & .407    & .017  & .937  & .664    \\
                  & df                      & 204   & 204     & 204   & 204   & 204     \\
Age               & Correlation             & -.042 & -.082   & -.187 & .170  & -.085   \\
                  & Signif. (2-tailed) & .551  & .241    & .007  & .015  & .222    \\
                  & df                      & 204   & 204     & 204   & 204   & 204     \\
Gender       & Correlation             & .030  & .056    & .005  & -.041 & .050    \\
                  & Signif. (2-tailed) & .668  & .423    & .947  & .558  & .473    \\
                  & df                      & 204   & 204     & 204   & 204   & 204     \\
\multicolumn{7}{c}{\textbf{Controlling Variable: Age and Gender}}                       \\
Openness          & Correlation             & .047  & -.003   & -.012 & .004  & -.009   \\
                  & Signif. (2-tailed) & .501  & .969    & .870  & .950  & .902    \\
                  & df                      & 202   & 202     & 202   & 202   & 202     \\
Conscientiousness & Correlation             & -.136 & -.079   & -.083 & .036  & -.026   \\
                  & Signif. (2-tailed) & .053  & .260    & .239  & .614  & .714    \\
                  & df                      & 202   & 202     & 202   & 202   & 202     \\
Extraversion      & Correlation             & .042  & .059    & -.032 & -.084 & .006    \\
                  & Signif. (2-tailed) & .554  & .399    & .646  & .233  & .930    \\
                  & df                      & 202   & 202     & 202   & 202   & 202     \\
Agreeableness     & Correlation             & .004  & .028    & -.021 & -.060 & .039    \\
                  & Signif. (2-tailed) & .958  & .686    & .770  & .394  & .576    \\
                  & df                      & 202   & 202     & 202   & 202   & 202     \\
Neuroticism       & Correlation             & -.033 & -.048   & -.153 & -.023 & .041    \\
                  & Signif. (2-tailed) & .639  & .495    & .029  & .741  & .557    \\
                  & df                      & 202   & 202     & 202   & 202   & 202     \\
\multicolumn{7}{l}{a Cells contain zero-order (Pearson) correlations.}                  \\ \bottomrule
\end{tabular}
\caption{Pearson's Partial Correlation (controlling variables: Gender and Age}
\label{tbl: PearControllingGenderAge}
\end{table}

\subsubsection{Key Findings and Discussion}

The above tables shows the output of Pearson's Partial Correlation. In Table~\ref{tbl: PearControllingGender}, the controlling variable is \emph{Gender}, a bivariate Pearson's correlation established that there was a strong, statistically significant linear relationship between \emph{Conscientiousness} and \emph{Anger}, ${r(204) = -.141}$, ${p < .05}$ , \emph{Neuroticism} and \emph{Fear} ${r(204) = -.166}$, ${p < .05}$. Pearson's partial correlation showed that the strength of this linear relationship was improved when \emph{Gender} was controlled for  \emph{Conscientiousness} and \emph{Anger} ${rpartial(203) = -.139}$, ${p=0.47}$ and it is still the same between \emph{Neuroticism} and \emph{Fear}  ${rpartial(203) = -.166}$ - ${p=.017}$ and still statistically significant. In Table~\ref{tbl: PearControllingAge}, the controlling variable is \emph{Age}. Pearson's partial correlation showed that the strength of this linear relationship was improved when \emph{Age} was controlled, in respect to the relationship between \emph{Conscientiousness} and \emph{Anger}  ${rpartial(203) = -.138}$ - ${p=0.49}$ and between \emph{Neuroticism} and \emph{Fear} ${rpartial(203) = -.152 - p=0.030}$  and still statistically significant. In Table~\ref{tbl: PearControllingGenderAge}, the controlling variable is \emph{Gender} and \emph{Age}. Pearson's partial correlation showed that the strength of this linear relationship was improved when \emph{Age} was controlled, in respect to the relationship between \emph{Conscientiousness} and \emph{Anger} and \emph{Neuroticism} and \emph{Fear} , ${rpartial(203) = -.138}$ - ${p=0.49}$ , and between and \emph{Neuroticism} and \emph{Fear} ${rpartial(203) = -.152}$ - ${p=0.030}$  and still statistically significant. The above findings suggests that \emph{Gender} and \emph{Age} as controlled variable combined (as per Table~\ref{tbl: PearControllingGenderAge}) would improve the linear relationship between Big Five and Emotions variables specially \emph{Conscientiousness}, \emph{Neuroticism} , \emph{Anger} and \emph{Fear} and improve strength of linear relationship between \emph{Extraversion} and \emph{Anger}, \emph{Disgust}, \emph{Fear}, \emph{Joy} and \emph{Sadness} although the linear relationship was not statistically significant. Those findings are aligned with the output from the Binomial Logistic Regression (see Section~\ref{BioGenderAge}), in the correlation of the \emph{Conscientiousness} and \emph{Age} and impact of \emph{Gender} in improving the association between variables.

\subsubsection{Summary}\label{genderSummary}

In the experiment presented in Section~\ref{bigFiveEmotionsGenderAge}, \emph{Age} was found to improve the association between \emph{Conscientiousness} and \emph{Anger}  and also, between \emph{Neuroticism} and \emph{Fear}. Therefore, \emph{Age} were selected as feature in the proposed model. While, \emph{gender} were found also, to improve the association between same traits, the \emph{gender} variable could not be used in the model as it is built using Neural Network and it is not possible to use \emph{dichotomy} variable as input to the model. However, it is noted for future work.

Based on the previous justification discussed on section \ref{genderSummary} and section \ref{featureSelectionProcess} , the features extracted for the model are \emph{anger}, \emph{disgust},\emph{joy}, \emph{sadness}, \emph{conscientiousness}, \emph{agreeableness}, \emph{neuroticism} and \emph{age}. We build upon these results in the next chapter to develop and refine the conceptual framework for the {\emph{PMSys}} engine.

\subsection{Rationale of using Random Forest Tree}

According to the literature, random forests (RF) were based on decision trees and combined with aggregation and bootstrap ideas and were first introduced by Breiman in 1999~\cite{Breiman2001}. It is a powerful non-parametric statistical methodology; it is working to improve the efficiency with regression problems as well two-class and multi-class classification problems, in a single and versatile framework. In 2015, Erwan Scornet et al. proved the consistency of RF in a paper published on \emph{The Annals of Statistics}~\cite{Scornet2015}. According to a 2014 research survey conducted by Khaled Fawagreh et al. to investigate the RF applications in \emph{ecology}~\cite{Cutler2007}, \emph{medicine}~\cite{Klassen2008}, \emph{astronomy}~\cite{Gao2009}, \emph{autopsy}~\cite{Labeaud2011}, \emph{traffic and transport planning}~\cite{Zaklouta2011}, \emph{agriculture}~\cite{Low2013} and \emph{bioinformatics and computational biology}~\cite{Boulesteix2012}, result shows in that RF has improved to be excellent due to its own characters in classification ~\cite{Fawagreh2014}. According to Zaklouta et al.~\cite{Zaklouta2011}, RFs performed better than K-d trees by improving the classification rate up to 97.2\% and 81.8\%. Other benefits as listed in the original paper about RF~\cite{Breiman2001}:

\begin{itemize}
\item Accuracy is as good as Adaboost and sometimes better;
\item It is faster than bagging or boosting;
\item It gives useful internal estimates of error, strength, correlation and variable importance;
\item It is simple and easily parallelised.
\end{itemize}

Therefore, the literature supports the choice of the RF as the primary classifier for the model. Nevertheless, the mixture of the dataset used as input for the classifier fits well with the best performance required for the RF as previous application reported reasonably similar variables data type (see Section~\ref{RFTree}).

\subsection{Key Findings}

\begin{table}[!ht]
\centering
\begin{tabular}{@{}ll@{}}
\toprule
\multicolumn{2}{c}{=== Summary ===}                     \\ \midrule
Correctly Classified Instances         965 & 68.5856 \% \\
Incorrectly Classified Instances       442 & 31.4144 \% \\
Kappa statistic                            & 0.5811     \\
Mean absolute error                        & 0.1972     \\
Root mean squared error                    & 0.3228     \\
Relative absolute error                    & 52.5898 \% \\
Root relative squared error                & 74.5455 \% \\
Total Number of Instances                  & 1407       \\ \bottomrule
\end{tabular}
\caption{Weka Summary model output}
\label{tbl:WekaSummary}
\end{table}

\begin{table}[!ht]
\centering
\begin{tabular}{@{}lllllllll@{}}
\toprule
\textbf{TP Rate} & \textbf{FP Rate} & \textbf{Precision} & \textbf{Recall} & \textbf{F-Measure} & \textbf{MCC} & \textbf{ROC Area} & \textbf{PRC Area} & \textbf{Class} \\ \midrule
0.716            & 0.079            & 0.752              & 0.716           & 0.734              & 0.648        & 0.9               & 0.829             & Down           \\
0.658            & 0.107            & 0.672              & 0.658           & 0.665              & 0.555        & 0.872             & 0.778             & Error          \\
0.662            & 0.202            & 0.522              & 0.662           & 0.584              & 0.428        & 0.835             & 0.58              & Idle           \\
0.707            & 0.031            & 0.883              & 0.707           & 0.785              & 0.731        & 0.886             & 0.849             & Slow           \\
\multicolumn{9}{c}{Weighted Avg.}                                                                                                                                       \\
0.686            & 0.105            & 0.707              & 0.686           & 0.692              & 0.591        & 0.873             & 0.759             &                \\ \bottomrule
\end{tabular}
\caption{Weka output (detailed accuracy by class)}
\label{WekaOutputDetailedAccuracyByClass}
\end{table}

As showing in Table~\ref{tbl:WekaSummary}, correlation coefficient 0.685856 implies 68.58\% of the variance in your data is explained by the model. Further details reported on Table~\ref{WekaOutputDetailedAccuracyByClass}, \emph{TP rate}, instances correctly classified as a given class show high average value at \emph{down} and \emph{slow} status. \emph{ROC area}, suggested high value across all classes with average weight of \emph{0.759} suggesting a high performance for the classifier.

\section{Model Verification: Observing Emotions in Real Time} \label{lbl:verificationData}

\subsection{Overview}

\begin{figure}[!ht]
\centering
\includegraphics[width=\textwidth,height=750px,keepaspectratio]{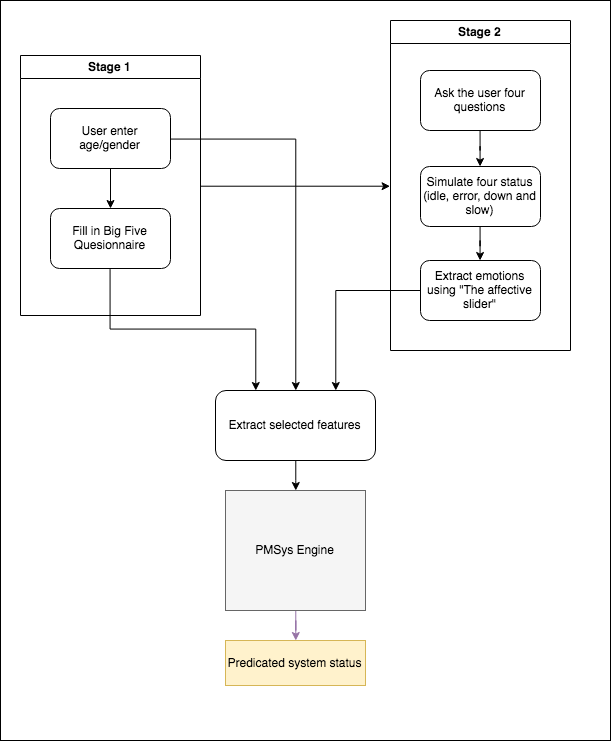}
\caption{Verification stages illustration}
\label{fig:flowchartVerificationProcess}
\end{figure}

The output of the model reported as 68.5\% accuracy using cross fitting same dataset; however, to verify the model performance with a new dataset, a new web-based application developed to collect a data similar to the dataset used in training the model (see Figure~\ref{fig:flowchartVerificationProcess}).

\subsection{Web-Based Verification Tool}

The web-based tool was built using Joomla, PHP, MySQL, JQuery and JavaScript, the main point of the verification model is to replicate the same scholarship system as a simulation to collect a new dataset in same system statues \emph{idle}, \emph{error},\emph{down} and \emph{slow}. The web-based verifier staged to two stages:

\begin{itemize}

\item Stage 1: Big Five Questionnaire;
\item Stage 2: Collect emotion in different system status.
\end{itemize}

In the live simulation all participants were given maximum one hour to complete the experiments, the experiment was robust and were given same conditions.

\subsubsection{Stage 1: Big Five Questionnaire}

\begin{lstlisting}[language=PHP,caption=Model function to calculate score of each personality trait,label={lst:calBigFIve}]
/**
* Function Calculate
* @param: $type_id INT,$test INT,$uid INT
* @param: Return float score of the user
* @throws: none
*/
function calculate($type_id,$test,$uid)
	{
		$db=JFactory::getDBO();	
		$sql='SELECT * FROM big5_elements WHERE type="'.$type_id.'" ORDER BY id';
		$db->setQuery($sql);
		$db->Query();
		$num=$db->getNumRows();
		$rows=$db->loadObjectList();
		$total=0;
		foreach ($rows as $row)
		{
			$score=$this->getScore($row->id,$test,$uid);
			if ($score)
			{


			if ($this->checkQuestionReverse($row->id,$test))
			{
				$score=$this->reverse($score);
			}
			else
			{
				$score=$score;
			}
				$total=$score+$total;
			}
	}
	return $total/$num;
	}
\end{lstlisting}

\begin{figure}[!ht]
\centering
\includegraphics[width=\textwidth,height=400px,keepaspectratio]{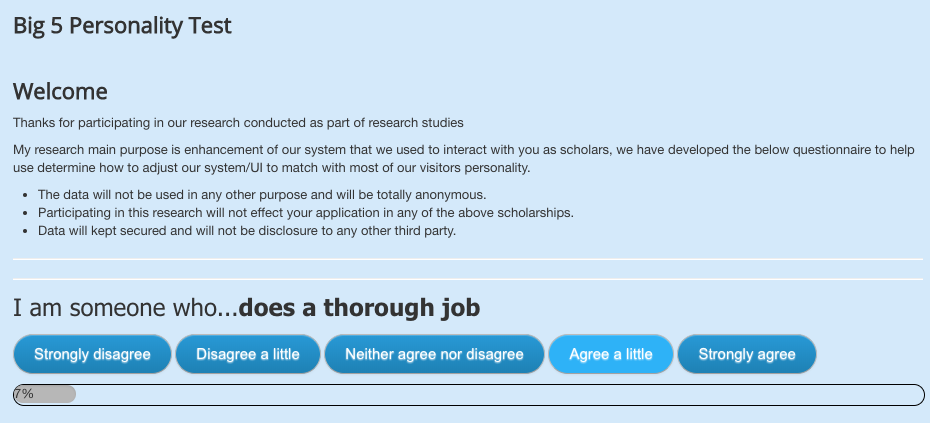}
\caption{Big Five Questionnaire (Web Version)}
\label{fig:webVerifierBigFiveQuestion}
\end{figure}

\begin{figure}[!ht]
\centering
\includegraphics[width=\textwidth,height=400px,keepaspectratio]{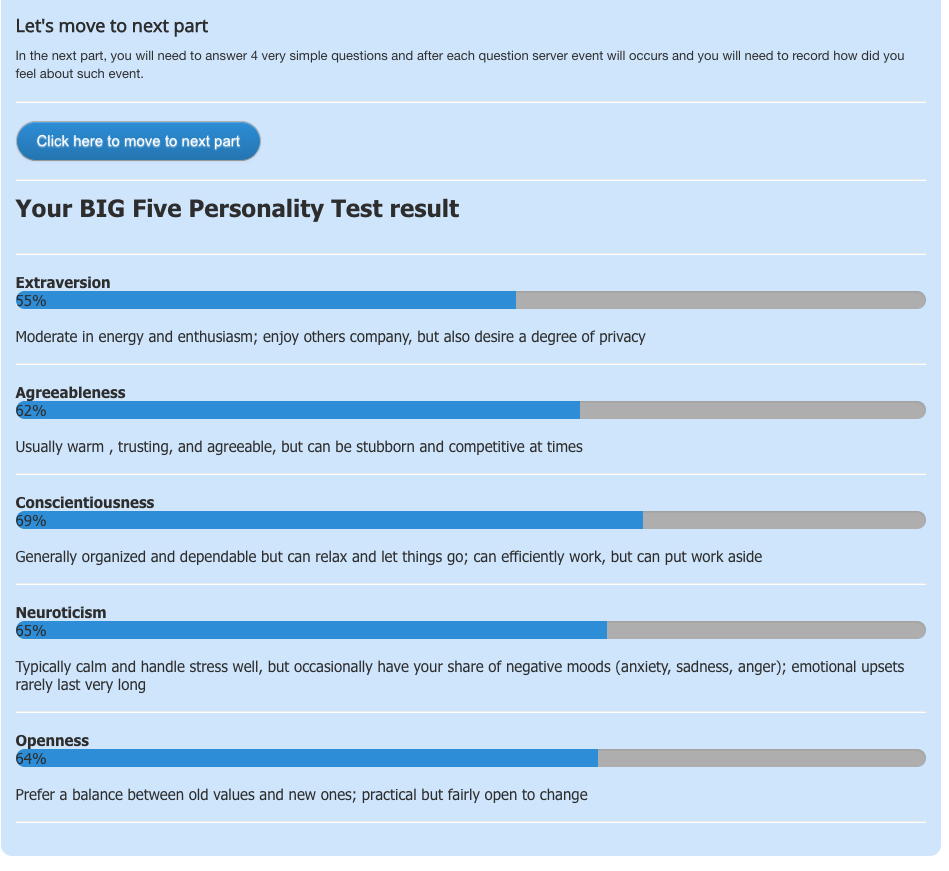}
\caption{Big Five Questionnaire Output sample (Web Version)}
\label{fig:webVerifierBigFiveQuestionOutput}
\end{figure}

As per the discussion in the literature review, the Big Five Questionnaire is used to extract the user's personality by providing the questionnaire in the Annexes; the literature proposed an equation to calculate each trait ~\cite{John1999}. The ``Big Five" questionnaire is developed in a web application to make it accessible to distribute, using PHP, JQuery and MySQL as the database to hold the date. Figure~\ref{fig:webVerifierBigFiveQuestion} is a screenshot of the questionnaire as web-version. Figure ~\ref{fig:webVerifierBigFiveQuestionOutput}, shows the output of the Big Five Questionnaire.

The architecture of the web verifier tool is based on the MVC design pattern to match the original EU scholarship system architecture, the code in Listing~\ref{lst:calBigFIve} refer to the model function responsible in calculating the score of each personality traits~\cite{John1999}.

\subsubsection{Stage 2: Collecting Emotions vs. System Status}

\begin{figure}[!ht]
\centering
\includegraphics[width=\textwidth,height=400px,keepaspectratio]{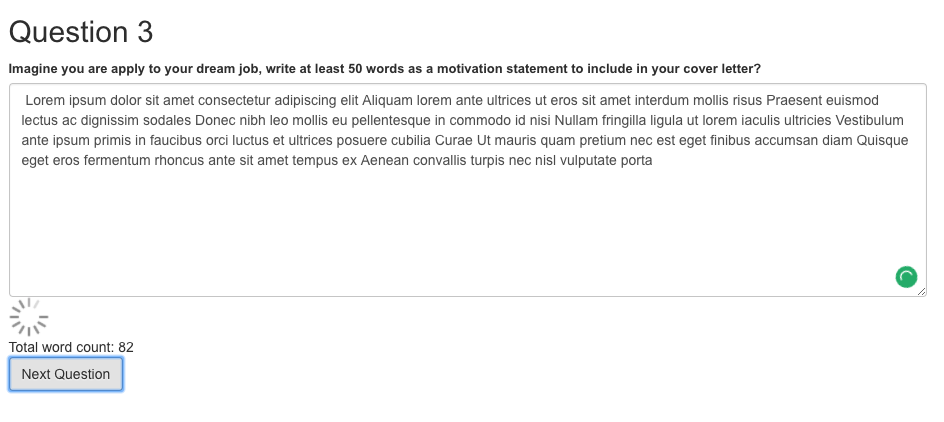}
\caption{Triggering slow event as part of the simulation}
\label{fig:webVerifierEmotionSlowQuestion}
\end{figure}

After storing the personality traits of the user based on the questionnaire, the next step is to collect the emotion from the user in different computer status. A web application has been developed particularly to ask the user four question and triggered a one of the four status  \emph{idle}, \emph{error},\emph{down} and \emph{slow}, after triggering the event a pop message is shown for the user to provide his/her emotion. According to the literature, Alberto et al. (2016)~\cite{Betella2016} suggested that  The ``Affective Slider"  is an away to capture the emotion and reactions of the user, by introducing a novel tool for the measurement of effect. ~\cite{Betella2016}. The similar methodology followed in capture the user's emotion, and reactions to the event occur.

\begin{figure}[!ht]
\centering
\includegraphics[width=\textwidth,height=400px,keepaspectratio]{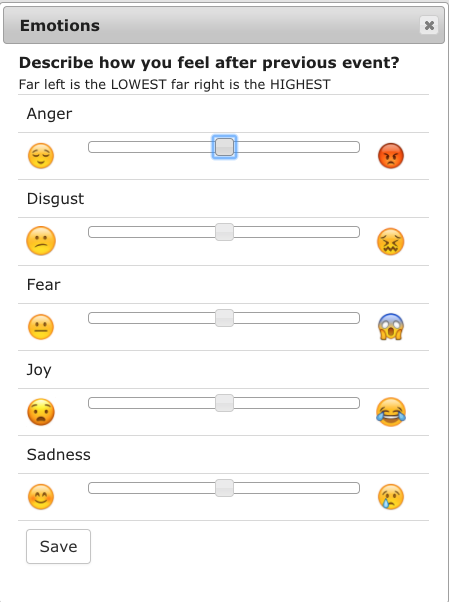}
\caption[Capturing emotions from the user]{Capturing emotions from the user (adapted from ``The affective slider methodology")}
\label{fig:webVerifierCaptureEmotions}
\end{figure}


Figure~\ref{fig:webVerifierEmotionSlowQuestion}, shows the simulation of the \emph{slow} status. After a user asked to complete a question the user is prompted to save and move to the next question. A spinning bar is represented to show that the transmission is slow and that the process is taking more than 10 seconds to save, and for each of rest of system status is a simulation to capture the user's reaction, and emotions. Figure~\ref{fig:webVerifierCaptureEmotions} shows a pop-up message appears after each event to capture how the user felt about the previous event and that based on the affective slider~\cite{Betella2016} supported by the findings in the literature as discussed in Section~\ref{selfassessment}.

\subsection{Dataset Verification}

The data stored in a MySQL database; for each user, the Big Five personality traits extracted from the questionnaire and the emotions alongside with the computer system status. The number of a participant in both stages is 47 participant. The number dropped from initially registered participants and that because some of the participants assumed the system was down for real while triggering the \emph{down}, event and they did not get back to complete the assessment. Table~\ref{tbl:testingData} shows sample of the data used to test the model using Weka tool.

\begin{table}[!ht]
\centering
\resizebox{\textwidth}{!}{%
\begin{tabular}{@{}llllllll@{}}
\toprule
Anger & Disgust & Joy   & Openness & Conscientiousness & Agreeableness & Neuroticism & serverStatus \\ \midrule
0.088 & 0.016   & 0.03  & 0.082    & 0.076             & 0.067         & 0.053       & Error        \\
0.054 & 0.051   & 0.049 & 0.058    & 0.056             & 0.08          & 0.048       & Error        \\
0.061 & 0.054   & 0.04  & 0.058    & 0.056             & 0.08          & 0.048       & Down         \\
0.055 & 0.05    & 0.041 & 0.058    & 0.056             & 0.08          & 0.048       & Slow         \\
0.05  & 0.05    & 0.05  & 0.058    & 0.056             & 0.08          & 0.048       & Idle         \\
0.09  & 0.04    & 0.045 & 0.073    & 0.084             & 0.076         & 0.038       & Error        \\
0.074 & 0.073   & 0.05  & 0.071    & 0.056             & 0.053         & 0.073       & Error        \\
0.069 & 0.031   & 0.05  & 0.058    & 0.067             & 0.06          & 0.07        & Idle         \\
0.05  & 0.05    & 0.05  & 0.058    & 0.067             & 0.06          & 0.07        & Slow         \\
0.076 & 0.031   & 0.064 & 0.058    & 0.067             & 0.06          & 0.07        & Down         \\
0.085 & 0.034   & 0.062 & 0.058    & 0.067             & 0.06          & 0.07        & Error        \\
0.06  & 0.065   & 0.036 & 0.064    & 0.053             & 0.062         & 0.068       & Error        \\
0.03  & 0.02    & 0.068 & 0.064    & 0.053             & 0.062         & 0.068       & Down         \\
0.01  & 0.03    & 0.047 & 0.064    & 0.053             & 0.062         & 0.068       & Slow         \\
0.19  & 0.015   & 0.057 & 0.064    & 0.053             & 0.062         & 0.068       & Idle         \\
0.01  & 0.002   & 0.077 & 0.067    & 0.062             & 0.064         & 0.068       & Error        \\
0.064 & 0.027   & 0.027 & 0.073    & 0.071             & 0.056         & 0.073       & Error        \\
0.075 & 0.063   & 0.018 & 0.073    & 0.071             & 0.056         & 0.073       & Down         \\
0.058 & 0.04    & 0.02  & 0.073    & 0.071             & 0.056         & 0.073       & Slow         \\
0.09  & 0.012   & 0.089 & 0.073    & 0.071             & 0.056         & 0.073       & Idle         \\ \bottomrule
\end{tabular}
}
\caption{Sample of the dataset used in the verification process}
\label{tbl:testingData}
\end{table}

\subsection{Key Findings}

\begin{table}[!ht]
\centering
\begin{tabular}{@{}lll@{}}
\toprule
\multicolumn{3}{c}{\textbf{=== Summary ===}}        \\ \midrule
Correctly Classified Instances   & 115    & 61.17\% \\
Incorrectly Classified Instances & 73     & 38.83\% \\
Kappa statistic                  & 0.4407 &         \\
Mean absolute error              & 0.3055 &         \\
Root mean squared error          & 0.3474 &         \\
Total Number of Instances        & 188    &         \\ \bottomrule
\end{tabular}
\caption{Weka summary - evaluating model using test dataset}
\label{wekaSummaryVerifyTestDataSet}
\end{table}

The Weka explorer was used to verify and evaluate the model against new dataset collected from using the web verifier tool as explained in the previous step, the table below shows the output of the Weka evaluation report, as shown in table ~\ref{wekaSummaryVerifyTestDataSet} the model correctly classified 61.17\% of the data. The dataset consists of 188 instances, that four record for each of the 47 participants. 

\begin{table}[!ht]
\centering
\begin{tabular}{@{}lllllllll@{}}
\toprule
TP Rate & FP Rate & Precision & Recall & F-Measure & MCC   & ROC Area & PRC Area & Class \\ \midrule
0.747   & 0.139   & 0.823     & 0.747  & 0.783     & 0.615 & 0.804    & 0.683    & Down  \\
0.535   & 0.026   & 0.927     & 0.535  & 0.679     & 0.598 & 0.772    & 0.799    & Error \\
0.4     & 0.191   & 0.154     & 0.4    & 0.222     & 0.14  & 0.451    & 0.144    & Idle  \\
0.4     & 0.133   & 0.207     & 0.4    & 0.273     & 0.2   & 0.603    & 0.19     & Slow  \\
\multicolumn{9}{c}{Weighted Avg.}                                                        \\
0.612   & 0.1     & 0.76      & 0.612  & 0.658     & 0.538 & 0.748    & 0.644    &       \\ \bottomrule
\end{tabular}
\caption{Model evaluation - Detailed Accuracy By Class}
\label{wekaDetailedVerifyTestDataSet}
\end{table}

Furthermore, Table~\ref{wekaDetailedVerifyTestDataSet} shows more detailed regarding the performance of the model, the highest value for \emph{ROC Area} is when the status is \emph{down} and the lowest value is when the status is \emph{idle}, however, the average \emph{ROC Area} is \emph{0.644} which shows a good performance for the model. Overall, the \emph{TP Rate}, \emph{FP Rate} and \emph{PRC Area} is considerable low in case of status \emph{idle} and \emph{slow}, which shows the performance of the model in those two cases are not good as expected.

\section{Summary}

This chapter presents the development process of the conceptual model grounded on the features extracted from the flow experiments in chapter~\ref{pmsys}. The model contributes to the identified gap in the literature towards producing a conceptual model in an attempt to digital profile the user in different computer status. The model achieved 68\% accuracy in predicting the system status. For instance, in case of users are posing in social networks (e.g. Facebook, Twitter) or any other median of communication, the PMsys engine will be able to tell (with 68\% accuracy) if the user complaints about the server being \emph{slow}, \emph{down}, \emph{error} or \emph{idle}. In case of idle means it is not relevant to the system issue. As previously stated the engine is currently being part of a KTP project to integrate it with in a intelligent chatbot custom service platform to improve user's communication and first level support.  With emerging approaches to language analysis, it is expected to be able to improve the accuracy of the engine by improving the personality traits and emotions; this will be discussed in further detail in Section~\ref{futureWork}.

\newpage
\chapter{Experimental Meta-Analysis}\label{disc}

\section{Introduction}
As each experiment and model has been discussed and presented separately in the previous chapter, this chapter will give an overview of all findings and discussion. The design of the experiments structure started by exploring the data with reference to the objective of the research question, to investigate the associations and correlations of the current dataset parameter.

\section{\nameref{subs:ProfilingComplex}}
The experiment indicates a strong correlation between the \emph{Openness to Experience} traits and the \emph{Accepted} group and that agrees with the literature. Costa~\cite{costa+mccrae:1992} and Srivastava~\cite{John1999} described people with high \emph{Openness to Experience} as wanting to learn and explore new ideas, and creative in general, and that agrees with the outcome of the experiment presented in Section~\ref{subs:ProfilingComplex} conducted in our dataset. However, the final selection and evaluation process in the scholarship system \footnote{The system used to extract the data}, had many other parameters such as (Evaluator personality, language qualifications, academic background..etc) and that was the main reason why the \emph{Openness to Experience} was not selected as part of the final developed model. Also, that was the reason why the \emph{final selection} parameter was not selected as part of the input to the model.

\section{\nameref{subs:sentimentStages}}
The experiment presented by dividing the usage of the system over stages to investigate the changes of the personality traits over time, the outcome, suggested that \emph{Extraversion}, \emph{Agreeableness} and \emph{Conscientiousness} were statistically significant in the an attempt to calculate the probability of predicate the stage, the logistic regression model, successful correctly classified 80.2\% of the cases., which indicate the personality traits can play a vital role in profiling the digital behaviour of a user in using the complex system. Furthermore, as indicated in a study conducted by McCrae in 2002~\cite{McCrae2002}, agrees with the outcome of the experiment that there is no change of personality over time, however, the experiment indicates a possibility to identify the personality of the user in the system based on his/her sentiment.

\section{\nameref{releaseIBM}}

IBM Watson was made available in 2014, to extract personality traits and emotions from text (as initially discussed in Section~\ref{IBMWatson}), the output reported was encouraging to use as part of the study, however, it was essential to verify the IBM Watson tool result without current Java tool (see Section~\ref{Javatool}). The experiment presented in Section~\ref{releaseIBM} was conducted to verify the result using current collected Big Five Questionnaire and Emotions reported by users, two methods used to investigate the difference between the two groups \emph{IBM Watson} and \emph{Data collected from the Questionnaire}, \emph{Independent samples of t-test} and \emph{The Mann-Whitney U test}, both confirmed that there is no significant statistical difference between both groups, therefore, in further experiments IBM Watson has been used as a tool to extract personality traits and emotions.

\section{\nameref{bigfiveEI}}
As per our review of the literature in Section~\ref{cognitiveScience}, there are two commonly used approaches to capture emotions \emph{Emotions Lexicon} and using \emph{temporal behaviour}. The experiment presented in Section~\ref{bigfiveEI} suggested a moderate correlation between \emph{Neuroticism}, \emph{self control} and \emph{well being}.  \emph{Well being} has a moderate correlation to \emph{agreeableness} and \emph{conscientiousness}. Furthermore, the experiment presented in Section~\ref{bigfiveBasicEmotions} also suggest a moderate correlation  between \emph{fear} and \emph{Conscientiousness},\emph{Neuroticism}, and moderate positive correlation between \emph{joy} and \emph{Conscientiousness}, \emph{Neuroticism}. However, it was decided to move forward with the \emph{Emotions Lexicon} approach as the main source of extracting emotions from the text as demonstrated in the literature in Section~\ref{IBMWatson}, the main source of data as part of this study is in the form of interactive text. The temporal behaviour experiment in Section~\ref{bigfiveEI} had to be conducted to compare between both approaches in case there is a signification statistical correlation, and however, since both reported reasonably same consistent result, then it is decided to move forward with basic emotions.

\section{\nameref{subs:sentimentStages}}
The previous experiment reported an encouraging result to in the association between sentiment values and the ``big five" personality traits. The classifier accurately identifies 80.2\% of the cases (See the section: Section~\ref{subs:sentimentStages}), which contributes to the understand more about user's behaviour in the system. Therefore, it was essential to extend the experiment to include \emph{Emotions lexicon} approach to investigate a further association. The Ordinal Regression suggested used to investigate further association, because of the continues variables and dependent variable the \emph{Stage ID}. The \emph{disgust and sadness} emotion had a statistically significant effect on the performance of the classifier. Multinomial Regression has been used to verify the output of previous of analysis, and it reported the variables \emph{Disgust, Sadness and Conscientiousness}  has a significant overall effect on the dependent \emph{Stage ID}\footnote{StageID: System divided two four stages to investigate the change of emotion in different system stages.} The suggested output from this experiments shows a good potential towards including other forms of system behaviour (I.e, \emph{server status}).

\section{\nameref{emotionSystemStages}}
In light of building a conceptual framework to improve user experience and computer system architecture design, the previous flow of experiment demonstrated written text reflects more than words agreeing with the literature ~\cite{pennebaker+king:1999}, emotions captured correlate with the activity of the user in the system (see Section~\ref{emotionSystemStages}). Moving towards developing a conceptual framework to predicate the system status based on user's behaviour. It was suggested to integrate the system status \emph{idle, down} with the personality traits and emotions extracted as explain in the experiment presented in Section~\ref{Emotionscentre}, the output suggested a strong correlation in predicating the server status as following personality traits \emph{Openness}, \emph{Extraversion}, \emph{Conscientiousness}, and \emph{Neuroticism}, and from basic emotions  \emph{Joy}, \emph{Sadness}, \emph{Anger} and \emph{Disgust}, and that confirms the literature founding as reported by Fast Funder in 2008, that it is possible to include cognitive science in the process of identifying the digital identity of the users~\cite{fast+funder:2008} and other studies~\cite{Hampson2012,ekman2013emotion,lambiotte+kosinski:2014,gou2014knowme,lambiotte2014tracking}. A model was built using \emph{Linear Discriminant Analysis} using the below-reported traits to predicate the system status \emph{idle, down}, and with good potential performance 75\% accuracy evaluated using cross-validation.

\section{\nameref{bigFiveEmotionsGenderAge}}
The literature suggested that \emph{gender} and \emph{age}, plays a vital role in personality traits and emotions, according to Weisberg in his study published in 2011 and Donnellan in 2008. ~\cite{Weisberg2011} ~\cite{Donnellan2008}, the findings of the experiment presented in Section~\ref{bigFiveEmotionsGenderAge}, agrees with the literature as the  The above findings suggests that \emph{Gender} and \emph{Age} as controlled variable improves the linear relationship between Big Five and Emotions variables specially \emph{Conscientiousness}, \emph{Neuroticism} , \emph{Anger} and \emph{Fear} and improve strength of linear relationship between \emph{Extraversion} and \emph{Anger}, \emph{Disgust}, \emph{Fear}, \emph{Joy} and \emph{Sadness}. However, in the feature selection process as presented in Section~\ref{featureSelectionProcess}, the \emph{gender} has been excluded from the features as it is not possible to include a dichotomy variable part of neural network.

\section{\nameref{lbl:verificationData}}
The developed model conceptual framework produced an accuracy of 68.5\% and evaluated using a new dataset in the verification process and produced an accuracy of 61.17\%, the numbers suggested a good start into continue researching in this direction to produce better performance by investigating a different other ranges of parameters that can contribute the the skeleton of the model, however, I believe the objective of this study is met by producing a good model supported by literature review to towards profiling the digital behaviour in the different system status.



\section{Summary}\label{discussionSummary}

We have summarised all of the experiments in order, based upon our robust design methodology. Furthermore, the dataset used as part of this study was extracted from a system that was not pre-defined for this study. Therefore, the initial stage was a range of exploratory statistics analysis applied to decided how to approach the data to achieve the objectives. Moreover, the in-depth literature review supported the direction of the methodology and identified the gap in the digital profiling of the user over a complex system using the personality and emotions as the main fundamental parameters to build the stricture of the experiments. The primary idea was to identify the features that may have any association with the key objectives of the study. The nature of the \emph{EU System} used here involves different parameters that may have a direct impact on the data, for instance, the \emph{Final Selection} parameters. It was important to investigate the impact of this parameter in the behaviour of the system and the personality type. However, the selection criteria and lack of enough data on the final selection process prevented the selection of this parameter, which moves to the future work plan. Further analysis revealed that the ``Big Five" personality traits  \emph{Conscientiousness}, \emph{Agreeableness} and \emph{Neuroticism} and the emotions namely, \emph{Disgust}, \emph{Joy} and \emph{Sadness} with the \emph{Age} associate together to build a model based on \emph{Random Forest Tree} predict the \emph{server status} with accuracy of 68\%. The real-time verification process of the model confirmed that model accuracy with 61\%. The accuracy provided at this stage is encouraging to investigate further by expanding the parameters involved in the \emph{human-computer interactions} life cycle. However, the {\emph{PMSys}} pilot version is currently integrated into a commercial industrial application to enhance the selection of responding in an intelligent agent and chatbot-based service, and the output of this integration will be used as a stimulus for further research and development activities. The following chapter will frame these challenges through an overall project summary, well as the potential future work..

\newpage
\chapter{Conclusions and Future Work}\label{conclusion}
\section{Main Conclusions}

The increased usage -- and wider impact -- of social networking platforms on our daily life has provided the motivation and foundation for developing a new conceptual model and thus deeper insight for profiling the types of users using these platform. by understanding the personality and emotion raised while using the system. Leading to signification improvement in the architecture of the complex computer system not only the design, but it is also delivering the information to the users. 

Reflecting on my experience as a professional software engineer for more than ten years, the motivation and underpinning rationale of this study was to provide more insight into how to improve the user experience and the architecture design of complex systems. It is always a challenge for the software developer to read how the system developed will help the user and how the user will behave in a different stage. This study provides the software developers and UX communities with a conceptual model to answer this question based on cognitive science, how the user will behave in different stages.

In this thesis, the work presented led to the development of a conceptual framework for predicting \emph{system status} based on \emph{personality traits} and \emph{emotions} captured from the interactive text with reasonably good accuracy. The broad classification of the literature survey conducted as part of the thesis demonstrated a potential led to the categories proposed for understanding gestures as a human interaction technique and was developed to enable us to gain a more theoretical perspective on the field of gesture interactions.

The experiments conducted during this research project has revealed encouraging findings at the intersection of cognitive science, human-computer interaction, psycholinguistics and user experience. Exploring the dataset extracted from the system triggered the flow of the experiment suggested in this study coupled with the literature. The literature supported that the emotions are dynamic factors were it could affect events and surrounding environments; researchers suggested that the weather can have a positive/negative impact on the emotions of the person. A recent study revealed the weather has a direct impact on people emotions; this is because the weather is part of the daily life routine. The structure of this experiment built on this bases, as the computer/mobile is currently part of our daily-life routing. It is crucial to investigate the relationship between the emotions and the digital behaviour of the user, the findings suggested from the experiments conducted is there is a very positive relationship between personality traits of the users and the emotions raised in the different stages during the usage of the system. Sentiment analysis has been used to identify the association between the personality of the user and the change in sentiment in different stages in using the system and was found to have a strong statically correlation, which led to expanding the investigation to include more emotion variables.  Another objective is to investigate the personality and motions association as part of this study. The findings revealed that there is a moderate correlation between personality traits and emotions, for instance, people with high \emph{well being} more likely to have \emph{agreeableness} and \emph{conscientiousness}. Furthermore, the literature suggested that \emph{age} and \emph{gender} correlates with the personality and emotions of people with a strong correlation and that was confirmed by the findings and included in the final produced model. The suggested conceptual model shows that it is possible to profile the digital behaviour of the user to the usage of the system. Furthermore, the proposed model produced to demonstrate that the system status raises emotions of the users to the personality type, gaining such information would benefit the software development community to take into account how different user with different personality will behave in different system status.

\section{Limitations}
This study involves extracting data from different sources, \emph{Facebook}, \emph{EU Scholarship system}, \emph{Help desk} and the \emph{Verification tool}; extracting the data from these sources had various challenges. Facebook constantly updates its API and permission model, making it hard for researchers to harvest data for research purposes, these changes and restrictions affected this study to explore more parameters from the user's profile. It is expected to be harder next upcoming years especially with the major data breach occurs in March 2018, involves 50 million users, causing Facebook to act accordingly and add more restriction and changes to extract data from Facebook even with user's consent. One more challenge was the \emph{EU Scholarship} system was developed in-house with not enough documentation of the flow of the data or database structure to investigate further interactions and parameters to include it in the model. Furthermore, the system did not include \emph{Login using Facebook\cite{ko2010social}} developed by EU team, therefore, it was essential to develop a matching algorithm as explained on section \ref{matchingAlgorithm}, however, the output of the matching algorithm led to excluding of 35\% of the available data set. Limitation of the Computer system status to \emph{idle}, \emph{down}, \emph{slow} and \emph{error} due to the lack of documentation to the \emph{EU Scholarship System} and there was no pre-defined installed plugin to monitor the user's behaviour, which limited the system's parameters as part of the engine.






\section{Commercialisation}

Further to the peer-reviewed publications presented in Section~\ref{mypubs}, aspects of this research (and thus the {\emph{PMSys}} system) is currently being commercialised as part of a two-year Innovate UK-funded Knowledge Transfer Partnership (KTP) project to apply the research in industry with a company based in Cardiff focusing on developing intelligent chatbot technologies for customer service. The proposed model is currently being integrated into the main chatbot engine to enhance the selection of the statements based on the user's mood and provide the agent with a possible system issue based on the user's digital behaviour. The company offers the chatbot services to various industry and one of the common requirements is to handle the technical inquiries and integrate and as the model is suggested to improve the user experience and usability it can save some time for the first line support by producing the potential problem with the user with an option for the agent to manually override the conversation and take over.


\section{Future Work}\label{futureWork}
The last five years of research for this project has opened up a wide range potential future research areas, especially using our model and the {\emph{PMSys}} system to further explore how personality traits and emotions can play a significant role in software development and user experience while using a complex system. These future work themes are as follows:

\begin{description}
\item[Exploring other software environments:] The output model was built in a dataset retrieved from a web complex system, it is suggested to integrate to another software environment to explore the change in emotions in different system status, such as offline applications, mobile applications, tablet applications, etc.

\item[New and larger datasets:] The recent events occurs regarding Facebook, suggested to explore other social networks (e.g. Twitter, LinkedIn) to extract a new dataset, number of companies are using Twitter as a main context of the communication between their customers and commonly used to report technical issues, Twitter offers a rich API~\cite{makice2009twitter} makes it a productive environment for researchers to collect and analyse large-scale longitudinal datasets~\cite{albishry-et-al:iccci2017,albishry-et-al:ssei2018,albishry-et-al:iccci2018}. Furthermore, since Twitter is largely an open, public platform the data can be used in investigating further and verifying the model outcome and improving the accuracy of the model.

\item[Topic modelling:] A number of research studies suggested a strong association between personality traits, emotions and topics~\cite{andre2000integrating,austin2008associations}. In upcoming studies, it is suggested to integrate topic classification as one more parameter to verify the potential association.

\item[Focus experiments:] This study were focused on data collected from Facebook and EU System, with some limitation to the type of the data collected as explained previously, to continue working in same direction it is important to collect data in more controlled environment to expand the input variables (i.e, facial expression, video observation, pre-defined system to monitor behaviour, eye tracker). 

\item[Deeper analysis of personality traits throgh emotion extraction] with emerging NLP tools and technologies, different companies are producing a NLP tool to extract personality traits and emotions from text. It is suggested to track the state-of-art tools and verify the efficient to use it in further analysis. Furthermore, extracting emotions emerged to include facial expression with high efficient quality tools (e.g. Affectiva\footnote{Affectiva: is a solution for massive scale engagement detection. Offer SDKs and APIs for mobile application, and provide an analytics API to track expressions over time - http://www.affectiva.com/}, Emotient\footnote{Emotient: is an ad campaign that tracks attention, engagement, and sentiment from viewers. The RESTful Emotient Web API can be integrated into application - http://emotient.com/} and EmoVu\footnote{EmoVu: is a facial detection products incorporate machine learning and micro expression detection. Provides a very powerful SDK, Mobile SDK, and an API for application integration - http://emovu.com/e/}).

\item[Expanding the system status functionality:] as the current study suggested only four status of the system \emph{idle}, \emph{down}, \emph{slow} and \emph{error}, it is planned to explore further sub types for each main classification, such as \emph{Error 404 not found}, \emph{Internal 500 server error}.

\item[Investigating the final selection parameter:], as discussed in Section~\ref{discussionSummary}, based on the outcome analysis from the profiling complex experiment ~\ref{subs:ProfilingComplex}, it is suggested to investigate the process of selection further to understand more about the personality of each group.

\end{description}


\bibliographystyle{abbrv}
\bibliography{thesis}

\end{document}